\documentclass{article}

\usepackage{a4wide}
\usepackage{amsmath}
\usepackage{booktabs}
\usepackage{graphicx}
\usepackage{float}
\usepackage{subcaption}
\usepackage{longtable}
\usepackage[numbers]{natbib}
\usepackage{placeins}
\usepackage{xurl}
\usepackage{authblk}

\title{Adaptive rerouting reshapes impacts of maritime chokepoint disruptions}
\author[1,2,*]{Mitja Devetak}
\author[3]{Jasper Verschuur}
\author[1,2,4,5]{Peter Klimek}
\affil[1]{Complexity Science Hub, Vienna, Austria}
\affil[2]{Supply Chain Intelligence Austria, Vienna, Austria}
\affil[3]{Faculty of Technology, Policy and Management, Delft University of Technology, Delft, the Netherlands}
\affil[4]{Institute of the Science of Complex Systems, CeDAS, Medical University of Vienna, Vienna, Austria}
\affil[5]{Division of Insurance Medicine, Department of Clinical Neuroscience, Karolinska Institutet, Stockholm, Sweden}
\affil[*]{Corresponding author: devetak@csh.ac.at}
\date{\today}

\begin{document}

\maketitle

\begin{abstract}
Maritime chokepoints concentrate shipping traffic. Disruptions to this traffic can have a widespread impact on the global economy. However, the way in which these impacts are shaped by the shipping sector's adaptive behavior is not well understood. Here, we introduce an empirically calibrated full-scale agent-based model of the global commercial shipping fleet, representing 35,954 active ships moving among 1,651 ports. We use the model to quantify how rerouting changes arrival losses under chokepoint closures. Static route exposure alone does not predict realized losses. In the adaptive model, rerouting reduces losses at some directly exposed ports, while delayed vessel cycles create losses at later port calls and in dependent regions. Cumulative net shipping-day losses therefore continue to rise with closure duration because longer routes keep ships delayed after the initial adjustment. Each additional closure day reduces global shipping arrivals by $3.0\%$ for Suez and $7.7\%$ for simultaneous Suez, Panama, and Malacca closures. These losses are unevenly distributed in exposed regions and ports. Disruptions with known duration show different loss profiles from unexpected shocks with unknown duration, revealing that end-date information can reduce avoidable short-run losses. The results show that chokepoint risk is a dynamic problem of routing, timing, and regional exposure and not a static property of maritime-network topology.
\end{abstract}

\section{Introduction}

Global shipping is a complex adaptive system of thousands of individual ships making port, route, and operational choices across space and time \cite{caschili_review_2012,kosowska-stamirowska_network_2020}. It is optimized to move goods to serve global value chains at low cost and high frequency. Naturally, it concentrates in a limited set of ports and maritime chokepoints \cite{verschuur_ports_2022,verschuur_systemic_2024}. These maritime chokepoints, such as the Suez Canal, Panama Canal, and Strait of Malacca, are central network components: closing one of them changes travel times, queues, and arrival patterns across dependent routes and ports \cite{verschuur_systemic_2024,qu_modeling_2024}. Recent events, such as the blockage of the Suez Canal in 2021, the drought affecting the Panama Canal in 2023–24, attacks by the Houthi rebels in the Red Sea, and heightened geopolitical tensions in the Strait of Hormuz, have had a significant impact on the shipping network.

Past events have shown that the effects of closing a maritime chokepoint cannot be inferred from topology alone. Topology does not capture the fact that rerouting ships across different chokepoints helps to mitigate the direct effects, while transport interdependencies can spread the impact to ports and regions that are not directly dependent on the chokepoint. Here we present a model to study such behavior. Prior complex network studies of maritime transport have largely focused on structural exposure, including port criticality, route dependence, network vulnerability, and chokepoint risk under node (port) or edge (route) disruption \cite{calatayud_vulnerability_2017,kojaku_multiscale_2019,larock_path-based_2021,wen_exploring_2022,verschuur_ports_2022,verschuur_systemic_2024}. This strand of literature identifies where disruption risk is concentrated, but generally treats the maritime network as fixed or with simplified routing assumptions. It therefore cannot fully capture the adaptive behavior of ships after a shock, and the system-wide implications of those decisions \cite{huang_modelling_2018,huang_hub-and-spoke_2022,qu_modeling_2024}.

A second strand of literature models maritime operations with richer agent behavior or through optimization methods, including port-level nautical services, routing, congestion, and empty-container repositioning \cite{henesey_agent-based_2003,fransen_empirical_2021,xiao_nautical_2013,guo_network_2023,abdelshafie_simulated_2023,sirait_selection_2023}. While these models often include more operational detail, the associated problems of routing, scheduling, and repositioning are difficult to scale to a larger number of ships. However, this is necessary in order to study the system-wide impact of disruptions at maritime chokepoints.

Recent work has also linked delayed cargo derived from ship tracking data to adaptive supply chain agents. However, in these models, the maritime shock mainly enters as delayed, blocked cargo, rather than as rerouting by a global fleet of ship agents \cite{qu_modeling_2024}. Existing global agent-based model (ABM) approaches have therefore often relied on hard-coded routes or static behavioral assumptions \cite{muller_decision_2018,tierney_liner_2019,sinha-ray_container_2003}. What is currently lacking is a global understanding of how maritime logistics responds to external disruptions in a way that combines ship-level empirical destination choices and adaptive rerouting decisions after network shocks, and the resulting implications for system-wide arrival timing at ports. 

\begin{figure}[t]
\centering
\setlength{\tabcolsep}{8pt}

\begin{tabular}{cc}

\subcaptionbox{Model schematic\label{fig:model_response_A}}[0.37\linewidth]{
  \includegraphics[width=0.95\linewidth]{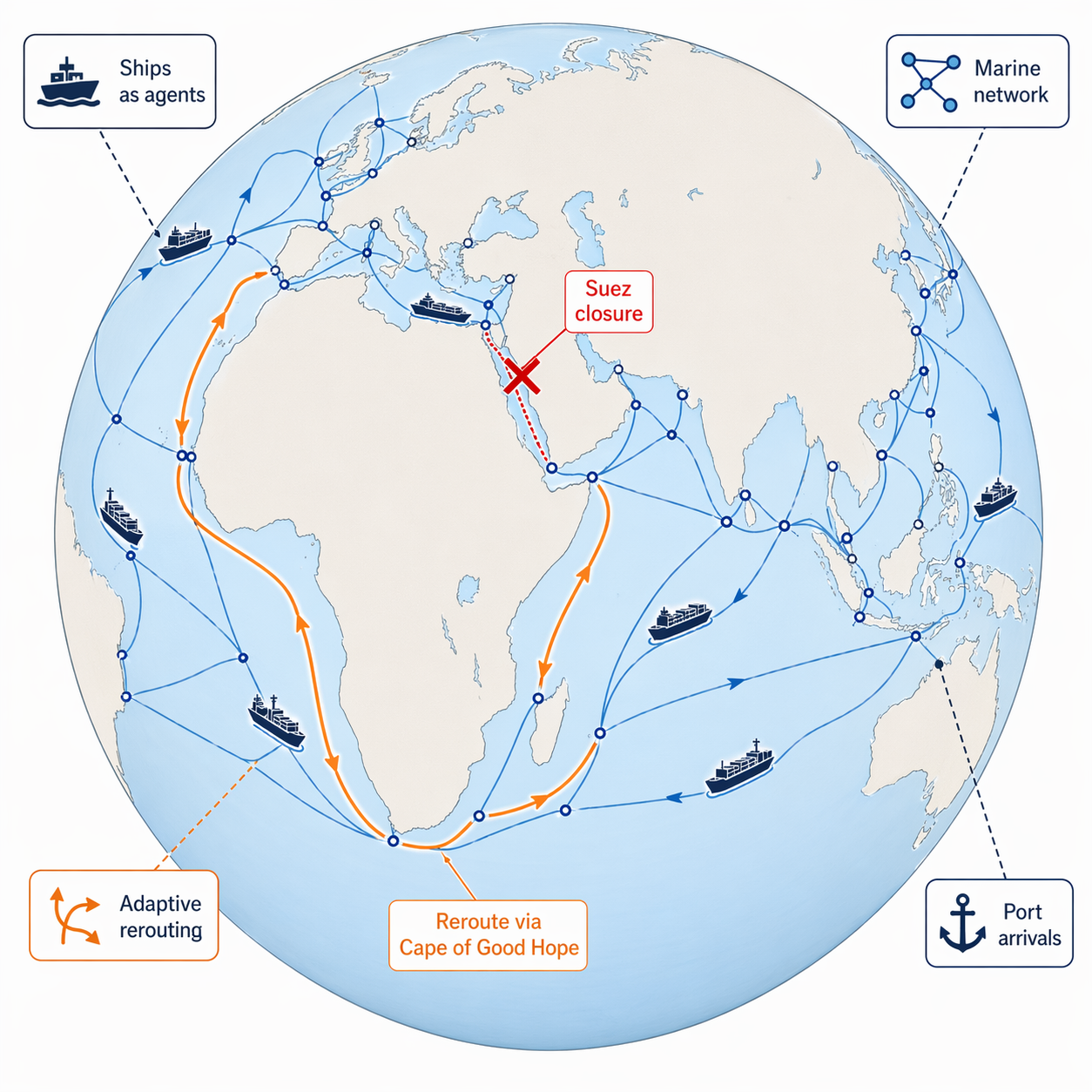}
}
&
\subcaptionbox{Next port choice model\label{fig:model_response_B}}[0.59\linewidth]{
  \includegraphics[width=0.95\linewidth]{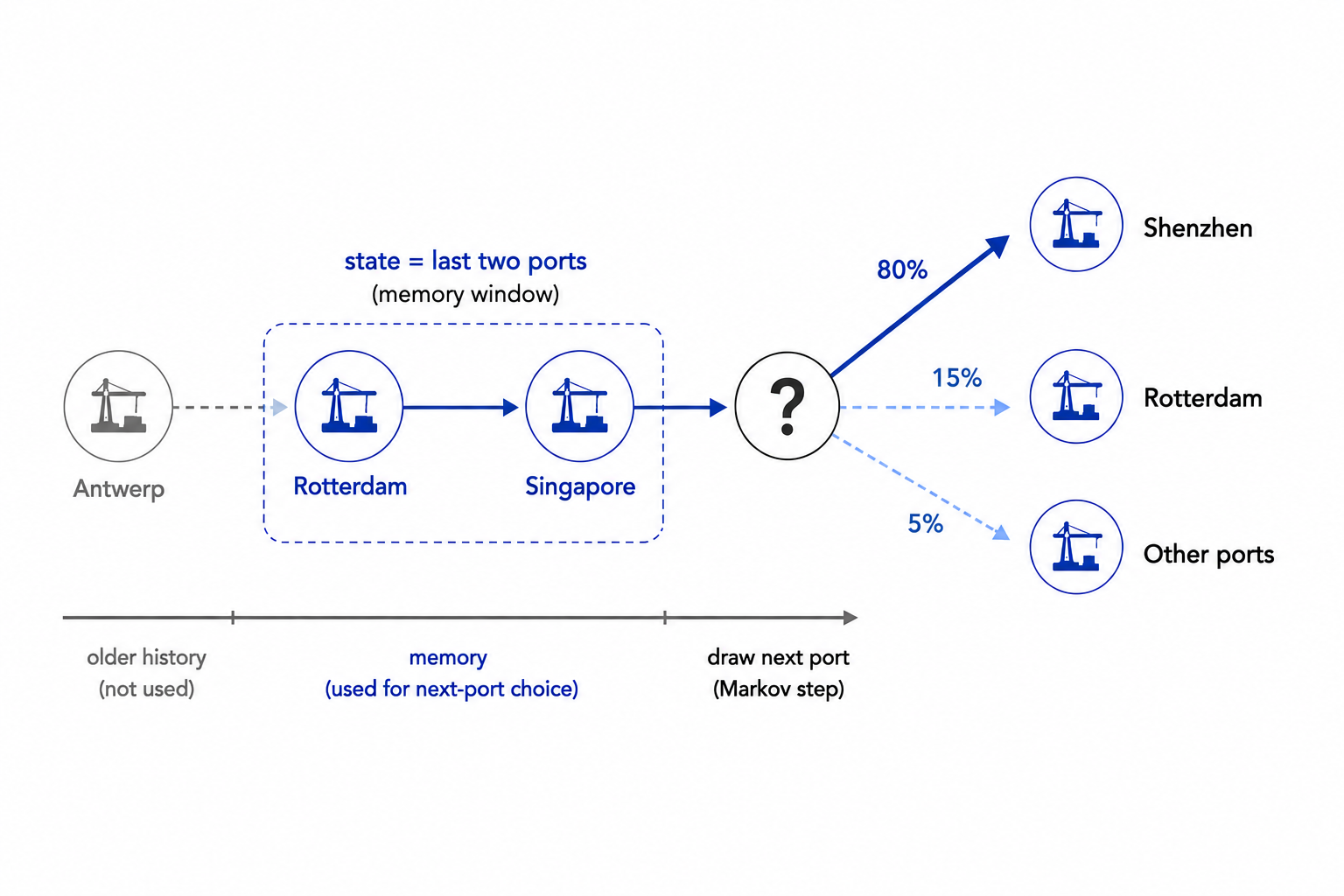}
}

\end{tabular}

\caption{Adaptive ship-level model. (A) Ships move on a weighted marine network between ports and recompute feasible routes when chokepoints close. The schematic illustrates rerouting around a blocked corridor and the port-service process. (B) Next port choice is generated by an empirical two-step Markov model. Each ship samples its next destination from transition probabilities conditional on its two most recent ports.}
\label{fig:model_response}
\end{figure}

Here, we aim to study the adaptive behavioral responses of global shipping by using a newly developed one-to-one agent-based shipping model. A schematic of this model is presented in Fig.~\ref{fig:model_response}A. In our model, the choice of destination is estimated using a Markov chain model that has been calibrated using observed port call sequences \cite{guo_trajectory_2018,marten2020scalable,spadon_learning_2025}. Ship-level agents recompute routing decisions after network changes and periodically while en route. This preserves empirical routing regularities while enabling ships to adapt when chokepoints close.

Our ABM can be used to model a wide range of disruptions at multiple chokepoints or ports. We illustrate the model by studying a set of chokepoint disruptions, specifically Suez, Panama, and Malacca. We vary closure duration from 1 to 50 days. To stress-test the system, we further consider extreme scenarios in which combinations of chokepoints are closed. Finally, within our framework, we test different information regimes regarding the start and end time of the disruption. We also examine a scenario involving closures of different durations at the Strait of Hormuz.

Our results suggest that adaptive capacity plays a key role in the short-term response to disruption in the global shipping network. Within this model, peak daily arrival shortfalls are limited after ships reroute and are weakly dependent on the duration of the closure. However, cumulative losses continue to grow because alternative routes are longer. The adaptive model realization differs from a static variant in two directions: rerouting can reduce direct losses on exposed routes or ports, but delayed vessel cycles can shift losses into later port calls and dependent regions. Co-occurring closures increase losses; however, for the reroutable chokepoint scenarios studied here, their effect is approximately additive. Together, these results reframe maritime chokepoint risk as a dynamic issue involving adaptive routing, arrival scheduling and regional vulnerability, rather than merely a static feature of network topology.

\FloatBarrier
\section{Results}

\subsection{A one-to-one model for adaptive shipping responses}

We developed a global shipping ABM to study disruptions of global maritime trade, see Fig.~\ref{fig:model_response}A for an illustration. In the model, ships choose their next port based on their most recent history of ports, see Fig.~\ref{fig:model_response}B, as calibrated from empirical data. The ABM represents 35,954 active ships and tracks arrivals in 1,651 ports over time.

The model is parsimonious in its modeled behavior (see the Appendix for a discussion on its limitations), yet reproduces observed mean daily port calls across ports well, with Pearson correlations of $r=0.990$ for the full fleet, $r=0.996$ for cargo ships, $r=0.971$ for dry bulk vessels, and $r=0.953$ for tankers. We further validate the model based on the 2021 Ever Given Suez Canal blockage, which can be interpreted as a natural experiment for the type of scenarios studied here. Across all ports, model predictions for the relative change between arrivals observed over 20-day pre- and post-blockage windows correlate with the observed changes with Pearson's $r=0.65$. All details are provided in the Appendix.

We measure disruption impacts using daily completed port calls at destination ports, which we refer to as arrivals throughout the main text \cite{cerdeiro_world_2020,kim_economic_2023,tumbarello_big_nodate,international_monetary_fund_nowcasting_2025}. This output captures realized maritime transport into the port system: when a ship waits, takes a longer route, or queues before service completion, the effect appears as a reduction or displacement of arrivals relative to normal traffic. We normalize each port or region by its pre-shock mean daily arrivals, so that a value of one denotes baseline traffic and deviations from one are comparable across scenarios. For every simulated vessel, the model returns its ordered port-call sequence, including the arrival and departure time at each port.

In this study, we operationalize port-level resilience primarily using two metrics. The first is maximum arrival shortfall, which measures the largest one-day normalized deficit in completed port calls relative to the baseline. It captures the worst day of the disruption in terms of ship arrivals. The second is net shipping-days lost, which measures the cumulative normalized arrival deficit over the evaluation window, net of later catch-up arrivals. For example, for a port with $100$ arrivals on an average day, a maximum arrival shortfall of $20\%$ means that the worst day had $80$ arrivals. A net shipping-days lost value of $2$ means that, after accounting for later surplus arrivals, the port missed the equivalent of two normal days of traffic, or $200$ arrivals.

\begin{figure}[t]
\centering
\setlength{\tabcolsep}{8pt}

\begin{tabular}{cc}

\subcaptionbox{Normalized arrivals\label{fig:model_example_A}}[0.37\linewidth]{
  \includegraphics[width=0.95\linewidth]{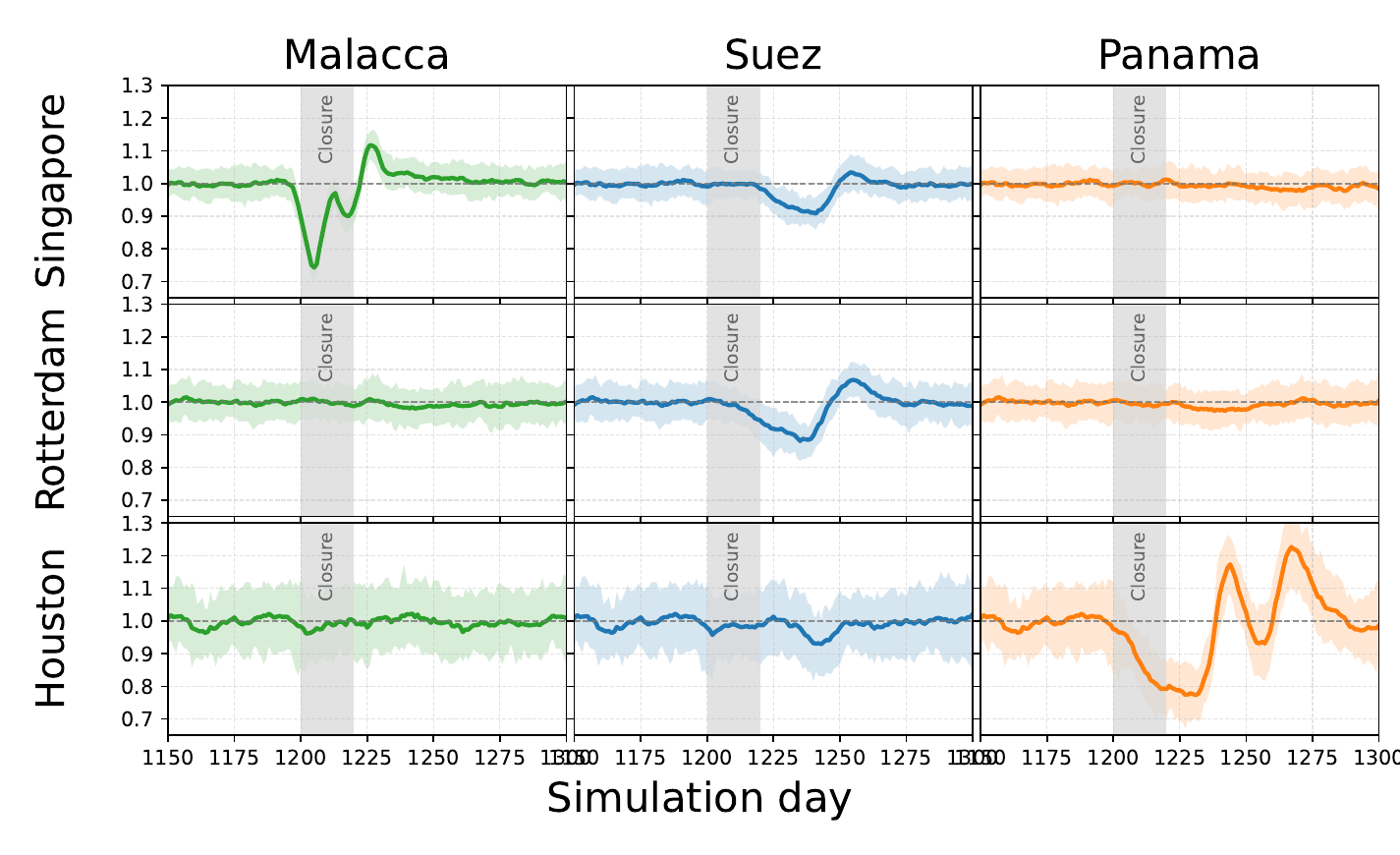}
}
&
\subcaptionbox{Global losses under Panama closure\label{fig:model_example_B}}[0.59\linewidth]{
  \includegraphics[width=0.95\linewidth]{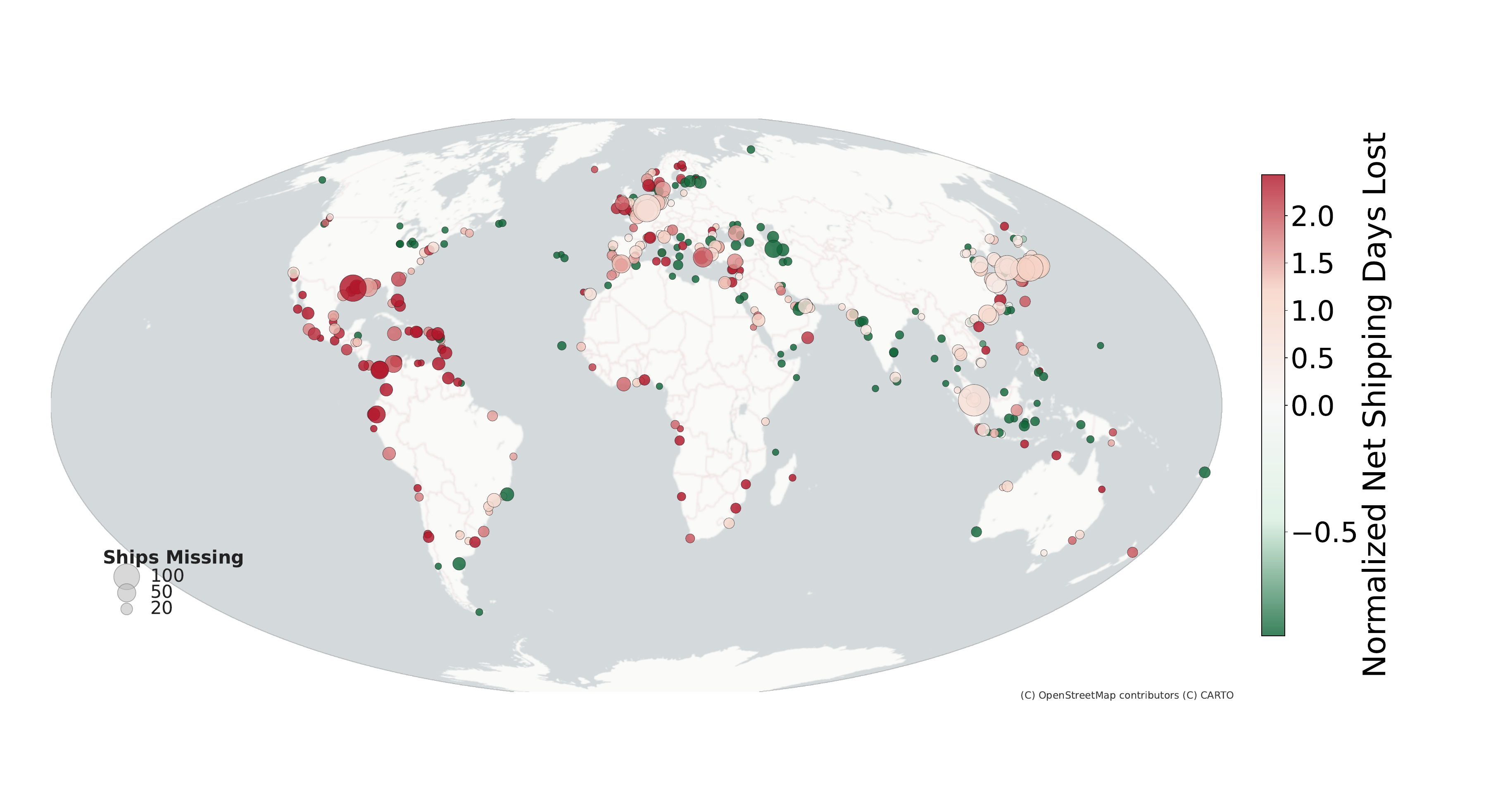}
}

\end{tabular}

\caption{Illustrative effects of single-chokepoint closures. (A) Normalized arrivals for Singapore, Rotterdam, and Houston under 20-day closures of Malacca, Suez, and Panama. The shaded interval marks the closure period, and values below one indicate fewer completed port calls than baseline. (B) Port-level net shipping-day loss under a 20-day Panama closure. Marker size is proportional to total missed arrivals, and color indicates signed net shipping-day loss after later catch-up arrivals are accounted for.}
\label{fig:model_example}
\end{figure}

Fig.~\ref{fig:model_example}A shows normalized arrival patterns for three ports under closures of the three chokepoints studied here. A 20-day closure of Malacca has no effect on arrivals to Rotterdam or Houston, but causes a sharp decline in arrivals to Singapore. The impacts can also extend beyond the closure window, as for Houston under a Panama closure, or appear only after the shock has subsided, as for Rotterdam under a Suez closure. These delayed effects reflect the length of the affected routes. Ship dynamics also generate double spikes in losses, as for Singapore under a Malacca closure, and post-shock recovery, as for Houston under a Panama closure. For Singapore, the first drawdown reflects the immediate loss of traffic from the north, while subsequent traffic reroutes around the Strait of Malacca or a change in course leads to a temporary increase in arrivals as the two synchronize. For Houston, the recovery pattern reflects second-order effects from a temporary increase in ships in the Gulf of Mexico, which raises traffic between ports in the region. 

Fig.~\ref{fig:model_example}B illustrates the spatial distribution of losses for a 20-day Panama closure. The largest net losses, shown in red, are concentrated along the coasts of the Americas, where the Panama Canal is an important route. Smaller losses also appear across ports outside the directly exposed region. Although these losses are less severe in relative terms, high-activity ports can still account for large absolute numbers of missed arrivals, as indicated by marker size.

\subsection{For reroutable chokepoints, cumulative losses grow with closure duration but peak arrival shortfalls remain limited}

We next compare closure durations and chokepoint combinations under the baseline information regime, in which ships receive no advance warning and no reopening forecast.

\begin{figure}[t]
\centering
\setlength{\tabcolsep}{8pt}

\begin{tabular}{cc}

\subcaptionbox{Global maximum arrival shortfall\label{fig:baseline_A}}[0.47\linewidth]{
  \includegraphics[width=0.90\linewidth,trim=8 8 8 8,clip]{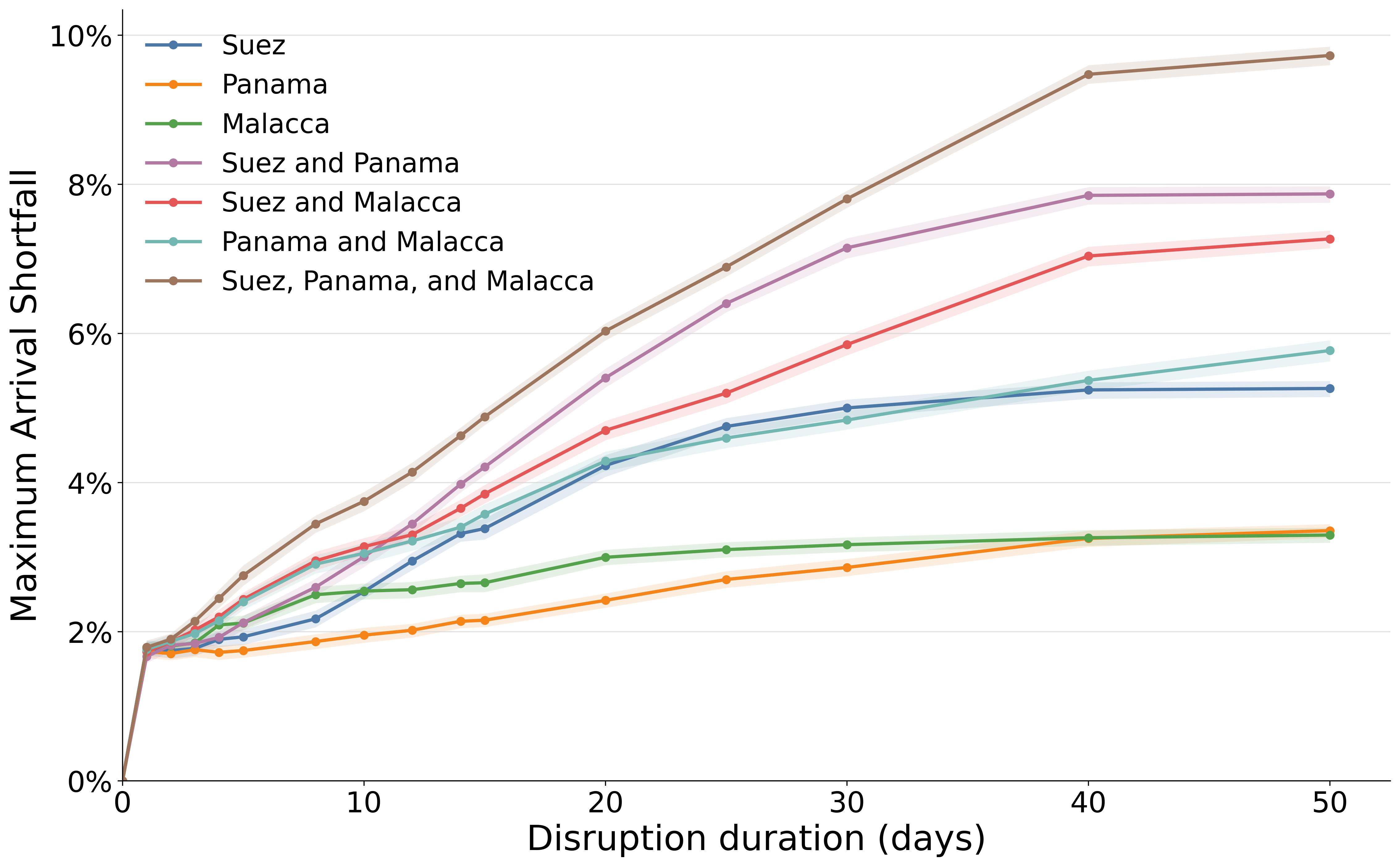}
}
&
\subcaptionbox{Global cumulative loss\label{fig:baseline_B}}[0.47\linewidth]{
  \includegraphics[width=0.90\linewidth,trim=8 8 8 8,clip]{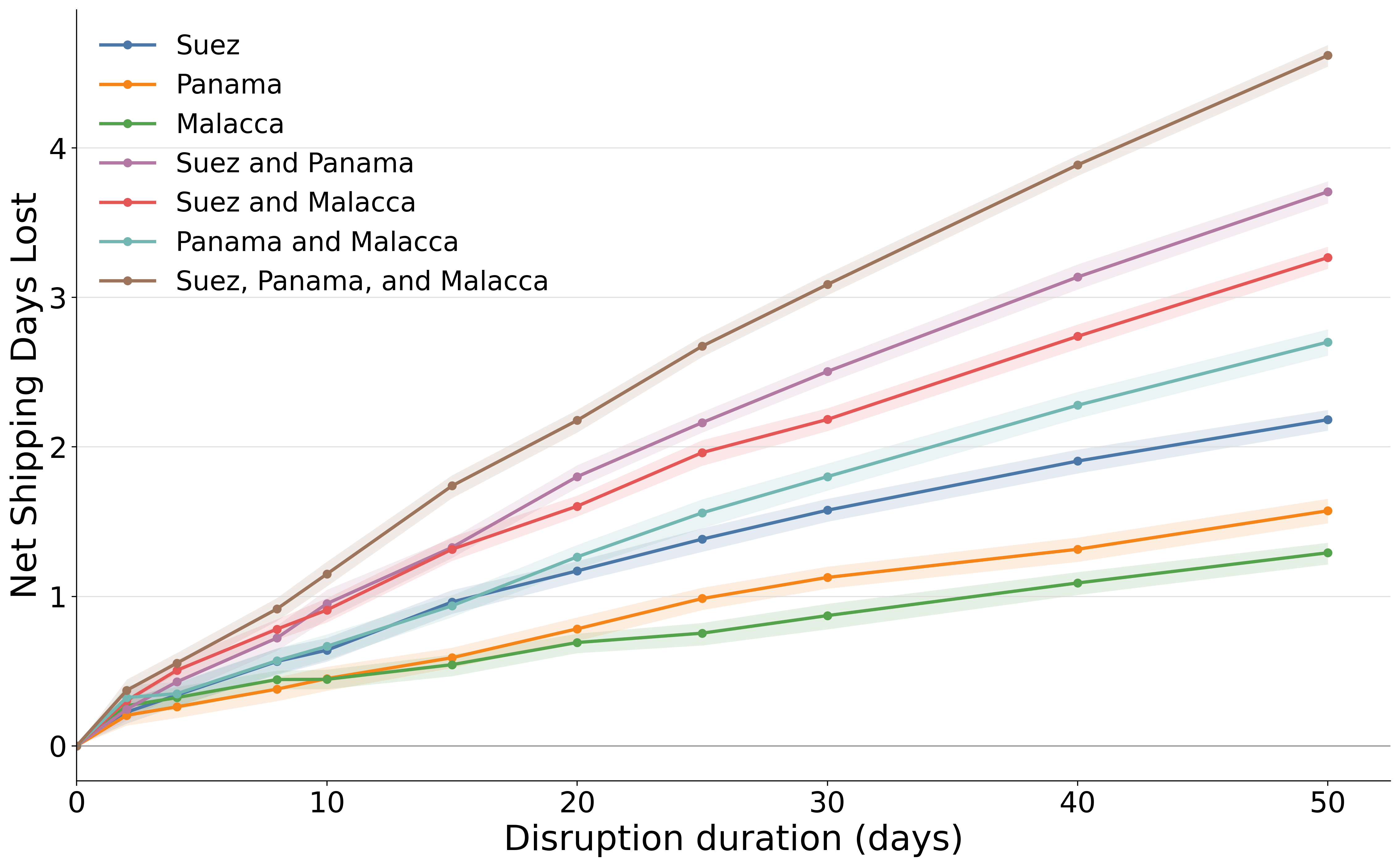}
}
\\[1.4em]

\subcaptionbox{Maximum arrival shortfall at 30 days\label{fig:baseline_C}}[0.47\linewidth]{
  \includegraphics[width=0.90\linewidth]{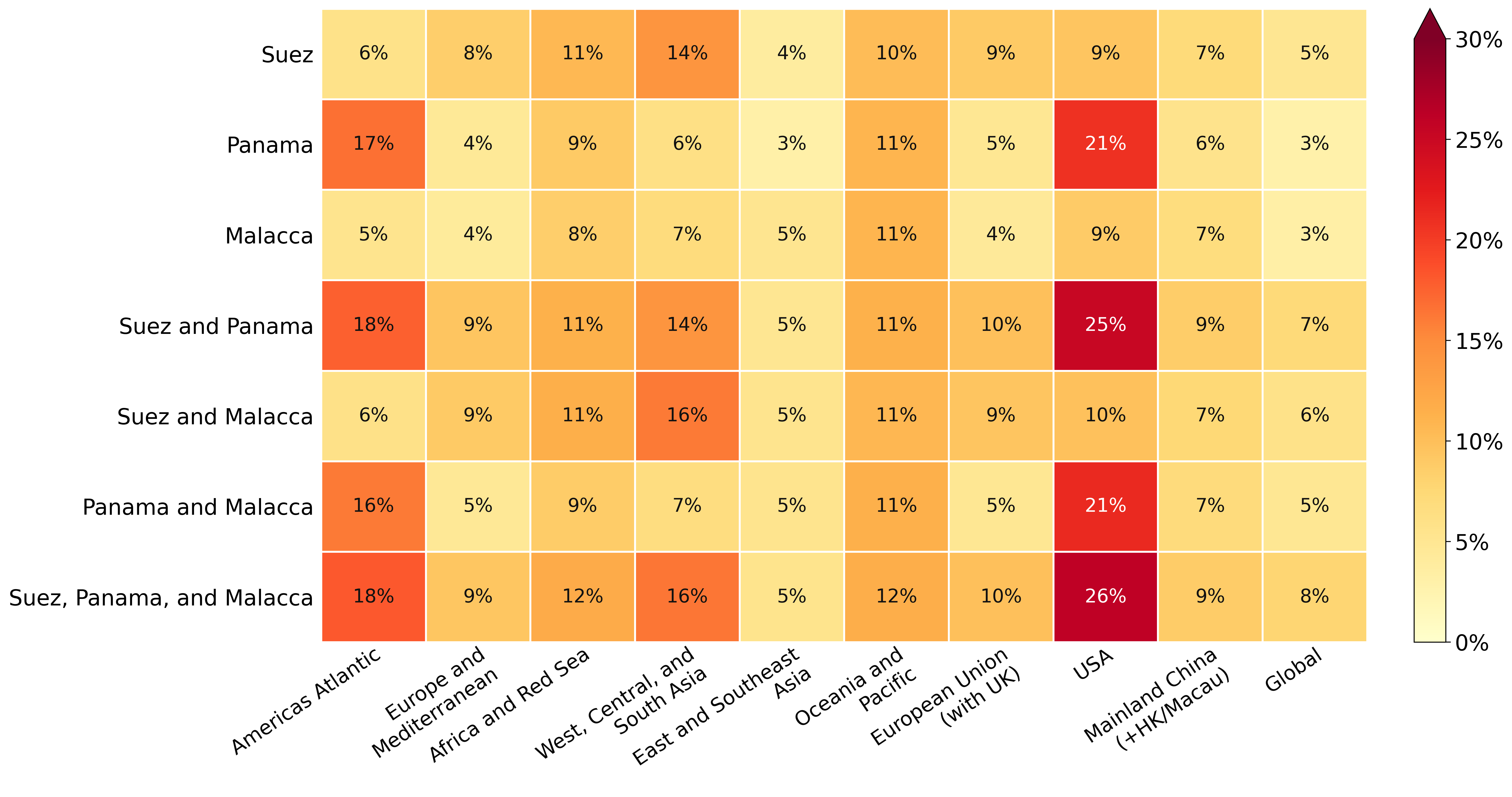}
}
&
\subcaptionbox{Cumulative-loss slope\label{fig:baseline_D}}[0.47\linewidth]{
  \includegraphics[width=0.90\linewidth]{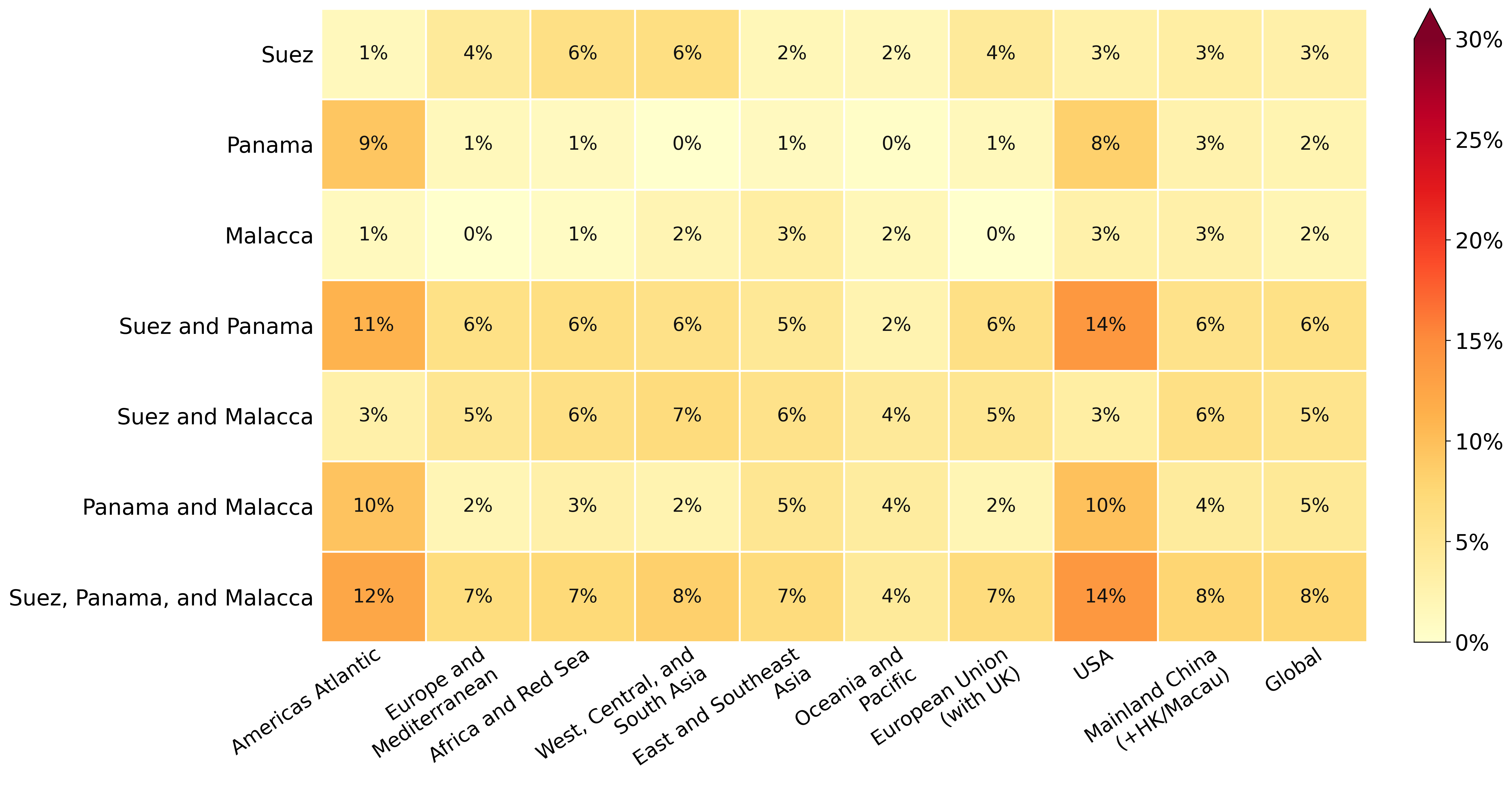}
}

\end{tabular}

\caption{Baseline disruption effects across chokepoint scenarios. (A) Global maximum arrival shortfall as closure duration increases. (B) Global net cumulative shipping-day loss across closure durations, with the post-adjustment loss rate slope indicated. Regional panels use aggregated United Nations M49 geographic subregion groupings, with three additional non-exclusive major economic zone aggregates shown separately for the European Union with the United Kingdom, the USA, and Mainland China with Hong Kong and Macau. Further details are given in the Appendix. (C) Regional maximum arrival shortfall for a representative 30-day closure. (D) Regional cumulative loss slope after adaptation, measured as additional net shipping-days lost per additional closure day.}
\label{fig:baseline_results}
\end{figure}

The global curves, aggregated over all ports, show that these two quantities respond differently to closure duration, Fig.~\ref{fig:baseline_results}A--B. Peak arrival shortfall occurs mainly when the fleet first encounters the disrupted network. Once affected ships have rerouted, extending the closure does not make the worst day much worse. This value differs between chokepoints depending on traffic volume, ease of rerouting, and route length. In particular, Suez, which is usually part of longer intercontinental routes and requires long reroutes, exhibits the largest peak arrival shortfall.

Fig.~\ref{fig:baseline_results}C shows the geographic distribution of maximum arrival shortfalls for the 30-day closure scenarios. The results reveal marked corridor dependencies, with shortfalls concentrated in regions that depend on the affected chokepoints. A Suez closure produces the largest worst-day shortfalls in West, Central, and South Asia ($14\%$). A Panama closure mainly affects the USA ($21\%$). A Malacca closure has its largest peak effects in Oceania and the Pacific ($11\%$) and in the USA ($9\%$). Despite the 30-day closure duration, maximum shortfalls remain limited and, in many cases, are almost indistinguishable from normal operational variation. This motivates the analysis of cumulative shortfalls, which capture the persistence of smaller disruptions over time.

Cumulative loss continues to grow because each additional closure day keeps affected ships on longer routes or in waiting states, delaying both current and later port calls. The slope of the curves in Fig.~\ref{fig:baseline_results}B gives this persistent post-adaptation loss rate. Interpreted in terms of arrivals, a slope of $3\%$ per closure day means that each additional closure day removes a net amount of traffic equal to about $3\%$ of one normal day of ship arrivals. Globally, these slopes are only a few percentage points per day: about $3.0\%$ for Suez, $2.2\%$ for Panama, $2.1\%$ for Malacca, and $7.7\%$ for the simultaneous closure of all three chokepoints.

Fig.~\ref{fig:baseline_results}D shows where these persistent losses are concentrated after the fleet has adapted. Under the simultaneous Suez, Panama, and Malacca closure, the highest loss rates occur in the USA, at $13.9\%$ per closure day, and in West, Central, and South Asia, at $8.4\%$ per closure day. These regional cumulative-loss rates are lower than the corresponding maximum arrival shortfalls in Fig.~\ref{fig:baseline_results}C, where the worst-day drawdown reaches $26\%$ in the USA and $16\%$ in West, Central, and South Asia. This separation shows that regional shocks can be severe on the worst day, but are partly spread over time after rerouting. At the global level, by contrast, the triple-closure maximum shortfall and cumulative-loss slope are nearly identical, at about $8\%$. This indicates that severe regional shortfalls are partly smoothed when aggregated across the world, while smaller rerouting losses continue to accumulate globally.

Simultaneous closures increase disruption, but they do not self-reinforce into a nonlinear cascade. At the global level, the cumulative loss slope of the triple closure is $7.7$ percentage points per closure day, close to the sum of the three single-closure slopes ($3.0+2.2+2.1 \approx 7.7$ percentage points). Pairwise closures show a similar near-linear pattern: Suez and Panama give $6.0$ percentage points, compared with an additive expectation of about $5.2$ percentage points, and other combinations are similar. In other words, simultaneous disruptions add persistent delay along exposed vessel cycles, but do not produce a modeled self-reinforcing collapse of global arrivals. Peak arrival shortfall also remains bounded by overlapping exposure rather than amplifying beyond it: for the 30-day closure, the global triple-closure drawdown is $8.0\%$, below the sum of the single-closure arrival shortfalls, and regional peaks are close to the dominant affected corridors without showing amplification. Regional curves, selected port-level panels, seed-level uncertainty bands, and endpoint values are reported in the Appendix.

An important result is that chokepoint closures have limited impacts outside regions that receive traffic from the affected routes. For short disruptions, the system does not have enough time to adjust before the closure is lifted, so missing arrivals remain concentrated in directly affected regions. We also find that longer routes generate small but persistent losses. These losses are not directly proportional to the volume of ships crossing the affected chokepoint, but also reflect higher-order effects from multi-step journeys and network-wide rerouting. At the global level, the losses sum linearly while on the regional level the losses are dominated by the most important chokepoint of the region.

\subsection{Information on disruption timing changes adaptive losses}

As the next experiment, we vary what shipping agents know about the closure window in the simultaneous Suez, Panama, and Malacca scenario. The default regime, discussed previously, is no warning and no reopening forecast: ships discover the closure only when it begins and do not know when it will clear. We compare this default with three information regimes: warning only, reopening forecast only, and full-window information. In the reopening-forecast-only case, the start is not anticipated, but the reopening date is announced when the closure begins; in the full-window case, both the start and reopening date are known before the shock. When an event is anticipated, shipping agents account for when they arrive at a specific edge to see if they will be able to pass or not.

\begin{figure}[t]
\centering
\setlength{\tabcolsep}{8pt}

\begin{tabular}{cc}

\subcaptionbox{Global maximum arrival shortfall\label{fig:information_effects_A}}[0.47\linewidth]{
  \includegraphics[width=0.90\linewidth,trim=8 8 8 8,clip]{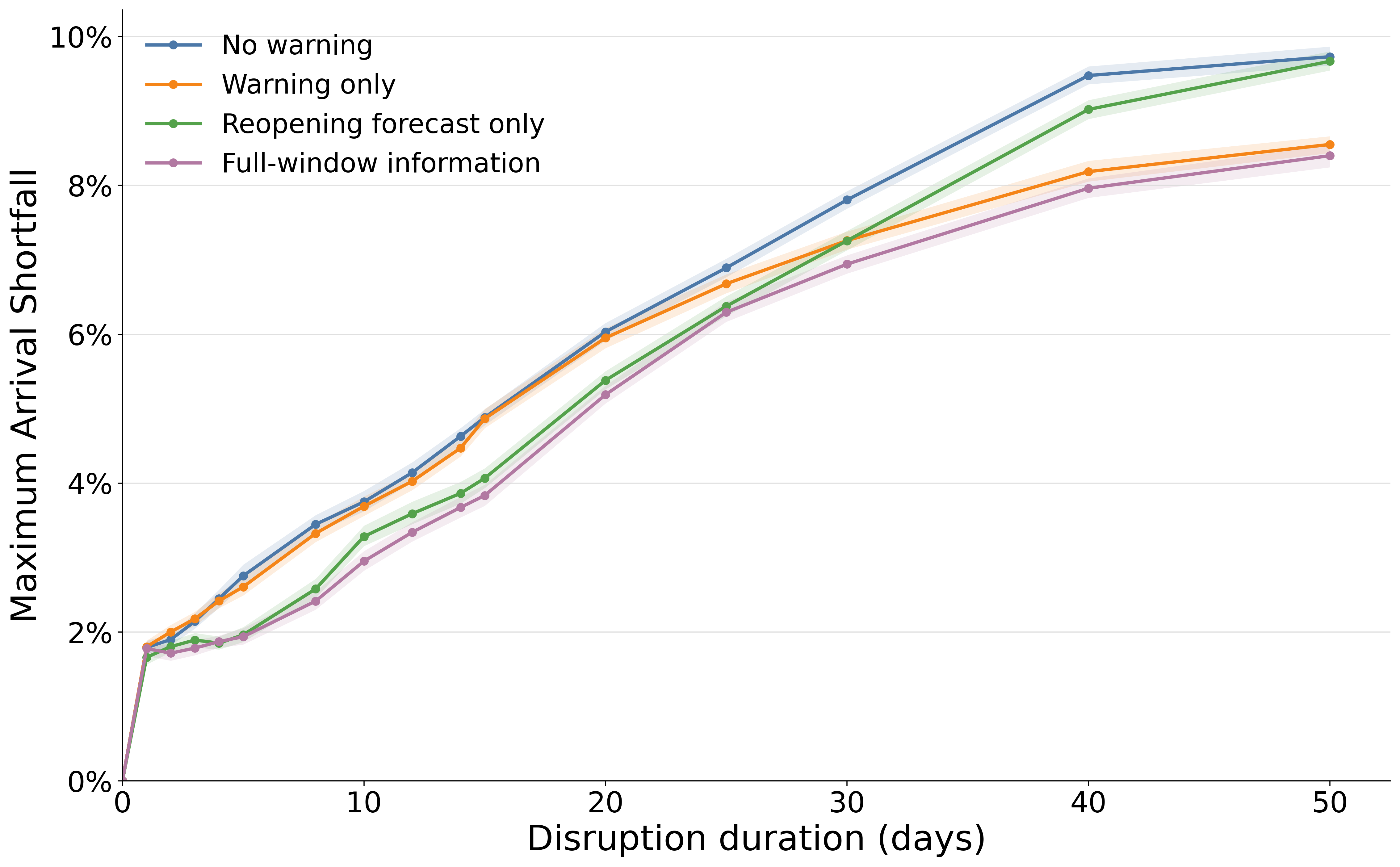}
}
&
\subcaptionbox{Global cumulative loss\label{fig:information_effects_B}}[0.47\linewidth]{
  \includegraphics[width=0.90\linewidth,trim=8 8 8 8,clip]{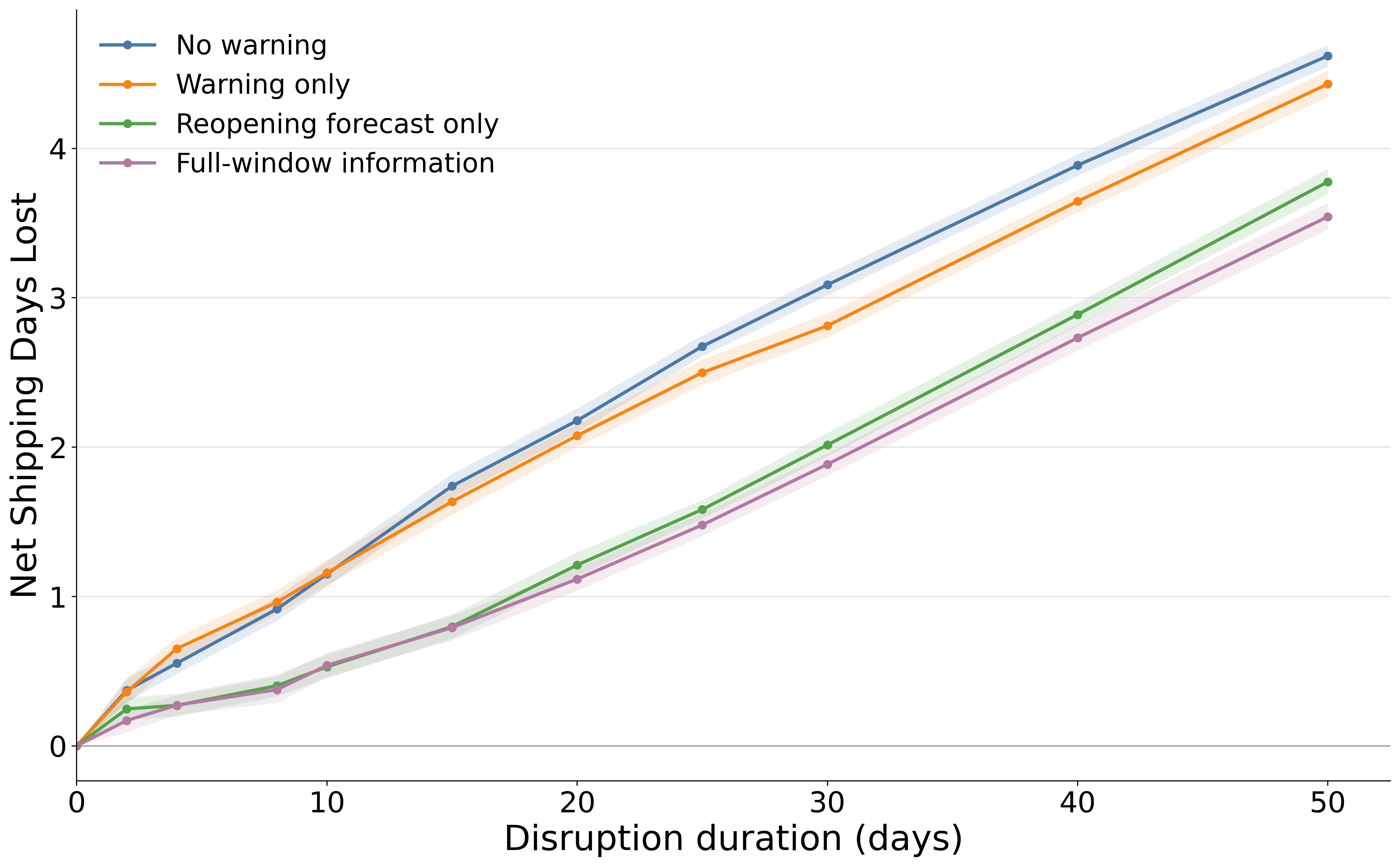}
}

\end{tabular}

\caption{Effect of timing information under simultaneous Suez, Panama, and Malacca closures. (A) Global maximum arrival shortfall and (B) global net cumulative shipping-day loss under four information regimes: no warning and no reopening forecast, warning only, reopening forecast only, and full-window information. The regimes vary whether ships know the closure start, the reopening date, or both when making routing decisions.}
\label{fig:information_effects}
\end{figure}

We find that information shifts the loss profiles, Fig.~\ref{fig:information_effects}. Relative to the default regime with no warning and no reopening forecast, knowing the reopening date is the most important distinction. For cumulative losses, the default case reaches $4.62$ shipping days at 50 days, and warning alone changes this to $4.43$. By contrast, reopening information reduces the loss to $3.78$, and full-window information reduces it further to $3.54$. The same mechanism also reduces maximum arrival shortfall for short and intermediate closures, because ships can wait through closures that are short enough instead of taking unnecessary detours. At long durations this peak effect changes: by 50 days, reopening information alone gives an arrival shortfall close to the default case, $9.67\%$ compared with $9.73\%$, while warning-only and full-window information reduce the peak to $8.55\%$ and $8.40\%$, respectively. Thus, end-date information is most important for cumulative loss and for avoiding short-closure peaks. Start-date information becomes more relevant for peak arrival shortfall when closures persist long enough that waiting through them is no longer the main response and instead ships can use the knowledge of a disruption to preemptively select alternative routes.

The cumulative loss curves should therefore be read as shifted rather than fundamentally different long-duration slopes. Once the fleet has adjusted, the curves with and without reopening information accumulate additional losses at similar rates; the difference is that reopening information avoids an early block of unnecessary waiting and detouring, shifting the trajectory downward. This remains visible even when maximum arrival shortfall curves have largely converged by the 30-day window. At 30 days, the cumulative losses under reopening information and under full-window information remain well below the default no-warning/no-forecast loss. 

The key result is that knowing the reopening date does not eliminate the persistent cost of operating on longer routes, but it delays the accumulation of net shipping-day losses by preventing avoidable early wait-or-detour decisions. Regional panels and endpoint values are reported in the Appendix.

\subsection{Structural exposure overconcentrates realized adaptive losses}

We compare the adaptive simulations with a static structural exposure benchmark to separate pre-shock route dependence from realized disruption. The benchmark assigns expected port-to-port movements to their undisrupted shortest routes and measures how many expected arrivals depend on the closed chokepoint. We use it as a route-dependence baseline; we remove flows whose routes cross the chokepoints we are investigating.

\begin{figure}[t]
\centering
\setlength{\tabcolsep}{8pt}

\begin{tabular}{cc}

\subcaptionbox{Port-level comparison\label{fig:static_adaptive_A}}[0.47\linewidth]{
  \includegraphics[width=0.90\linewidth]{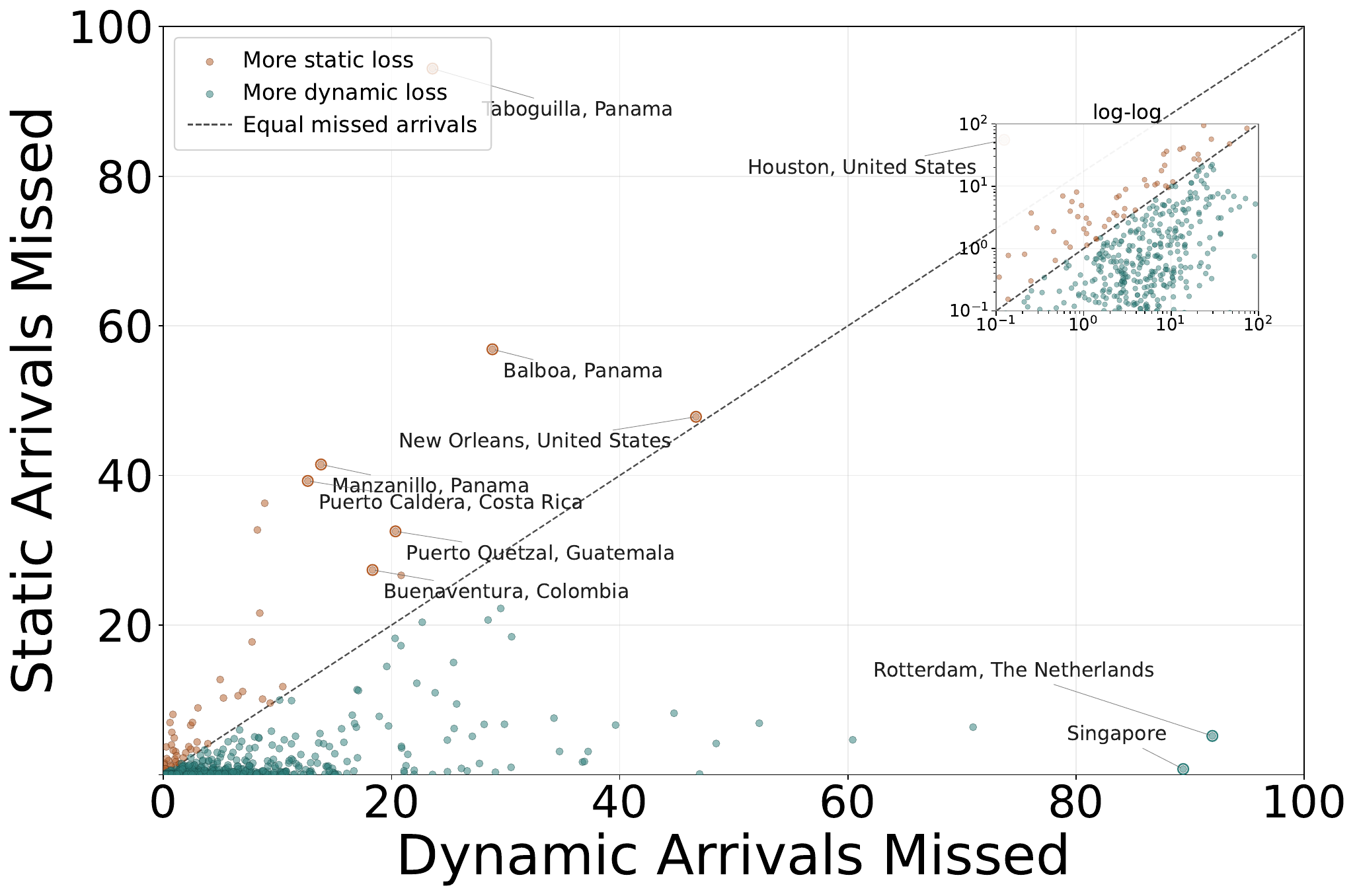}
}
&
\subcaptionbox{Regional adaptive/static total-loss ratio\label{fig:static_adaptive_B}}[0.47\linewidth]{
  \includegraphics[width=0.90\linewidth]{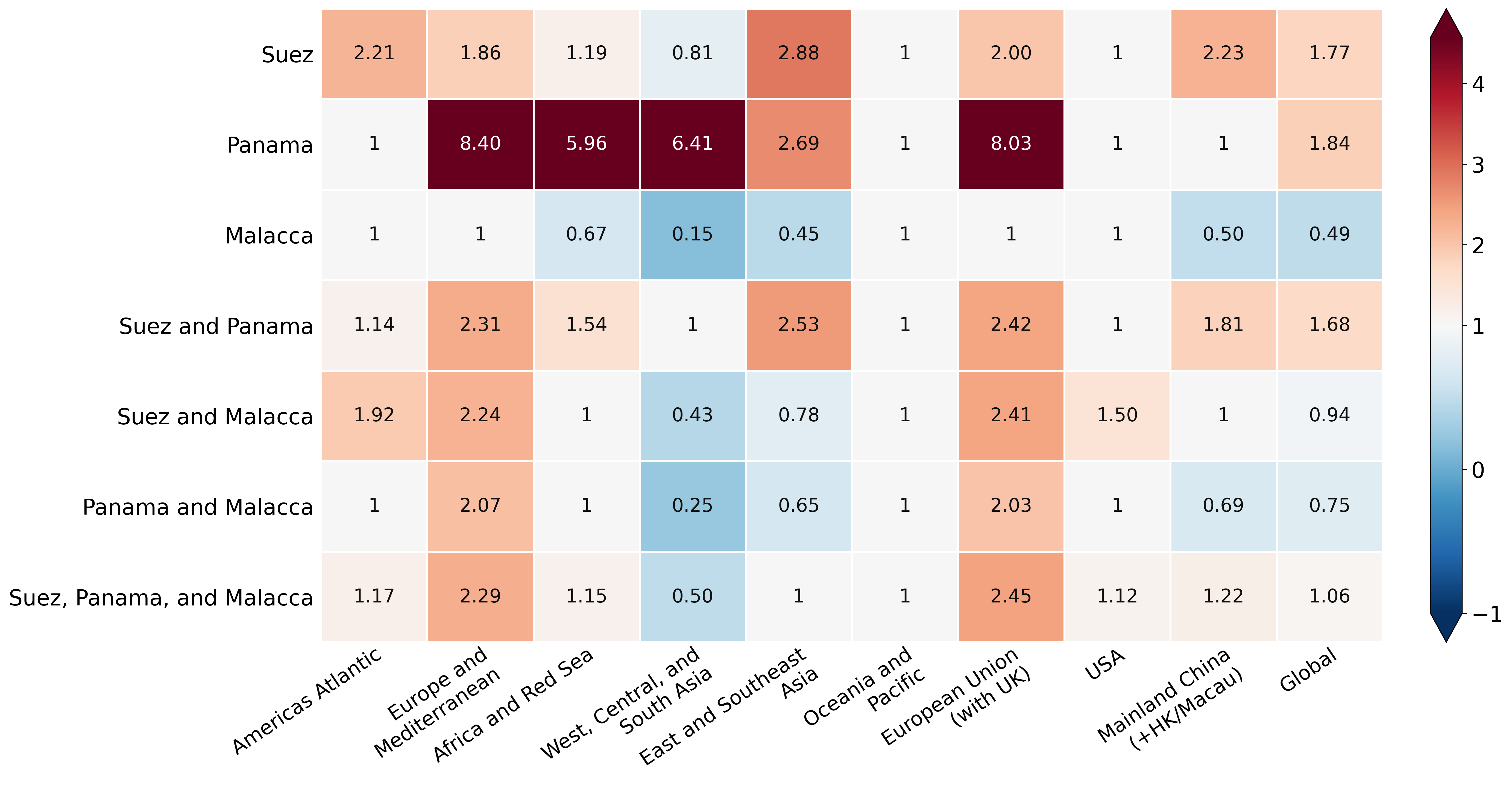}
}

\end{tabular}

\caption{Structural exposure benchmark versus adaptive realized loss. (A) Port-level comparison of adaptive missed arrivals and exposure-implied missed arrivals for a 30-day Panama closure. The dashed line marks equality. Points above the line have larger exposure-implied loss; points below the line have larger adaptive realized loss. The log-log view shows the same comparison across smaller affected ports. (B) Regional adaptive/static total-loss ratio for 30-day closures. Values below one indicate that structural exposure overstates realized adaptive loss. Values above one indicate that realized adaptive loss exceeds structural exposure because delays propagate into later port calls. Cells shown as one are not statistically distinguishable from parity. Duration-slope comparisons and port-level details are reported in the Appendix.}
\label{fig:static_adaptive}
\end{figure}

To illustrate the uncovered mechanisms at the port level, the 30-day Panama closure shows a large mismatch, see Fig.~\ref{fig:static_adaptive}A. Structural exposure puts most exposure-implied loss in the ports whose normal routes depend most directly on the strait, especially ports in Panama and in the Gulf of Mexico. In the adaptive model these ports still lose arrivals, but the losses are smaller than their direct exposure would imply because part of the traffic uses the Strait of Magellan and part of the deficit is recovered within the evaluation window. This can already be seen in the linear comparison for the main affected ports. The log-log comparison shows that the mechanism is more widespread: losses also appear across additional downstream ports, including Rotterdam, Shanghai and Singapore, because ships delayed on one leg also fail to complete later calls on schedule. The adaptive outcome is therefore less concentrated at the main exposed ports and more diffuse across the vessel cycle.

At the regional level, the total-loss ratios in Fig.~\ref{fig:static_adaptive}B show that redistribution can occur even when aggregate magnitudes are similar. Ratios below one mean that direct route exposure overstates realized loss, whereas ratios above one indicate losses shifted to later port calls that are omitted by the benchmark. Malacca provides the clearest rerouting case, where structural exposure substantially overstates how much loss actually materializes globally. Suez and Panama show the opposite pattern because the adaptive model picks up delayed later calls that the structural benchmark misses. The simultaneous closure lands closer to parity globally, although the picture remains uneven across regions.

\subsection{An application to the Strait of Hormuz}

Some chokepoints cannot be circumvented by rerouting due to their geographical location. Nevertheless, our model can be used to study the global impact of closing such non-reroutable chokepoints, such as the Strait of Hormuz. We consider three durations for 14, 28, and 56 days, each case with an unexpected start date but a communicated reopening date. We find that a 14-day closure has little to no impact on global shipping, with losses limited to fractions of a shipping day and often indistinguishable from zero. However, in contrast to the reroutable chokepoints examined previously, losses do not scale linearly with closure duration. Rather, they increase superlinearly, suggesting that disruptions accumulate when vessels cannot substitute toward alternative routes. For example, a 56-day closure of the Strait of Hormuz results in a global net loss of 3.43 shipping days for tankers, compared to 1.07 shipping days for a 28-day closure. Total global losses follow the same trend, rising from 0.74 to 2.54 shipping days when the closure duration doubles from 28 to 56 days. These losses are concentrated in Asian ports, where much of the tanker traffic passing through the Strait of Hormuz is directed. This also illustrates that the model can be meaningfully disaggregated by vessel type, rather than producing a single loss estimate. The full regional and vessel-type breakdown is reported in the Appendix, together with world maps showing the spatial distribution of the results.

\FloatBarrier
\section{Discussion}

\subsection{Implications for global shipping}

This study introduces a calibrated and adaptive agent-based model for stress-testing global shipping to chokepoint disruptions. Applied to simulated closures of the Suez Canal, Panama Canal, and Strait of Malacca, the model shows that adaptive rerouting limits many direct exposure losses while generating delayed downstream arrival losses along later itineraries in the same region. The results distinguish peak daily arrival shortfalls from cumulative losses and show that knowledge of the disruption end time can reduce net lost shipping days. The adaptive model therefore reframes static exposure: rerouting often lowers port level losses, but redistributes losses across specific regions and routes.

The one-to-one representation is central because it exposes response patterns that aggregate or static models cannot generate. A closure does not create a single proportional dip in arrivals. It can produce delayed drawdowns after reopening, catch-up spikes, and second-order missing voyages when an absent ship also fails to generate later port calls as shown in Fig.~\ref{fig:model_example}A. Because the model is agent-based, we only specify relatively simple vessel-level behavioral rules, while system-level effects emerge from their interaction. This is why dynamic and static estimates differ in both directions. Adaptive routing lowers local cumulative losses relative to static exposure, especially when short feasible detours exist, but static exposure is not an upper bound. For short disruptions or downstream regions, dynamic losses can equal or exceed static estimates because the model tracks the missing vessel cycle and not just the blocked leg. The comparison with static exposure should therefore be read as a diagnostic of where adaptation relocates loss, not as a single correction factor.

The scenario experiments show how these adaptive losses change with duration and co-occurring closures. Duration does not affect the basic impact regime. Peak drawdown is set largely when ships first encounter the disrupted network and remains bounded after rerouting. Cumulative losses then grow approximately linearly because each additional closure day keeps affected ships on longer routes or in waiting states. A 20-day closure is therefore roughly twice a 10-day closure in cumulative-loss terms. Co-occurring chokepoint closures are also close to additive at the global level in the reroutable scenarios studied here. They increase persistent delay along exposed vessel cycles, but do not produce a modeled self-reinforcing collapse of global arrivals. At the regional level, effects depend on exposure overlap: unrelated closures add little, while overlapping closures are often dominated by the main affected route system and not by the sum of independent peaks. This separation between duration, co-occurrence, and exposure also makes information central, because ships’ routing choices depend on whether the closure is expected to be short enough to wait out or long enough to avoid.

The information regimes can also be interpreted as stylized classes of real disruptions. Forecastable hazards such as hurricanes are closest to the known-start and known-end case, because both the onset and approximate clearing window can enter route choice before ships commit to a path. Unexpected canal blockages such as the Ever Given are closer to cases in which the start is not anticipated, but the end becomes known once the obstruction is cleared or a reopening time is announced. Open-ended geopolitical closures, including a possible Strait of Hormuz disruption, correspond most closely to the unknown-start and unknown-end case, where ships initially react to a changed network without reliable information about how long the closure will persist. The model results therefore suggest that the operational value of information depends less on warning alone than on whether the disruption duration is known well enough to distinguish waiting from rerouting.

\subsection{Future applications}

Although the experiments above focus on chokepoint closures, the same framework can be used to study a wider class of disruptions and operational changes. The model has three main points of intervention. First, shocks to the physical transport network can be represented by changing edge availability or travel time. This covers canal closures, draft restrictions, longer passages through contested corridors, or weather-related route constraints. Second, shocks to port operations can be represented by changing queuing and service parameters. Port expansion increases effective service capacity, while port damage, labor disruption, inspections, or congestion reduce throughput by increasing service times or lowering available capacity. Third, persistent changes in trade flows can be represented by changing the estimated routing model. Shifts in demand, sanctions, new trade agreements, or stable rerouting behavior would appear as changes in the Markov transition probabilities that determine next port choice.

\clearpage
\section{Materials and Methods}
\subsection*{Data}

Historical port call data were derived from vessel Automatic Identification System (AIS) data provided by the UN Global Platform. The data span from January 1, 2019, to April 27, 2025. Port call inference, port assignment, vessel-type classification, and treatment of missing or erroneous ship dimension records followed the approach described in earlier work \cite{verschuur2021global, verschuur_ports_2022}. We restricted the analysis to commercial vessels engaged in trade and grouped vessels into three operational classes: cargo, tanker, and dry bulk. General cargo, container, and roll-on/roll-off vessels were aggregated into the cargo class.

We excluded vessels missing both IMO and MMSI identifiers, leaving 70,457 unique vessels, and further restricted the analytical sample to ships with at least 10 observed port calls. This left 59,725 ships for the Markov analysis. For size-stratified analyses, we excluded 27 vessels with missing records and split ships by vessel type and area-based size category. For simulation, we defined the active fleet using the reference year 2024 and retained ships with at least 10 port calls in that year, yielding 35,954 ships: 14,621 cargo ships, 10,303 dry bulk vessels, and 11,030 tankers. Additional sample construction details and summary tables are provided in the Appendix.

\subsection*{Markov modeling of port sequences}

We modeled next port choice using finite order discrete time Markov chains \cite{guo_trajectory_2018,marten2020scalable,spadon_learning_2025,fu_multi-scale_2024}. For each ship-type or ship-type/size slice, we fit models of order $k \in \{0,1,2,3\}$. Each configuration defines an evaluation port set $S$, either the 50 most frequent ports or the full observed set. Let $s_1,\dots,s_T$ be a sequence after filtering to $S$. An order-$k$ model specifies
\[
P(s_t \mid s_{t-1},\dots,s_{t-k}) =
\frac{N(s_{t-k},\dots,s_{t-1},s_t) + \alpha\,\pi(s_t)}
{\sum_{x\in S} N(s_{t-k},\dots,s_{t-1},x) + \alpha},
\]

where $N(\cdot)$ denotes transition counts extracted for that slice, $\alpha=1.0$, and $\pi(\cdot)$ is the uniform distribution on $S$. By convention, the order-0 model predicts the marginal distribution on $S$. For the comparisons of models, $\alpha > 0$ is useful to avoid degenerate losses. In the ABM we take $\alpha = 0.0$.

When evaluation is restricted to the most-frequented ports, raw trajectories may contain intermediate ports outside $S$. In that case, ports outside $S$ are dropped while preserving the remaining chain if it still yields transitions in $S$. At scoring time, unseen histories are handled by back-off from order $k$ to lower orders and finally to the order-0 marginal on $S$. Models were trained on trajectories up to and including 2023. Predictive performance was evaluated using ship-level 20-fold cross-validation and forward temporal validation on later data. We also compared pooled and vessel-type-specific specifications. Full details of sequence alignment, estimation windows, resampling, and multiplicity correction are provided in the Appendix.

Predictive performance is scored transition by transition on held-out ship histories. For evaluated transitions $i=1,\dots,N$, with realized next port $x_i$ and history $h_i$, the per-transition predictive log-likelihood is
\[
\mathrm{PLL} = \frac{1}{N}\sum_{i=1}^{N}\log p(x_i \mid h_i), \qquad x_i \in S.
\]
The score uses evaluated transitions for which the realized next port lies in the evaluation set $S$. Predictive loss is $-\mathrm{PLL}$, so lower loss means higher assigned probability for the realized next port. We evaluate models on both the top-50-port set and the full observed port set, retain ship-level PLL values for bootstrap resampling, report perplexity as $\exp(-\mathrm{PLL})$, and apply Holm correction to the order comparisons and pooled-versus-type-specific comparisons.

\subsection*{Agent-based model}

We embedded the estimated routing process in an agent-based model in which agents correspond to individual ships moving on a weighted marine network. The environment is an undirected weighted network based on SeaRoute, with the addition of global ports \cite{eurostat_searoute,verschuur2021global}. Edge weights represent traversal time in days under a constant speed approximation.

We calibrated cruising speed by estimating empirical ship speeds from observed inter-port movements. Sea route distances between consecutive port calls were computed using SeaRoute \cite{eurostat_searoute} and divided by the elapsed time between calls to obtain end-to-end speeds. The empirical distribution was stable across vessel classes and voyage distances, and was centered near 10 nautical miles per hour. The simulator therefore assigns all ships a common cruising speed of 10 nautical miles per hour.

Each port has a single service capacity parameter and a single shared queue used by all vessel classes. Capacity is calibrated from the historical 90th percentile of the total number of ships present on a day at that port, where a ship is considered present if it is between port calls, including the days of arrival and departure. When arriving demand exceeds available service slots, ships wait in the shared queue until service becomes available. Port service times are sampled from exponential distributions calibrated from historical port stay data.

In the simulation, ships are initialized from Markov-consistent order-2 histories sampled from the stationary distribution over observed two-port histories. The most recent sampled port is used as the current port, an initial destination is sampled from the vessel-type Markov model, an A* route is computed, and the ship is placed along that route. Initial movement times are drawn uniformly from the first 30 model days. The long pre-shock simulation, approximately $800$ timesteps, allows the system to relax before disruption outcomes are evaluated. We verified that the model had reached dynamic equilibrium by that point. The model is normalized so that a timestep is equivalent to $24$ hours.

After service at a port, a ship samples its next destination from the fitted vessel-type-specific routing model conditional on its stored two-port history using a two-step Markov model. If the sampled destination equals the current port, we say that the ship idles at that port for a stochastic dwell period, also calibrated from data. If the sampled destination differs from the current port, the ship computes the shortest path on the marine network using the A* algorithm with geographic distance as the heuristic. Ships recompute routes immediately after network changes and also during movement on a periodic cadence of approximately 20 model days. Additional implementation details are provided in the Appendix.

\subsection*{Shock scenarios and simulation design}

We modeled chokepoint disruptions as temporary removals of edges from the marine network. The scenario set includes closures of the Suez Canal, Panama Canal, and Strait of Malacca, both individually and jointly. For chokepoints represented by multiple edges, all relevant segments were removed in order to impede passage.

Runs start at model day $0$ from the synthetic initial condition described above. Temporary closures are scheduled at $1200$, well after the system has reached a dynamic equilibrium. We swept closure duration over $1,2,4,5,6,8,10,12,15,20,30,40,50$ days and report selected durations in the main text for readability. By default, each scenario was replicated over 50 random seeds. For the joint three-chokepoint case, we also considered information regime scenarios that vary whether ships only react once closures begin or can anticipate the future closure window during route planning.

Information about a closure duration enters the ABM via the A* search. Starting from time $t$, we can compute when a ship will potentially be present at a given node, say $t'> t$. Hence, if at $t'$ some edges are expected to be closed, we simply do not allow this to become a feasible path. Similarly, if an edge is closed but expected to reopen before a ship arrives, the ship will be able to pass even though at time $t$ the edge is closed. Note that one of the options is for the ship to wait for the edge reopening if that is known. A ship that does not know about a closure assumes none and a ship that does not know about a reopening assumes it will never reopen. This latter assumption is conservative; it can be relaxed and is operationally close to assuming reopening no earlier than about one month, because the reroutes are shorter than that.

The model is heavily based on random sampling. To make the results more comparable all simulations share the same list of random seeds. This yields matched stochastic realizations for comparisons that hold the information regime specification fixed. Comparisons across different information regimes do not in general share identical pre-shock trajectories, because anticipation can alter route choice before the closure starts.

\subsection*{Calibration and validation}

We evaluated the baseline model by comparing simulated and empirical average arrivals across global ports. We also performed an event-matched validation using the March 2021 Suez Canal disruption associated with the Ever Given. In this validation, the Suez corridor was made unavailable for six and a half days. The end of the disruption was known. Simulated and empirical changes in arrivals were then compared across regions over multiple pre-shock and post-shock windows using Pearson correlation. Full validation details are provided in the Appendix.

\subsection*{Output processing and disruption metrics}

The primary model output is daily completed port calls by port, which we refer to as arrivals in the main text. In the simulation, time is recorded in model days counted from the start of the run, and a ship is counted on the day on which it finishes loading and unloading at port. Day bins are used as a simulation timescale. The model is calibrated and normalized such that one day corresponds to one real-life day.

Because these series are noisy at daily resolution, each seed-specific time series was smoothed separately for pre-shock and post-shock periods using a 7-day moving-average window. When a window would cross the shock boundary, a smaller window was used that did not cross the boundary.

Each time series of port arrivals was then normalized by its pre-shock mean daily arrivals $\bar{A}_i$ computed over model days 800 to 1000, so that
\[
x_i(d)=\frac{A_i(d)}{\bar{A}_i}.
\]
Regional time series were computed by aggregating ports within region and then applying the same smoothing and normalization procedure.

We summarized disruption impacts with two metrics \cite{paululybinadaria_metric_nodate}. Maximum arrival shortfall was defined as

  \[
  \max\left(0, 1-\min_{d\in\mathcal{T}}x_i(d)\right),
  \]

which is positive when fewer arrivals occur than usual.

We also computed a net cumulative loss measure. Specifically, for each normalized series $x_i(d)$, deficits below $1-\sigma_i$ were summed and surpluses above $1+\sigma_i$ were subtracted, where $\sigma_i$ is the standard deviation of the pre-shock normalized series. The reported unit is net shipping days,

  \[
  L_i=
  \sum_{d\in\mathcal{T}:x_i(d)<1-\sigma_i} \left(1-x_i(d)\right)
  -
  \sum_{d\in\mathcal{T}:x_i(d)>1+\sigma_i} \left(x_i(d)-1\right),
  \]
  
which is the absolute loss minus the absolute gain in later periods.

For ratio, slope, and arrival shortfall analyses, we used simple null hypotheses tied to the relevant baseline value. For adaptive--static ratios, the null hypothesis was equality between the adaptive and static estimates, $H_0:R=1$. For duration response slopes, the null hypothesis was no increase in net loss with disruption duration, $H_0:\beta=0$; when dynamic slopes were compared with static per-day losses, the corresponding null was $H_0:\beta=s_i$, where $s_i$ is the static per-day loss estimate. For maximum arrival shortfall, the null hypothesis was no disruption relative to baseline, $H_0:D=0$. Uncertainty was summarized using the seed-level distribution for ratios and normal 95\% intervals for slope and shortfall estimates. Null hypotheses were treated as unsupported when the corresponding interval excluded the null value.

\subsection*{Static counterfactual}

As a structural exposure benchmark, we computed chokepoint exposure from the two-step Markov model. Using transition probabilities $P^{(c)}_{ijk}$ from previous port $i$ and current port $j$ to destination port $k$ for vessel class $c$, we computed the stationary distribution $\pi^{(c)}_{ij}$ over two-port histories. Each expected next-leg flow $\pi^{(c)}_{ij}P^{(c)}_{ijk}$ was assigned to the shortest travel-time path from current port $j$ to destination port $k$ on the undisrupted marine network. For chokepoint $q$, vessel class $c$, and destination port $k$, structural exposure is

\[
E^{(c)}_{kq}
=
100
\frac{
\sum_{i,j:j\ne k}
\pi^{(c)}_{ij} P^{(c)}_{ijk}
\mathbf{1}\{q \in r(j,k)\}
}{
\sum_{i,j:j\ne k}
\pi^{(c)}_{ij} P^{(c)}_{ijk}
}.
\]

Here, $r(j,k)$ is the undisrupted shortest travel-time route from current port $j$ to destination port $k$. This benchmark has no ship agents, queues, service times, disruption dynamics, or rerouting. It measures the baseline share of expected arrivals whose fixed shortest route depends on a chokepoint.

\clearpage
\appendix
\section*{Appendix}
\addcontentsline{toc}{section}{Appendix}

\section{Validation}

\begin{figure}[H]
\centering
\setlength{\tabcolsep}{3pt}
\begin{tabular}{cc}
\begin{minipage}[t]{0.47\linewidth}
\centering
\includegraphics[width=0.90\linewidth,trim=8 8 8 8,clip]{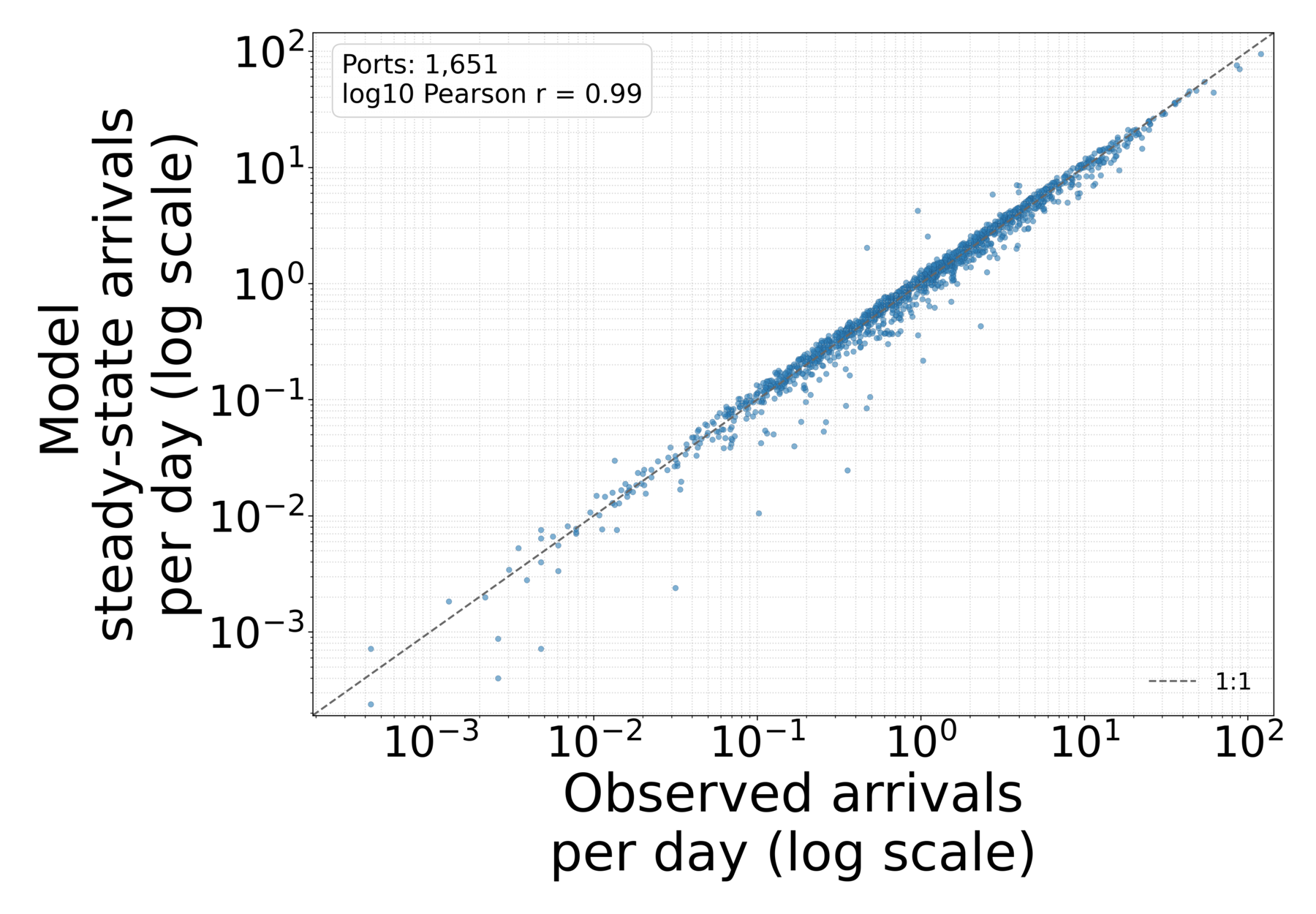}
\scriptsize (A) Total
\end{minipage} &
\begin{minipage}[t]{0.47\linewidth}
\centering
\includegraphics[width=0.90\linewidth,trim=8 8 8 8,clip]{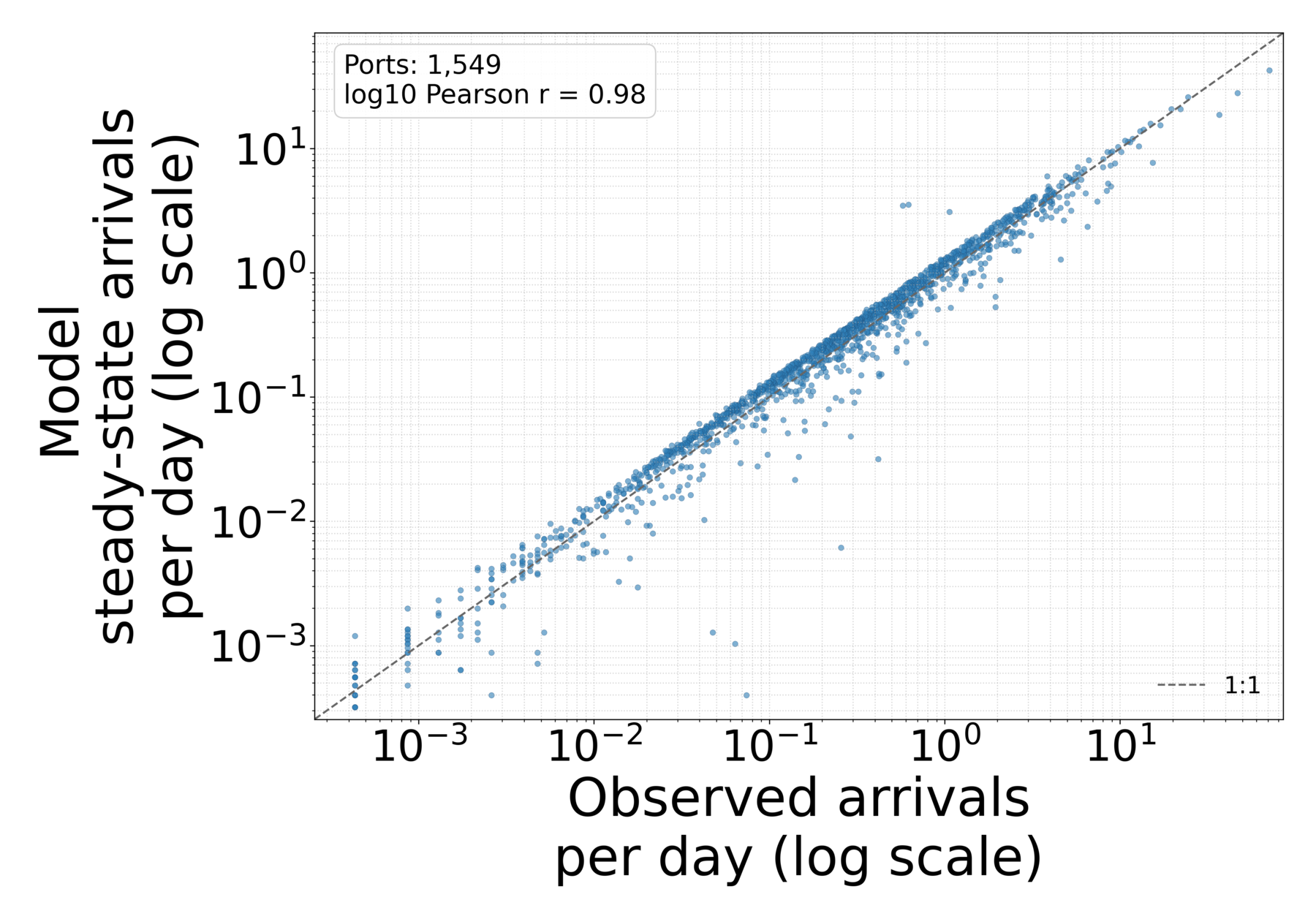}
\scriptsize (B) Tankers
\end{minipage} \\
\begin{minipage}[t]{0.47\linewidth}
\centering
\includegraphics[width=0.90\linewidth,trim=8 8 8 8,clip]{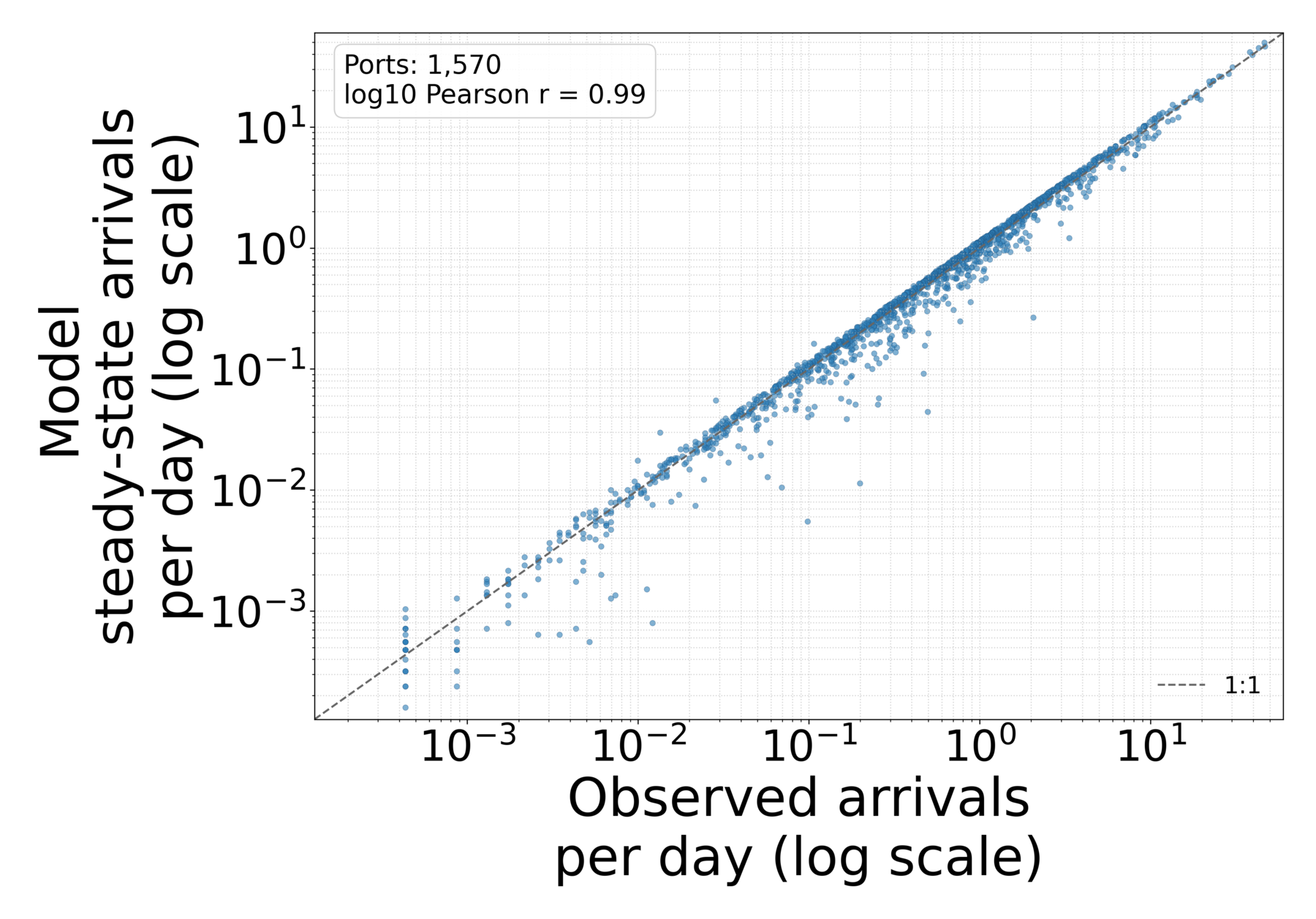}
\scriptsize (C) Cargo
\end{minipage} &
\begin{minipage}[t]{0.47\linewidth}
\centering
\includegraphics[width=0.90\linewidth,trim=8 8 8 8,clip]{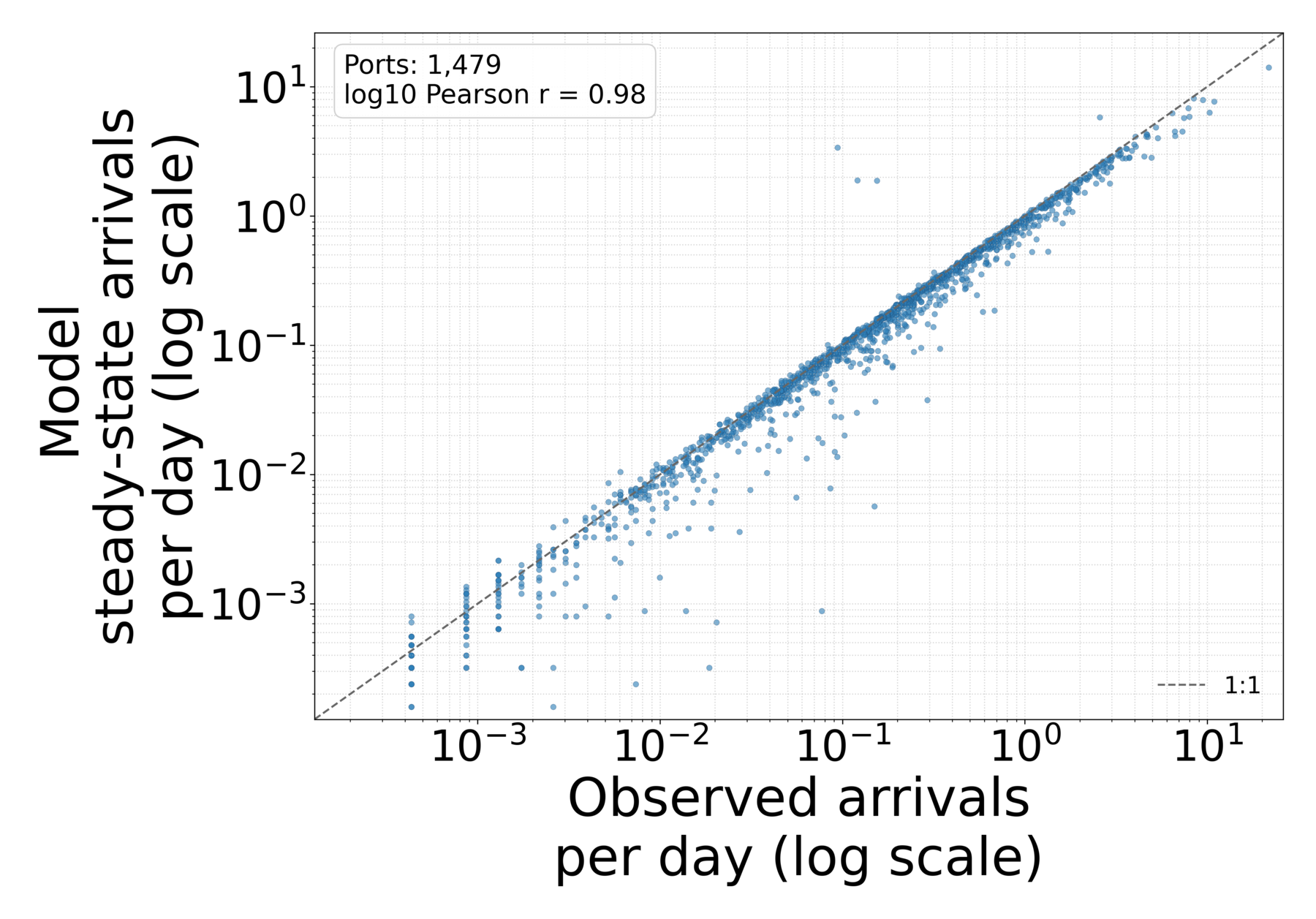}
\scriptsize (D) Dry Bulk
\end{minipage}
\end{tabular}
\caption{Steady-state validation comparing observed and simulated mean daily
port arrivals for all vessels and by vessel class.}
\end{figure}

For the Ever Given validation, we compare the relative change in arrivals
before and after the March 2021 Suez Canal blockage. For each pre-blockage
baseline window and post-blockage comparison window, we compute port-level
arrival ratios in the real data and in the event-matched simulation, then report
the Pearson correlation across ports. High positive correlations indicate that
ports with larger real-world arrival changes also have larger simulated changes.
A random post-shock window with no actual disruption should have correlation
near zero because the port-level changes are dominated by noise rather than a
shared shock signal.

\begin{figure}[p]
\centering
\includegraphics[width=0.86\linewidth]{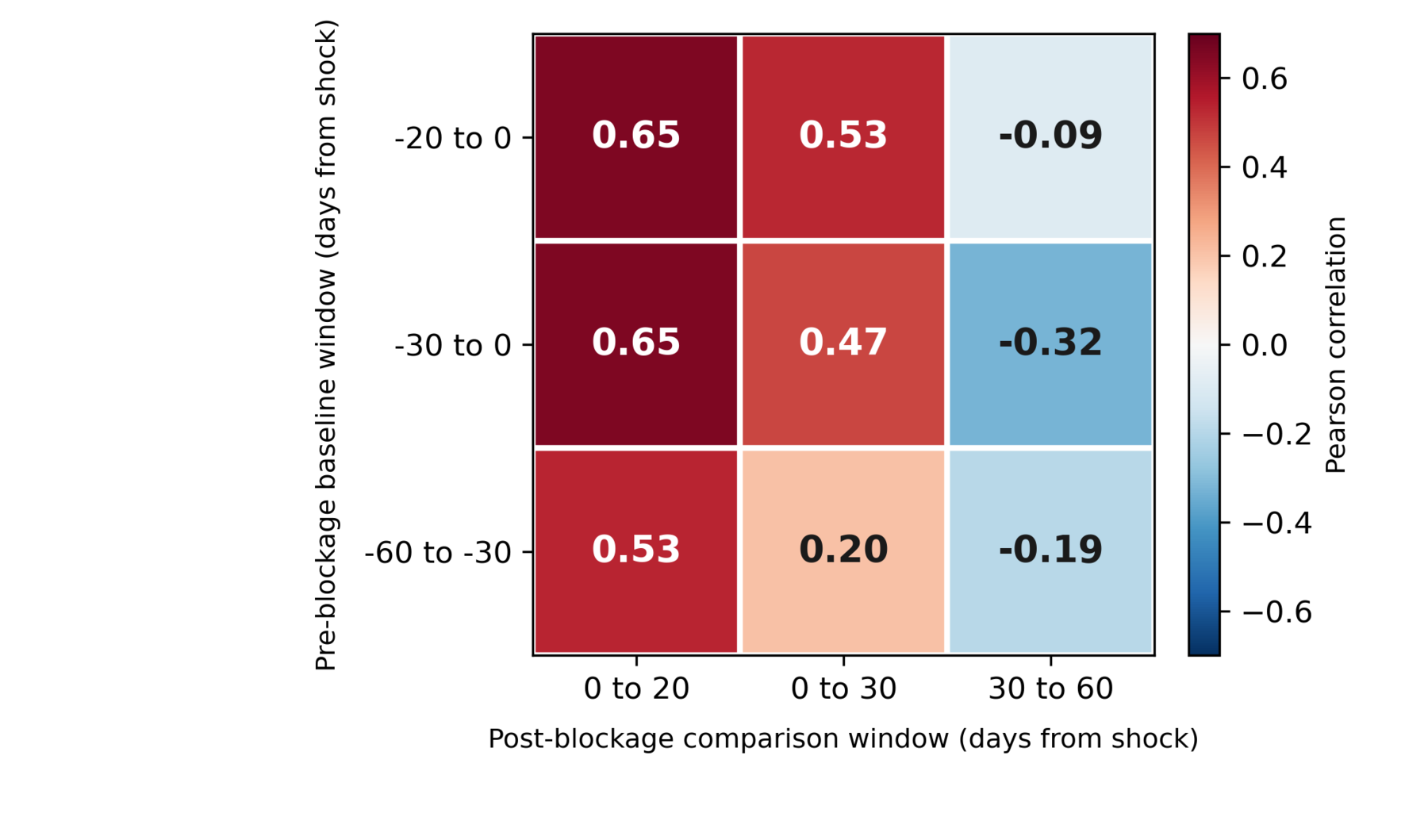}
\caption{Ever Given validation heatmap. Cells report the Pearson correlation
between simulated and observed port-level relative arrival changes, comparing
the indicated pre-blockage baseline window with the indicated post-blockage
window.}
\label{fig:si-ever-given-validation}
\end{figure}
\FloatBarrier

\begin{table}[ht]
\centering
\caption{Steady-state validation correlations for the regenerated calibration
plots.}
\label{tab:assoc_steady_state}
\begin{tabular}{lrrrrr}
\hline
Scope & $n$ ports & Pearson-linear & Pearson-log & Spearman & Kendall \\
\hline
Cargo    & 1570 & 0.9958 & 0.9903 & 0.9915 & 0.9398 \\
Dry Bulk & 1479 & 0.9710 & 0.9825 & 0.9873 & 0.9328 \\
Tanker   & 1549 & 0.9532 & 0.9829 & 0.9863 & 0.9219 \\
\hline
Total    & 1651 & 0.9896 & 0.9887 & 0.9904 & 0.9315 \\
\hline
\end{tabular}
\end{table}

\section{Hormuz}

We use the Strait of Hormuz simulations as an additional geopolitical corridor
scenario. The simulations use closure durations of 14, 28, and 56 days and are
run with an unexpected start and a communicated reopening date. Point color gives normalized net shipping days lost, and point size gives the
corresponding estimated number of missed ship arrivals. Marker sizes use the
56-day panel as the shared scale across all three Hormuz maps, and the color
scale is shared across all three panels.

\begin{figure}[p]
\centering
\includegraphics[width=0.79\linewidth]{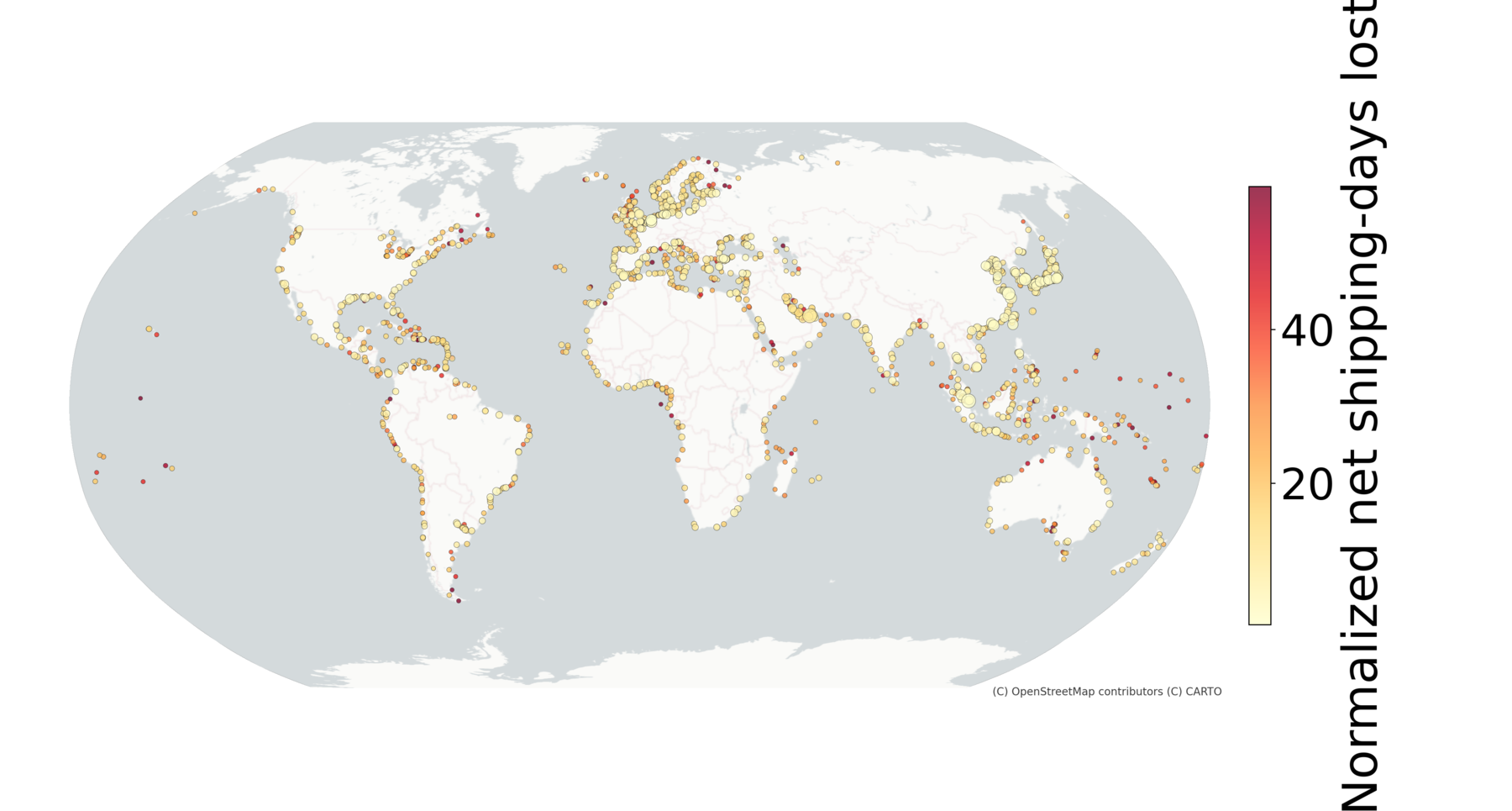}\\[-0.4em]
\includegraphics[width=0.79\linewidth]{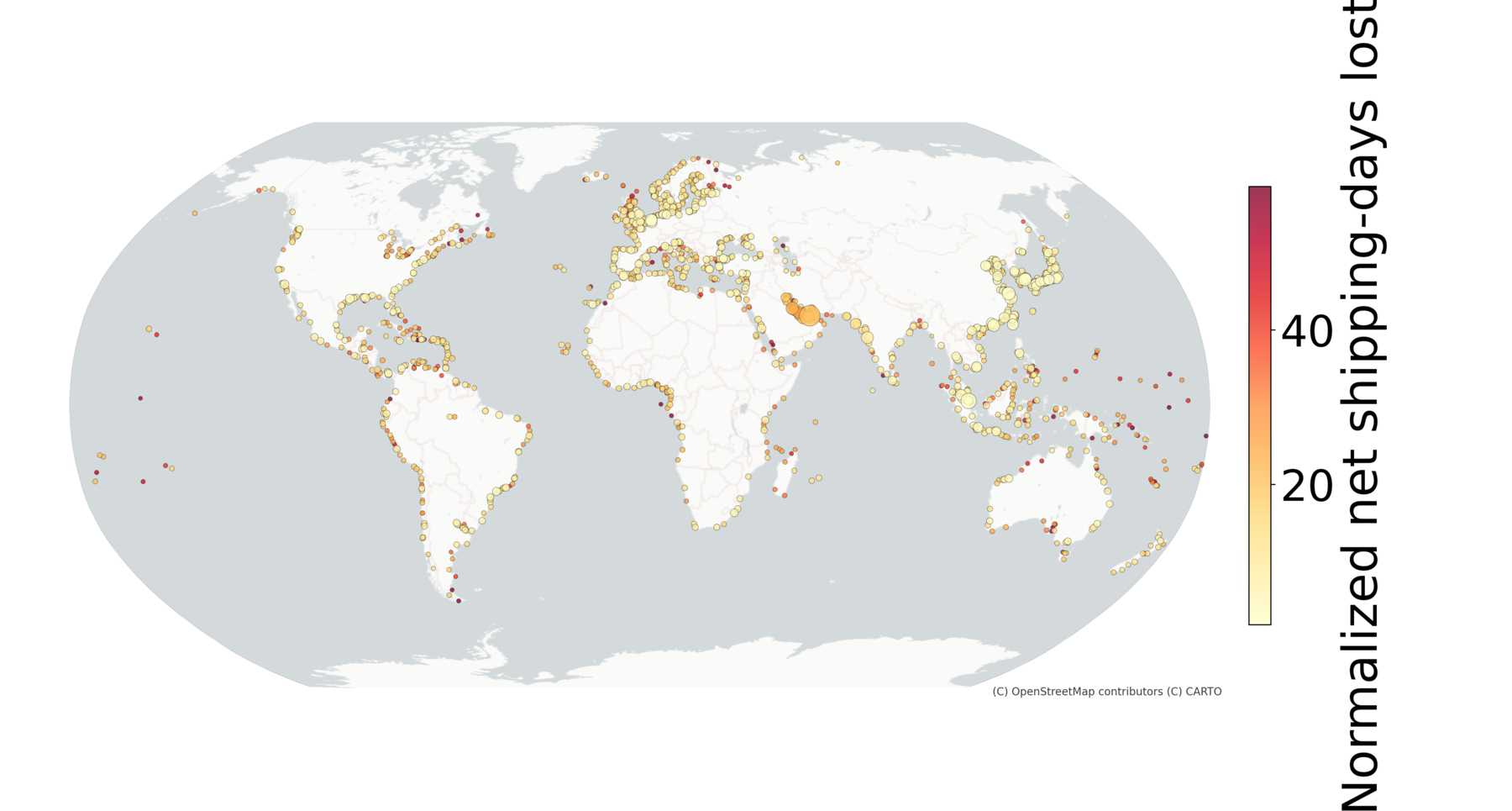}\\[-0.4em]
\includegraphics[width=0.79\linewidth]{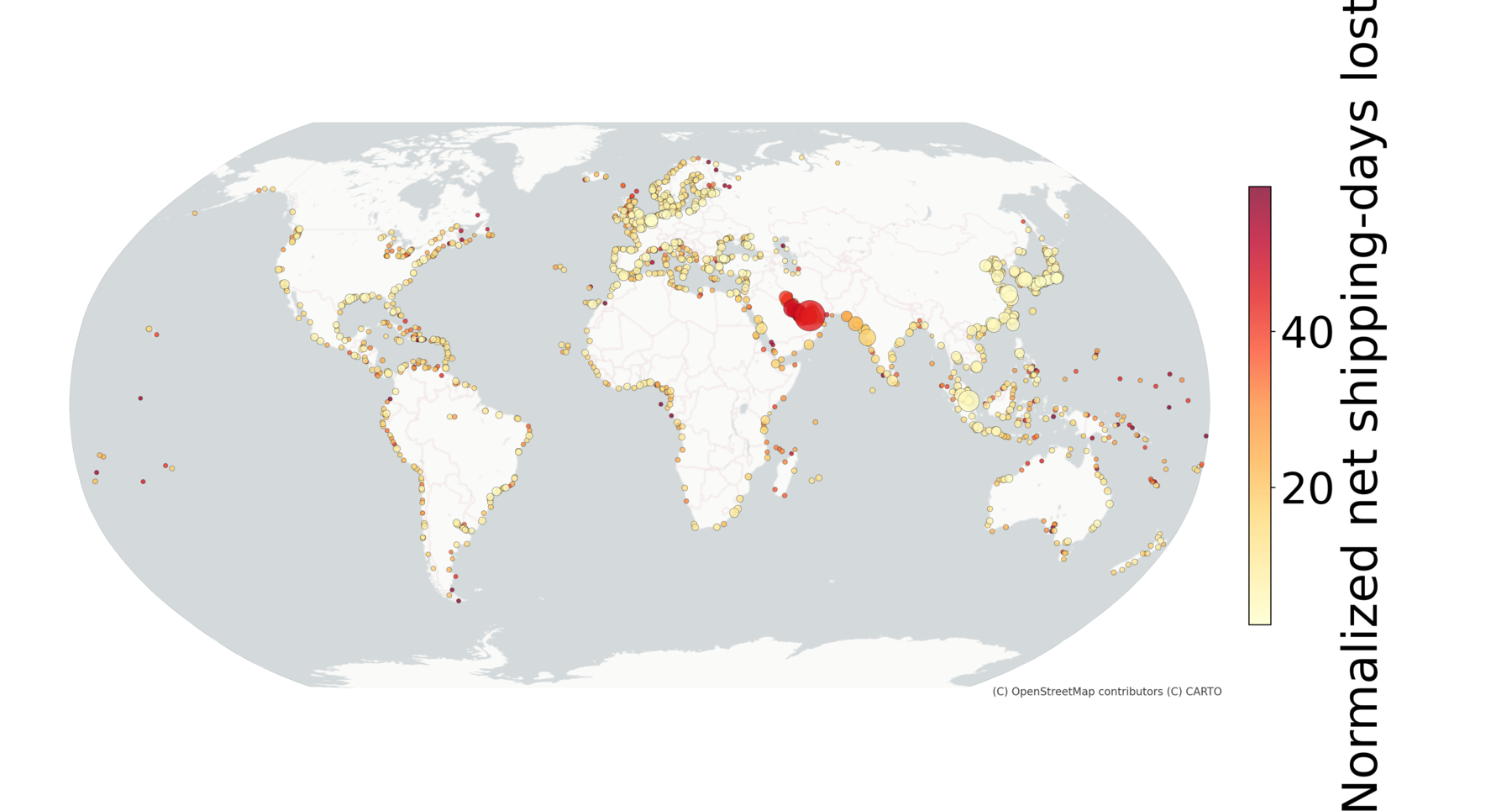}
\caption{Port-level normalized net shipping-day loss for Strait of Hormuz
closures of 14 days, 28 days, and 56 days from top to bottom, shown in the
Robinson projection. Point size gives estimated missed ship arrivals.}
\label{fig:hormuz-maps}
\end{figure}
\FloatBarrier

\begin{table}[p]
\centering
\small
\caption{Baseline-referenced net shipping-days lost for the Strait of Hormuz closure experiments by reporting region, closure duration, and vessel type. Values are computed from seed-level regional arrival time series aggregated before smoothing and normalization. Cargo aggregates container, general cargo, and ro-ro arrivals. Negative values indicate net catch-up or surplus within the evaluation horizon.}
\label{tab:si_hormuz_region_type_net_loss}
\resizebox{\textwidth}{!}{%
\begin{tabular}{lrrrrrrrrrrrr}
\toprule
 & \multicolumn{4}{c}{14 days} & \multicolumn{4}{c}{28 days} & \multicolumn{4}{c}{56 days} \\
\cmidrule(lr){2-5} \cmidrule(lr){6-9} \cmidrule(lr){10-13}
Region & Cargo & Dry bulk & Tanker & Total & Cargo & Dry bulk & Tanker & Total & Cargo & Dry bulk & Tanker & Total \\
\midrule
Americas Atlantic & 0.21 & -0.39 & 0.76 & 0.24 & 0.29 & 0.46 & 1.50 & 0.71 & 1.14 & 1.37 & 1.77 & 1.42 \\
Europe and Mediterranean & -0.71 & 1.59 & 1.61 & 0.20 & -0.82 & 0.74 & 1.99 & 0.17 & -0.52 & 2.15 & 3.06 & 0.80 \\
Africa and Red Sea & 0.11 & 0.69 & 2.57 & 0.76 & 1.03 & 0.89 & 3.73 & 1.63 & 2.76 & 2.86 & 6.64 & 3.69 \\
West, Central, and South Asia & -0.67 & 0.42 & -0.44 & -0.42 & 0.89 & 0.64 & 1.43 & 0.98 & 6.79 & 5.98 & 9.99 & 7.44 \\
East and Southeast Asia & 1.08 & -0.23 & -0.27 & 0.45 & 1.61 & -0.06 & 0.22 & 0.92 & 3.07 & 1.22 & 2.82 & 2.71 \\
Oceania and Pacific & 0.05 & 0.43 & -0.64 & 0.12 & 1.14 & 0.46 & -0.76 & 0.55 & 0.85 & 2.48 & 0.89 & 1.65 \\
European Union (with UK) & -0.68 & 1.51 & 1.79 & 0.26 & -0.56 & 0.80 & 2.20 & 0.39 & -0.24 & 2.11 & 3.31 & 1.01 \\
USA & -0.22 & -0.17 & 1.32 & 0.34 & 0.51 & -0.13 & 1.73 & 0.80 & 1.26 & 0.75 & 1.98 & 1.44 \\
Mainland China (+HK/Macau) & 0.92 & 0.17 & -0.82 & 0.38 & 1.67 & 1.01 & 0.43 & 1.25 & 3.35 & 2.03 & 4.47 & 3.27 \\
Global & 0.24 & 0.16 & 0.45 & 0.29 & 0.68 & 0.33 & 1.07 & 0.74 & 2.18 & 2.06 & 3.43 & 2.54 \\
\bottomrule
\end{tabular}%
}
\end{table}

\FloatBarrier

\section{Regional Results}
\begin{figure}[H]
\centering
\setlength{\tabcolsep}{2pt}
\renewcommand{\arraystretch}{0.78}
\begin{tabular}{ccc}
\begin{minipage}[t]{0.32\linewidth}\centering
\includegraphics[width=\linewidth,trim=8 8 8 8,clip]{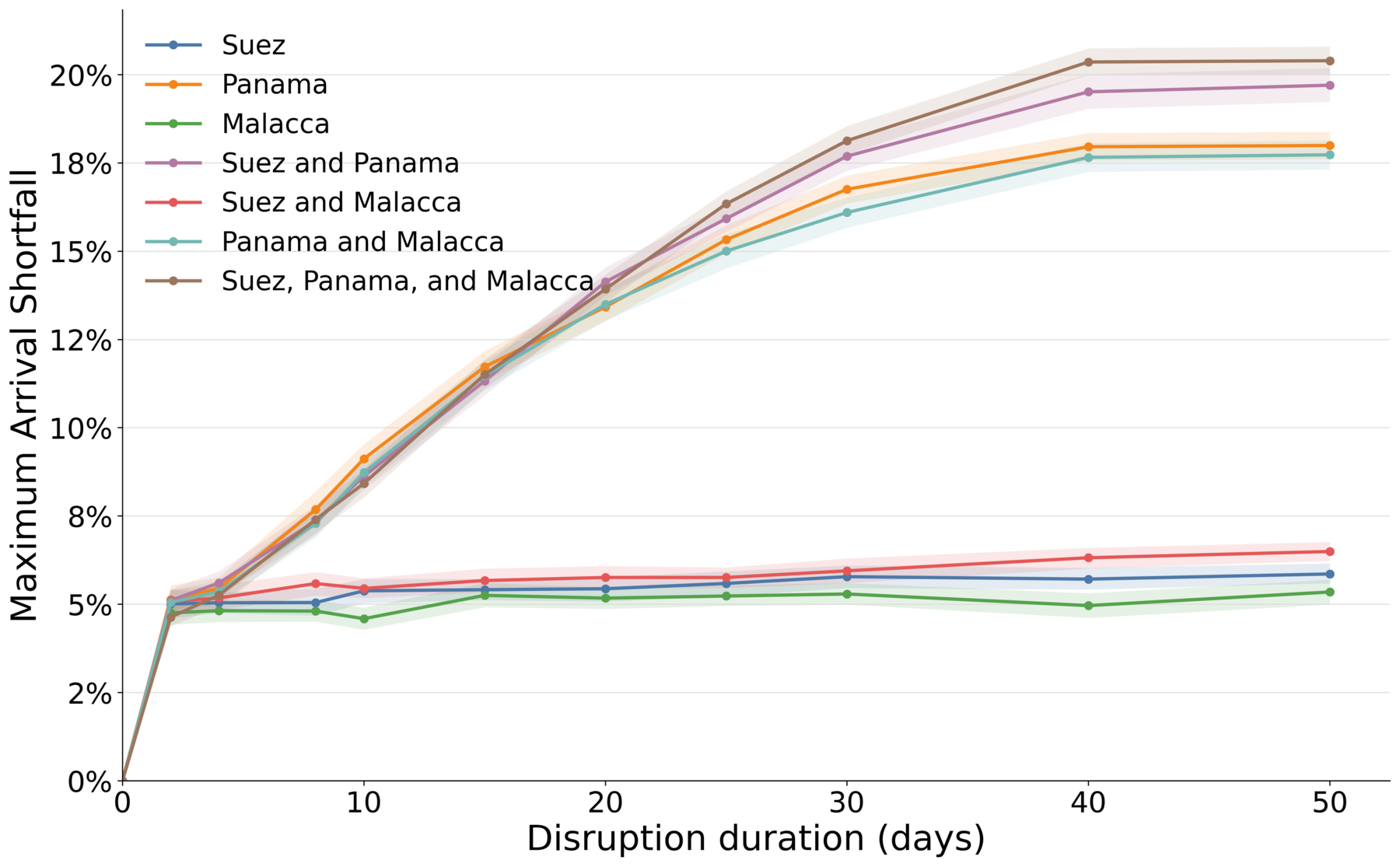}
\scriptsize Americas Atlantic
\end{minipage} &
\begin{minipage}[t]{0.32\linewidth}\centering
\includegraphics[width=\linewidth,trim=8 8 8 8,clip]{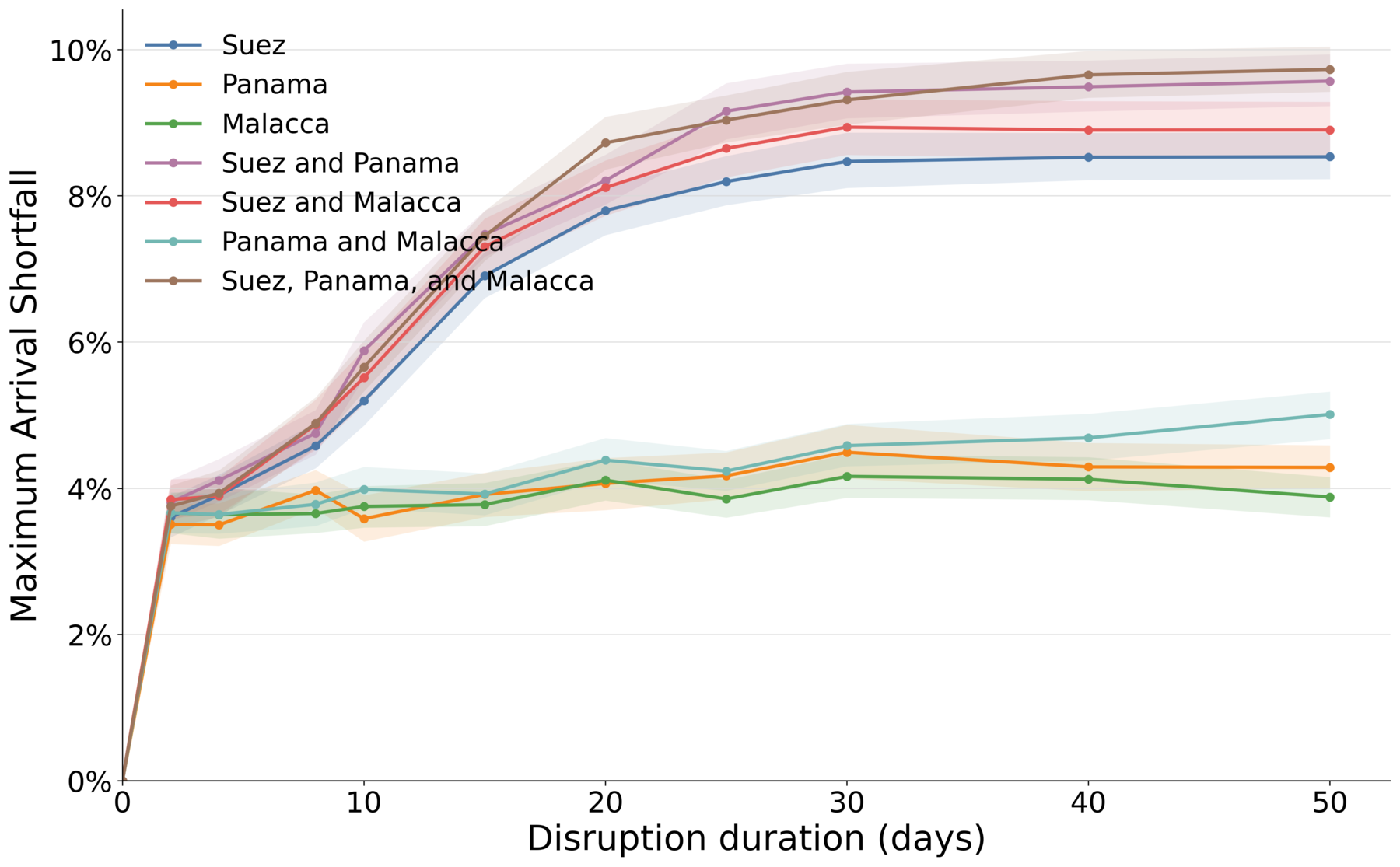}
\scriptsize Europe and Mediterranean
\end{minipage} &
\begin{minipage}[t]{0.32\linewidth}\centering
\includegraphics[width=\linewidth,trim=8 8 8 8,clip]{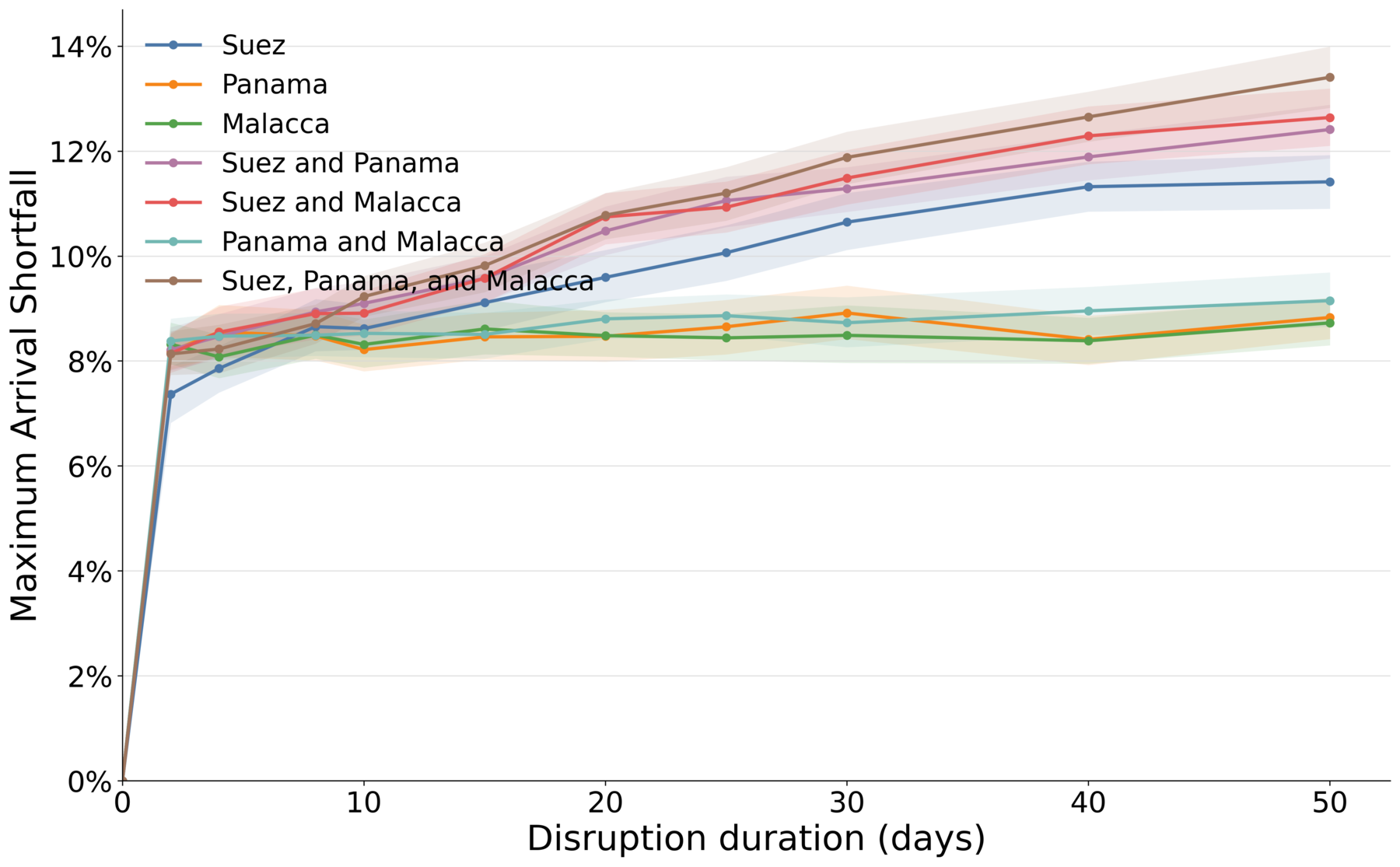}
\scriptsize Africa and Red Sea
\end{minipage} \\
\begin{minipage}[t]{0.32\linewidth}\centering
\includegraphics[width=\linewidth,trim=8 8 8 8,clip]{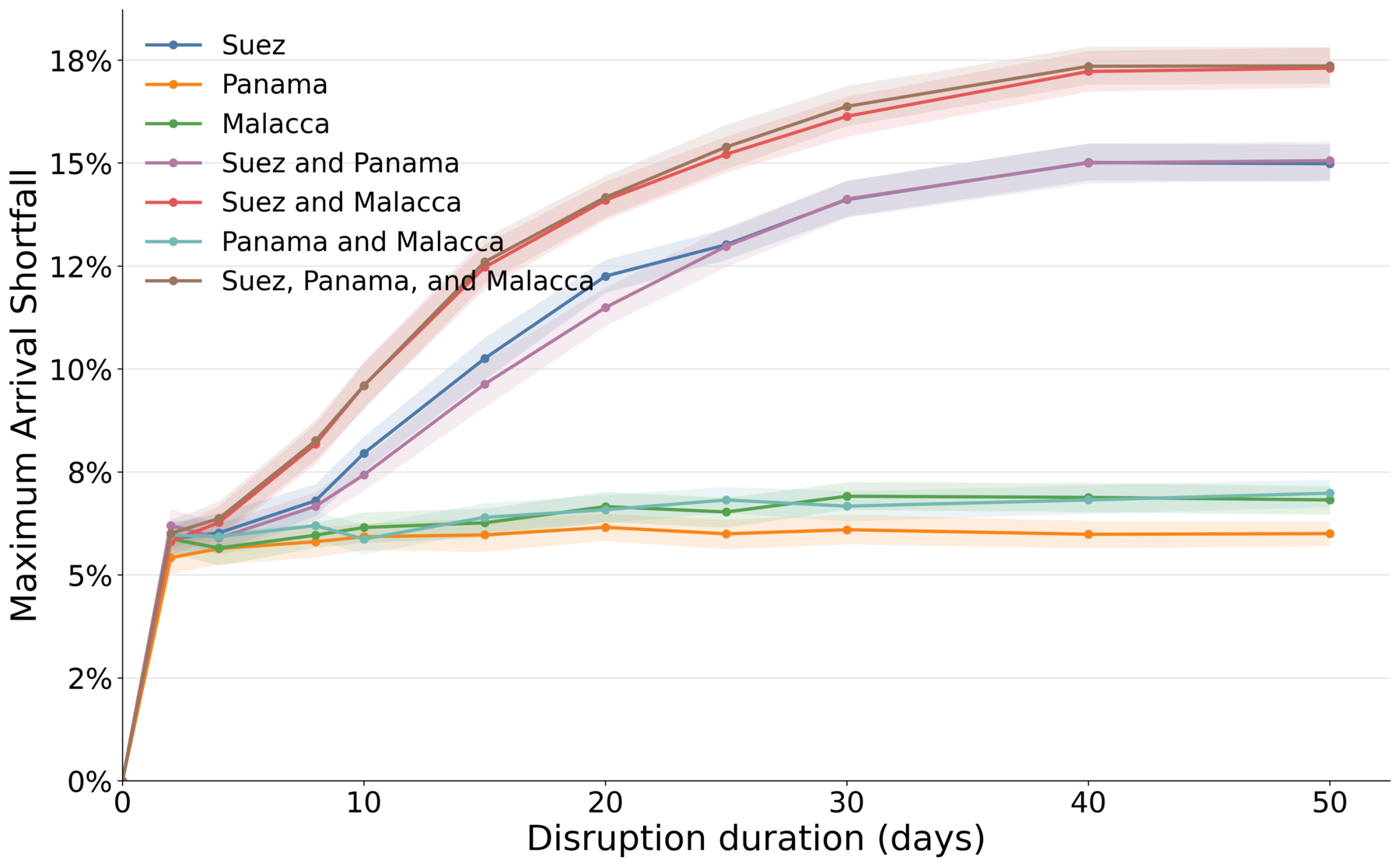}
\scriptsize West Central and South Asia
\end{minipage} &
\begin{minipage}[t]{0.32\linewidth}\centering
\includegraphics[width=\linewidth,trim=8 8 8 8,clip]{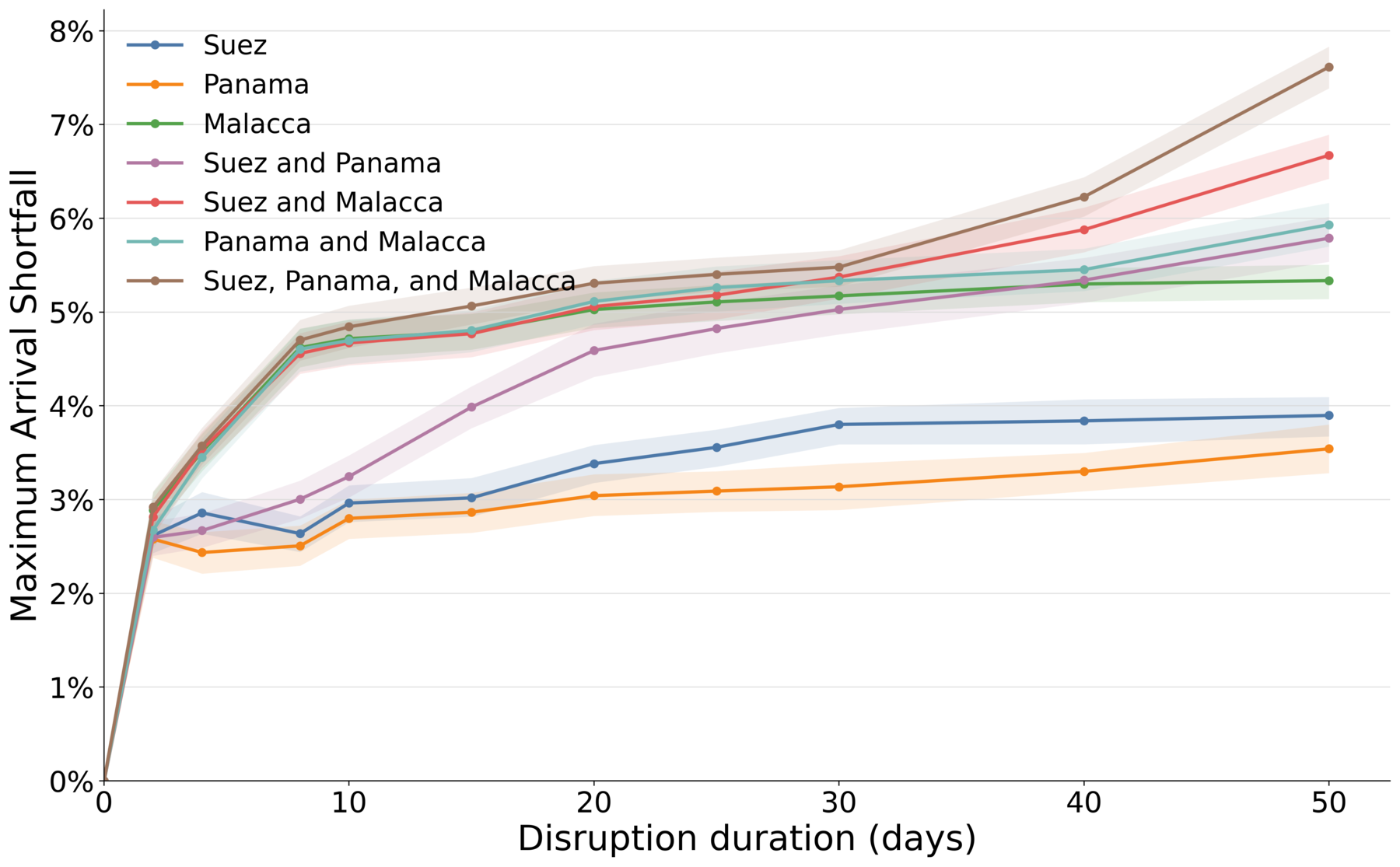}
\scriptsize East and Southeast Asia
\end{minipage} &
\begin{minipage}[t]{0.32\linewidth}\centering
\includegraphics[width=\linewidth,trim=8 8 8 8,clip]{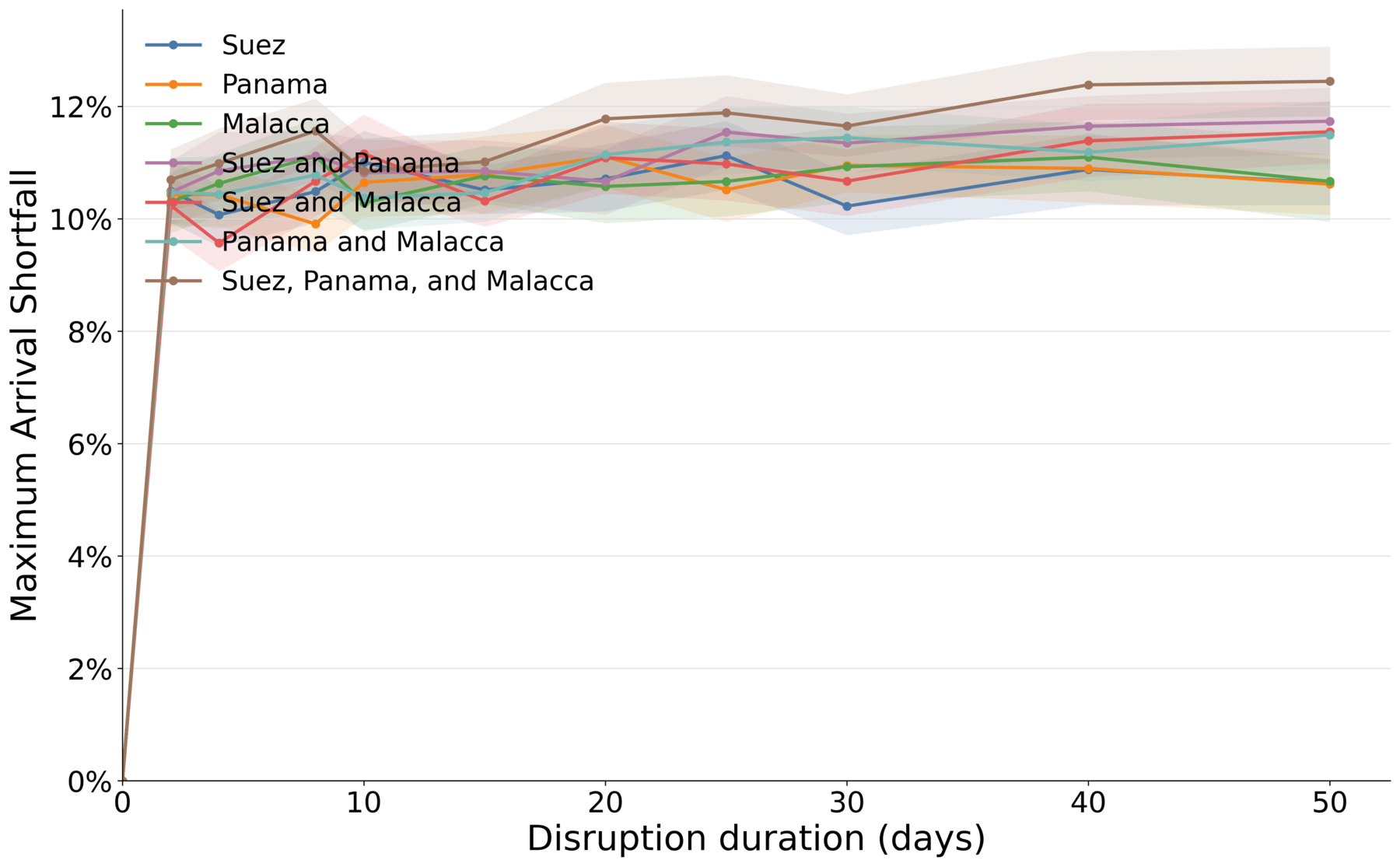}
\scriptsize Oceania and Pacific
\end{minipage} \\
\begin{minipage}[t]{0.32\linewidth}\centering
\includegraphics[width=\linewidth,trim=8 8 8 8,clip]{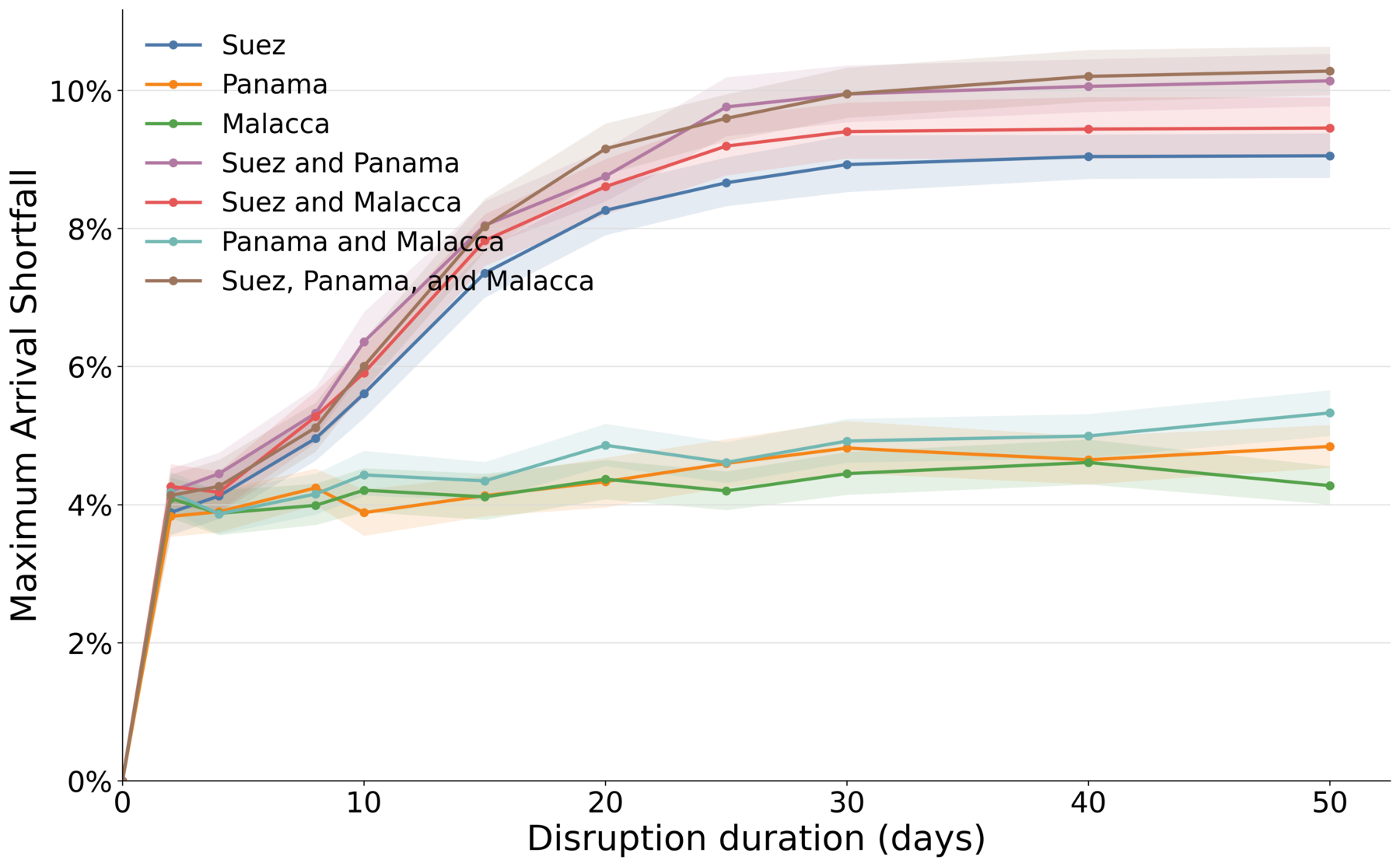}
\scriptsize European Union with UK
\end{minipage} &
\begin{minipage}[t]{0.32\linewidth}\centering
\includegraphics[width=\linewidth,trim=8 8 8 8,clip]{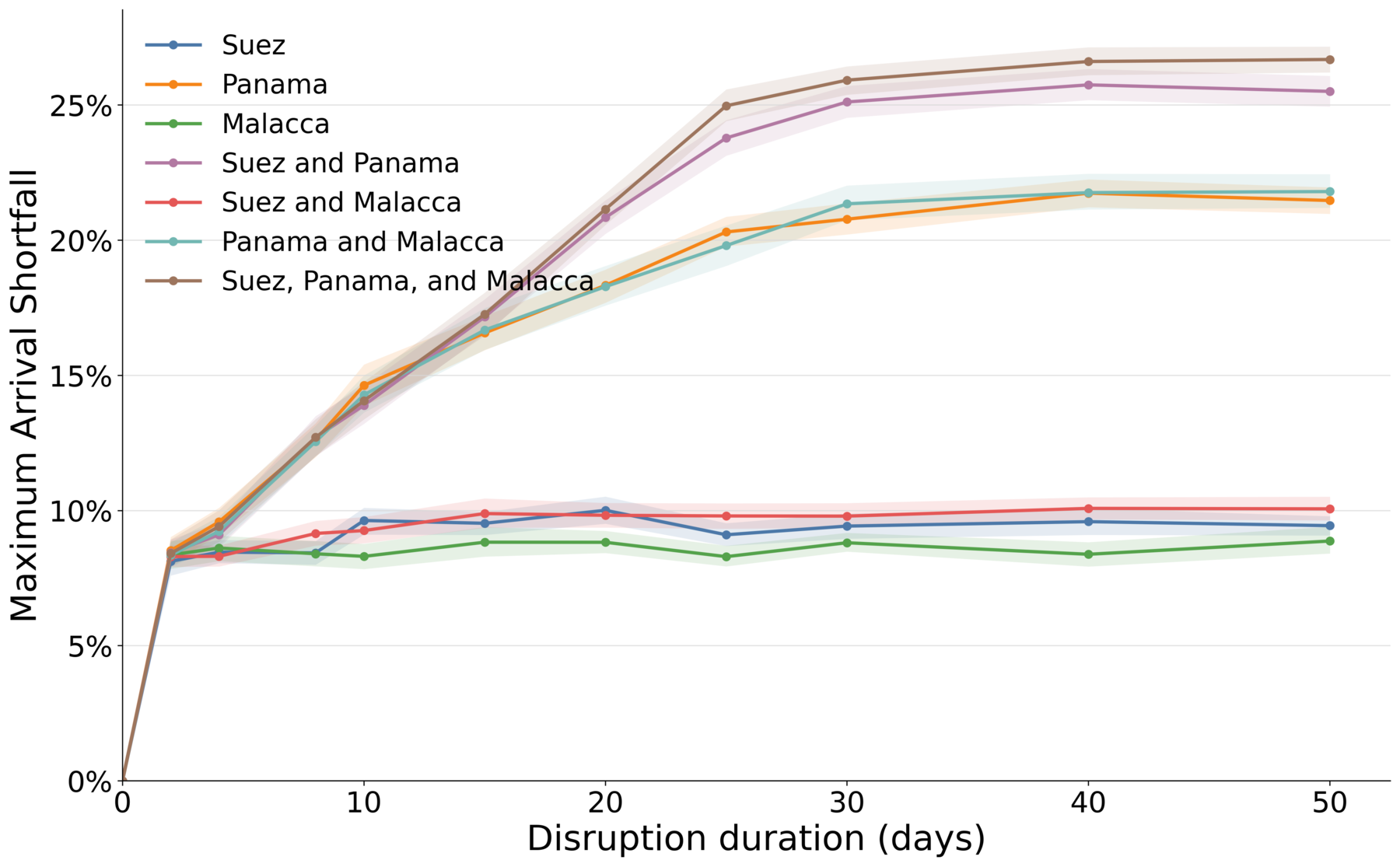}
\scriptsize USA
\end{minipage} &
\begin{minipage}[t]{0.32\linewidth}\centering
\includegraphics[width=\linewidth,trim=8 8 8 8,clip]{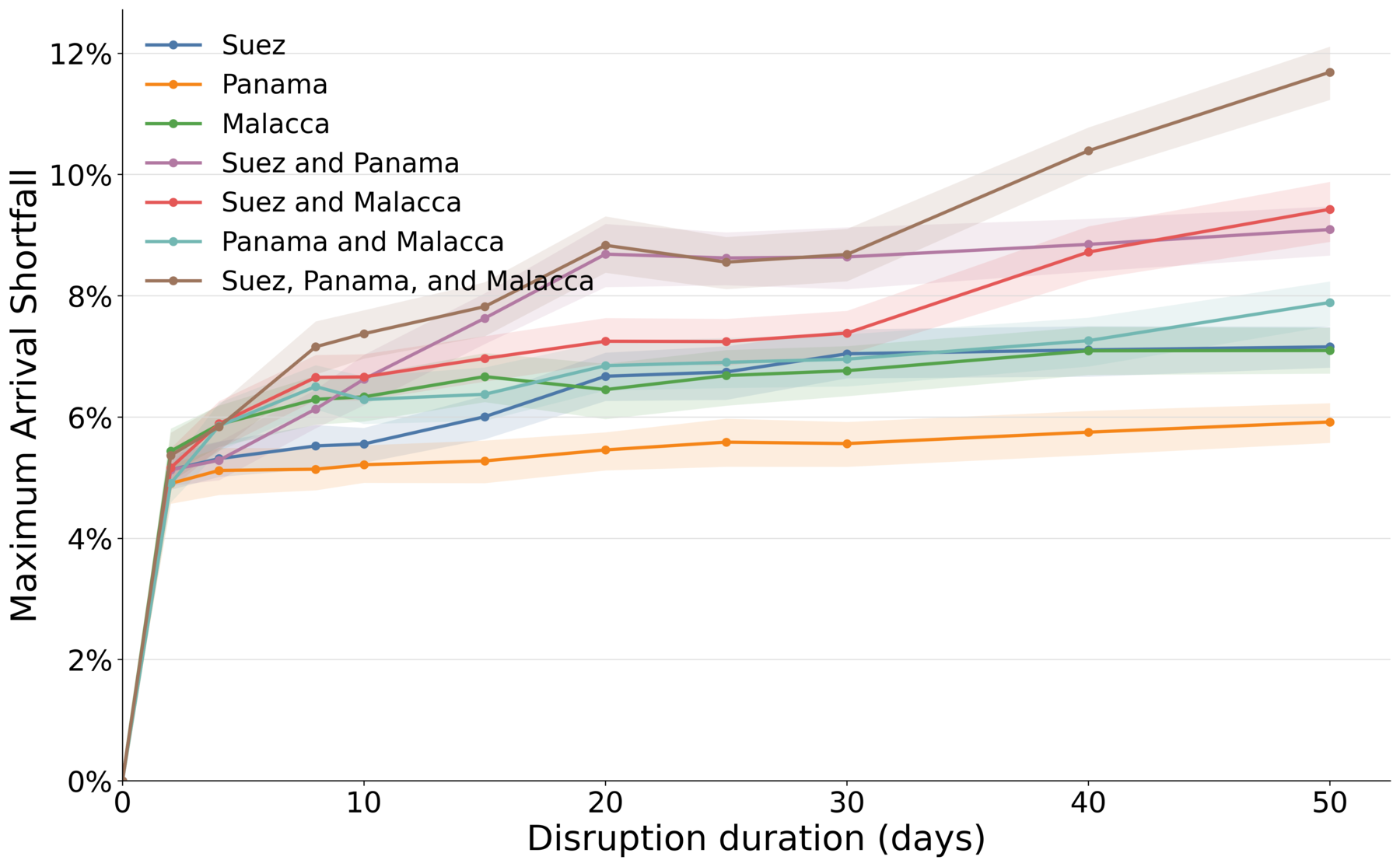}
\scriptsize Mainland China HK Macau
\end{minipage}
\end{tabular}
\caption{Regional maximum arrival shortfall across closure durations.}
\label{fig:regional_tandem_maximum_shortfall}
\end{figure}
\FloatBarrier

\begin{figure}[H]
\centering
\setlength{\tabcolsep}{2pt}
\renewcommand{\arraystretch}{0.78}
\begin{tabular}{ccc}
\begin{minipage}[t]{0.32\linewidth}\centering
\includegraphics[width=\linewidth,trim=8 8 8 8,clip]{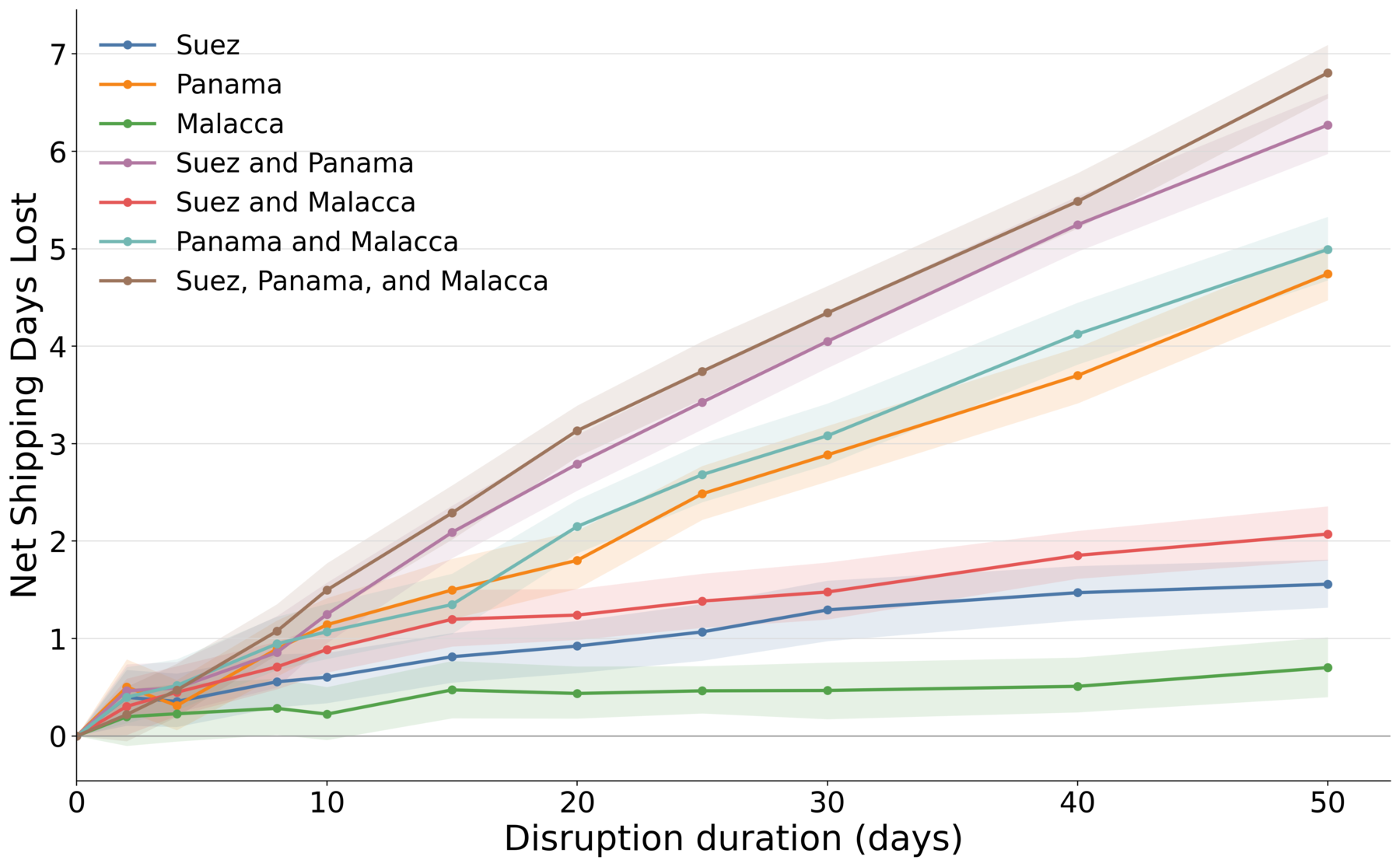}
\scriptsize Americas Atlantic
\end{minipage} &
\begin{minipage}[t]{0.32\linewidth}\centering
\includegraphics[width=\linewidth,trim=8 8 8 8,clip]{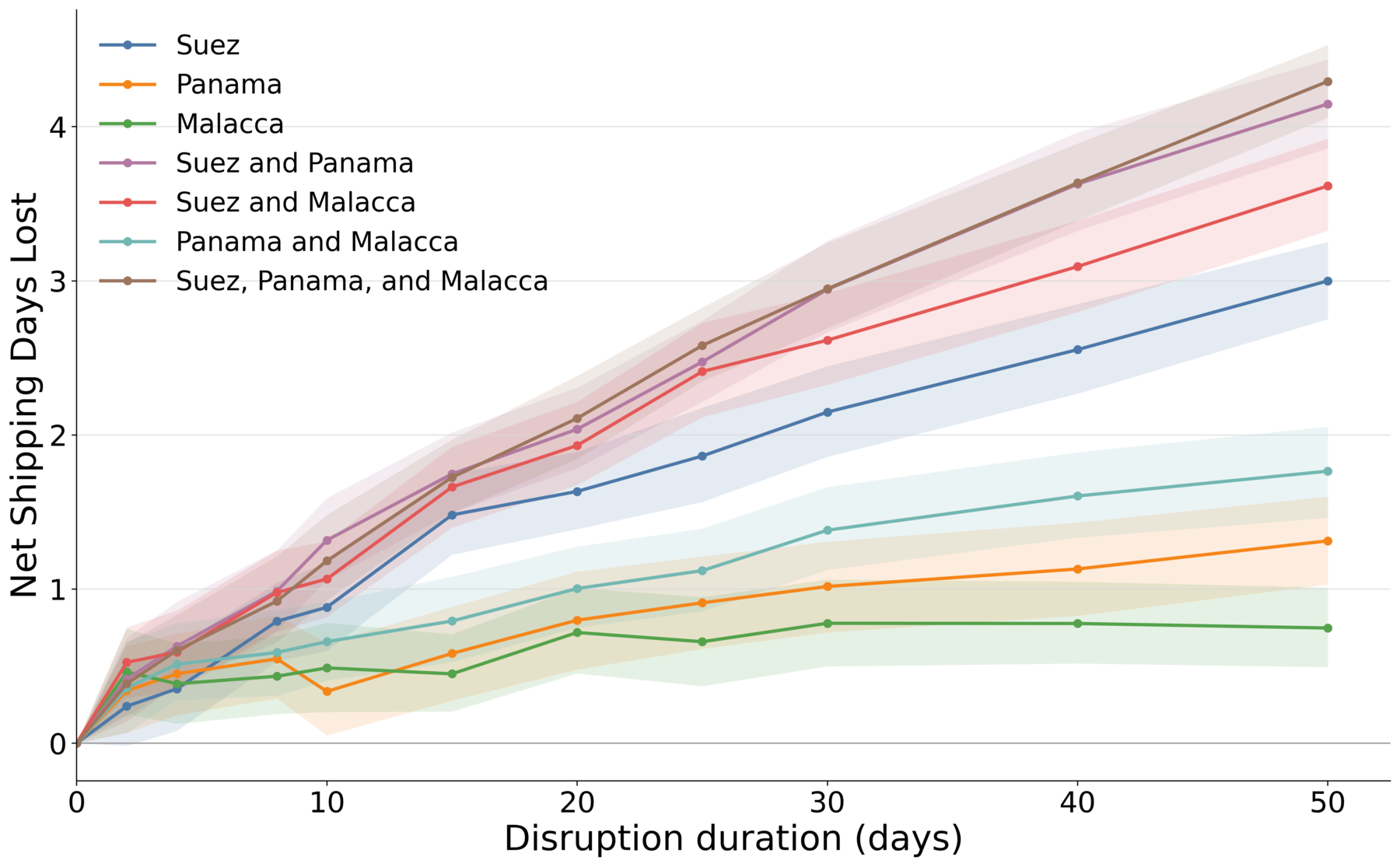}
\scriptsize Europe and Mediterranean
\end{minipage} &
\begin{minipage}[t]{0.32\linewidth}\centering
\includegraphics[width=\linewidth,trim=8 8 8 8,clip]{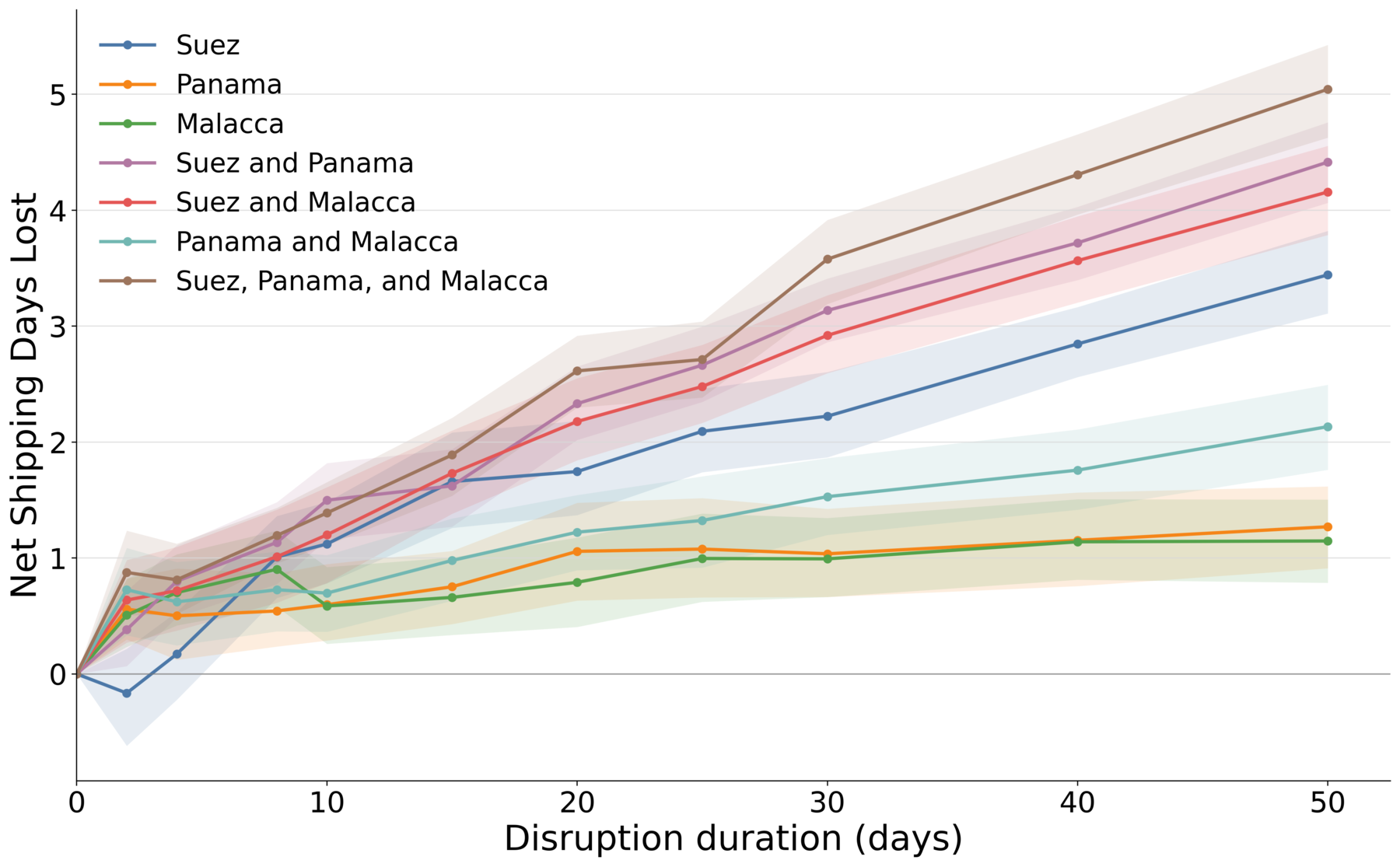}
\scriptsize Africa and Red Sea
\end{minipage} \\
\begin{minipage}[t]{0.32\linewidth}\centering
\includegraphics[width=\linewidth,trim=8 8 8 8,clip]{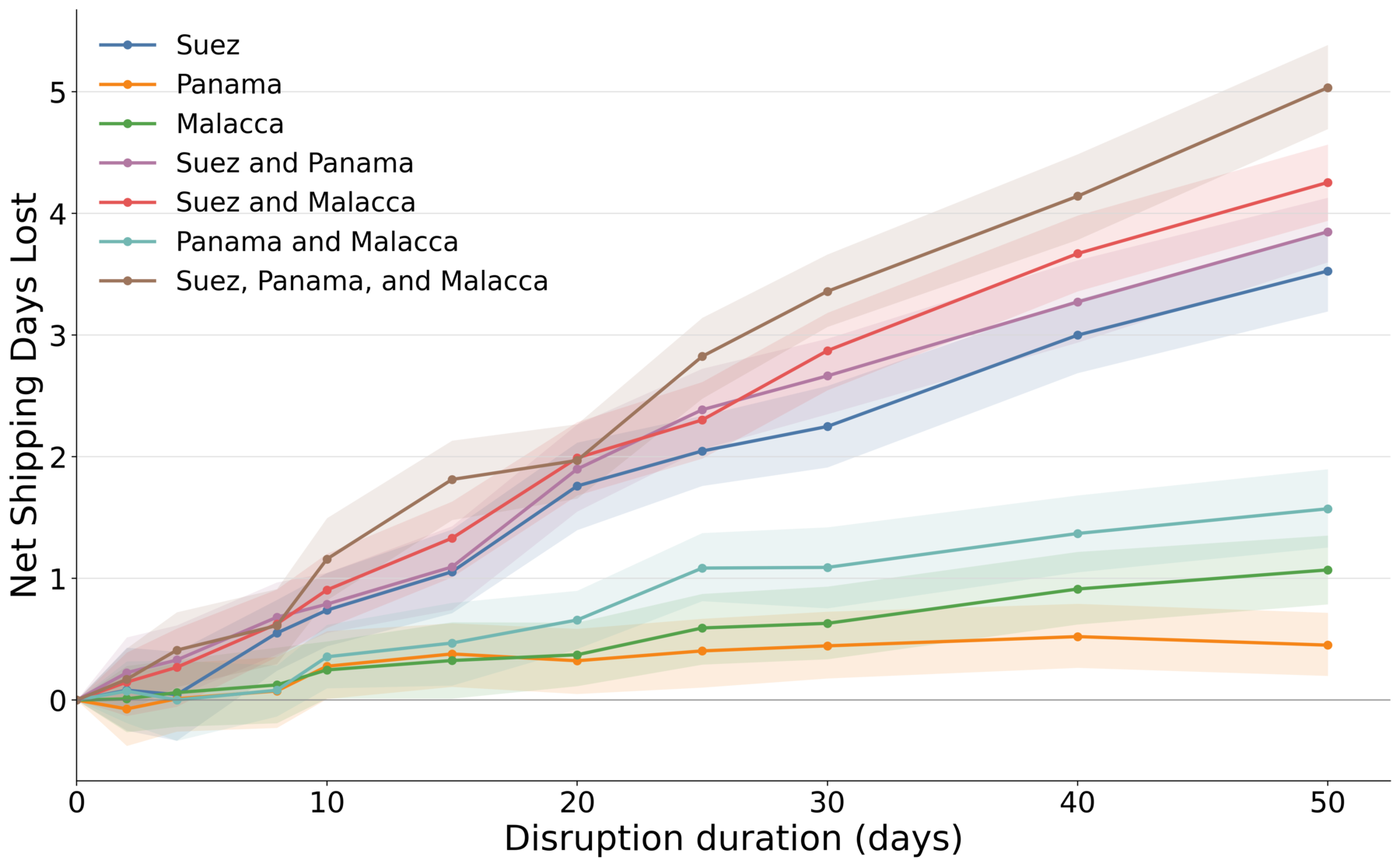}
\scriptsize West Central and South Asia
\end{minipage} &
\begin{minipage}[t]{0.32\linewidth}\centering
\includegraphics[width=\linewidth,trim=8 8 8 8,clip]{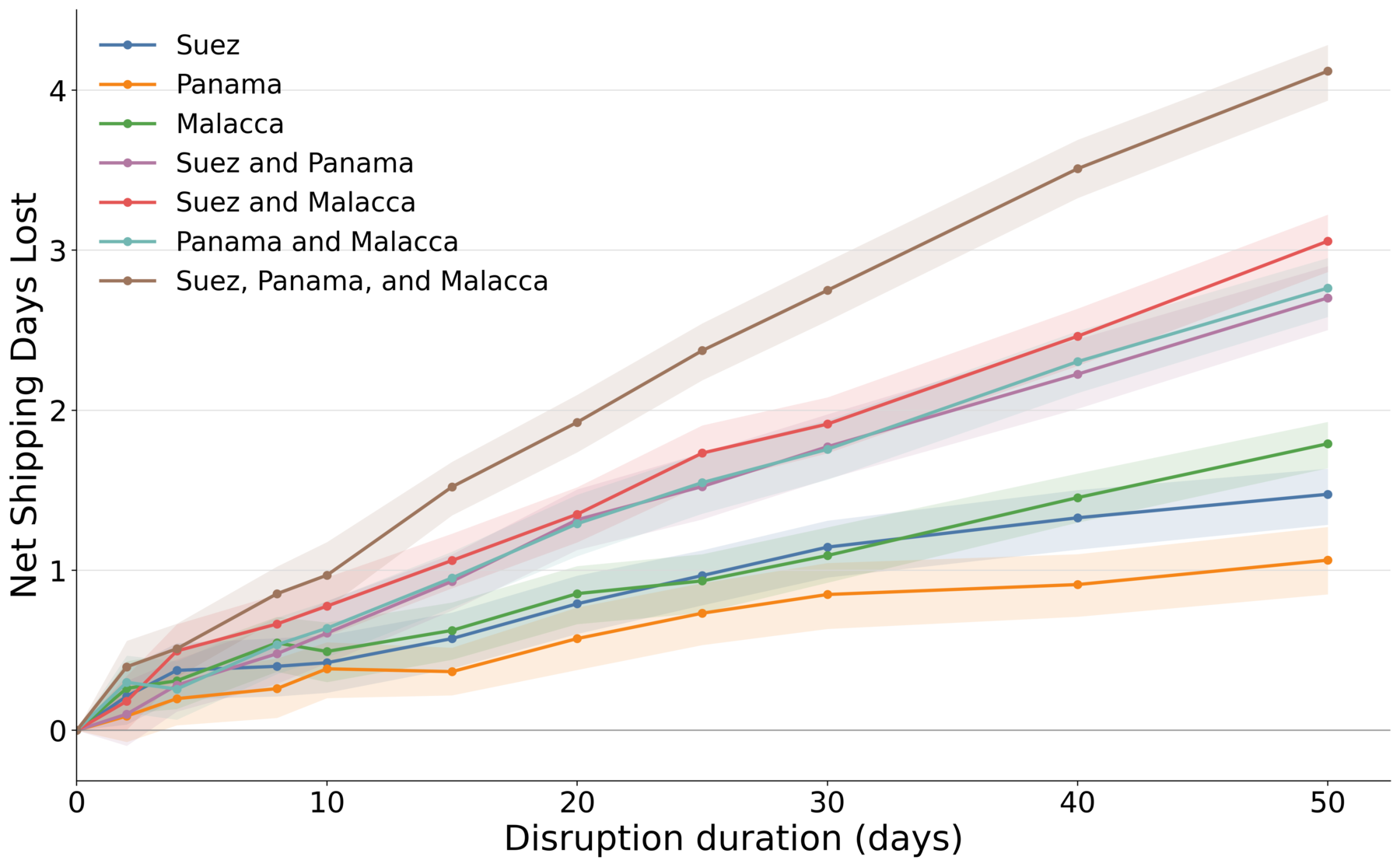}
\scriptsize East and Southeast Asia
\end{minipage} &
\begin{minipage}[t]{0.32\linewidth}\centering
\includegraphics[width=\linewidth,trim=8 8 8 8,clip]{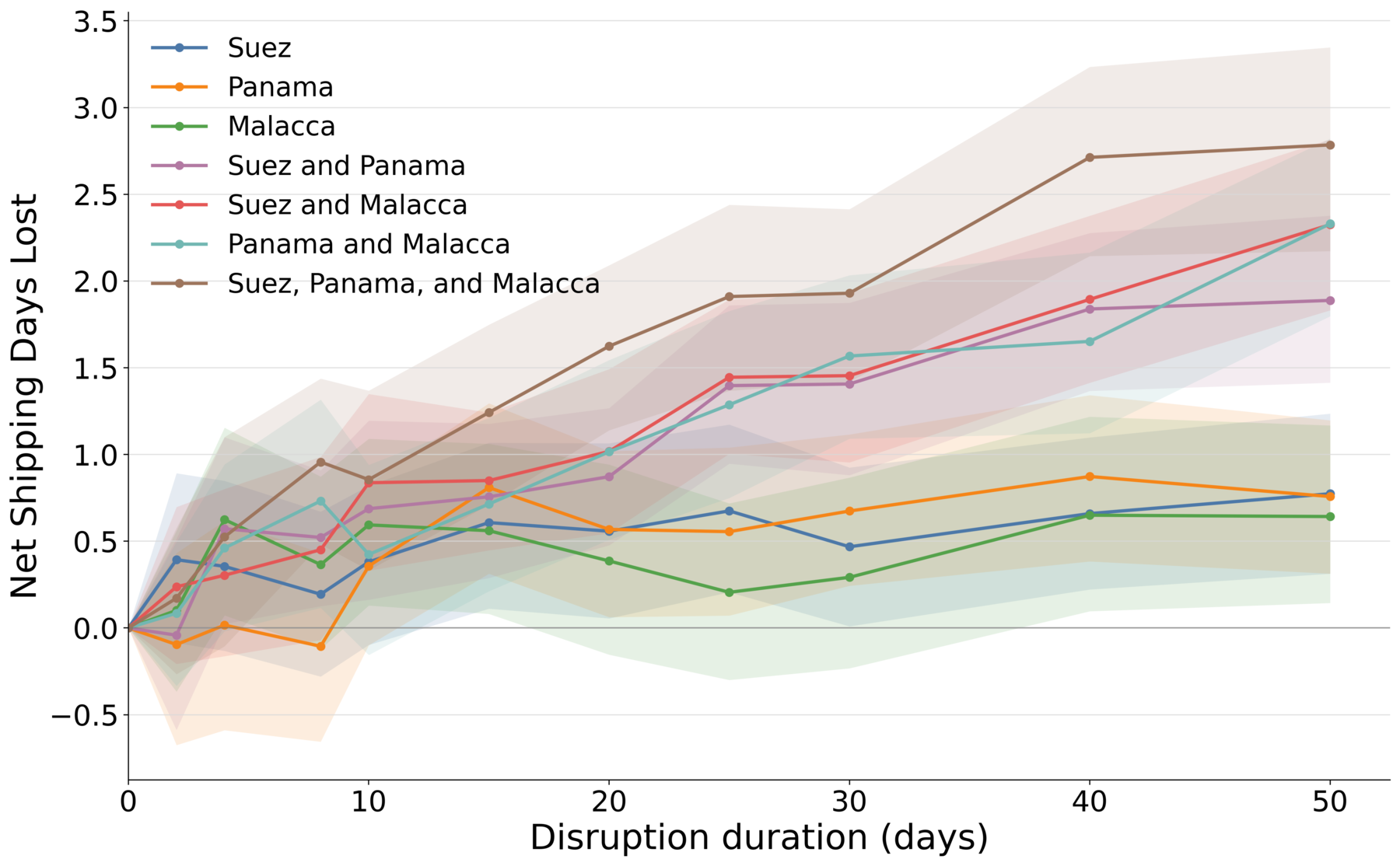}
\scriptsize Oceania and Pacific
\end{minipage} \\
\begin{minipage}[t]{0.32\linewidth}\centering
\includegraphics[width=\linewidth,trim=8 8 8 8,clip]{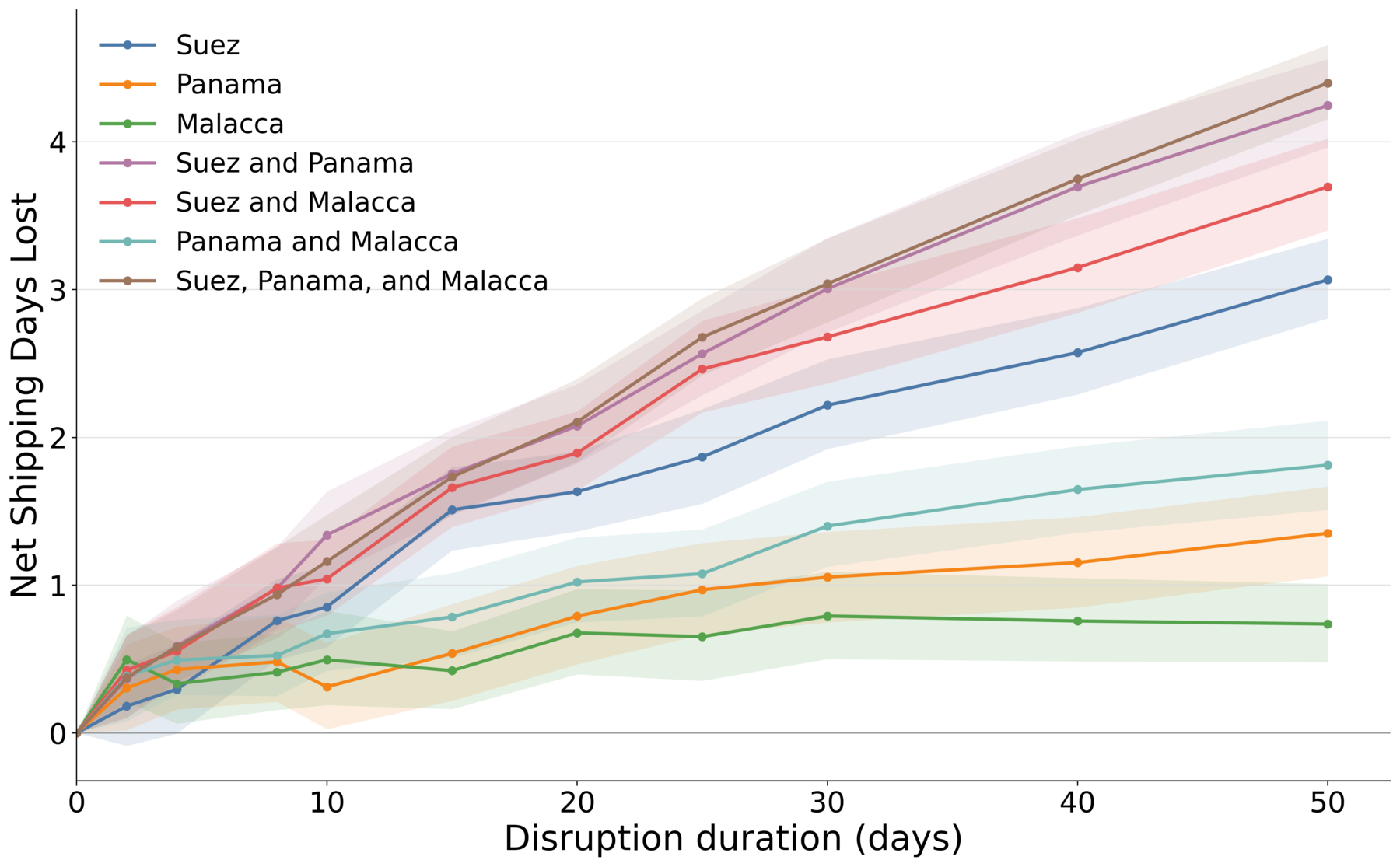}
\scriptsize European Union with UK
\end{minipage} &
\begin{minipage}[t]{0.32\linewidth}\centering
\includegraphics[width=\linewidth,trim=8 8 8 8,clip]{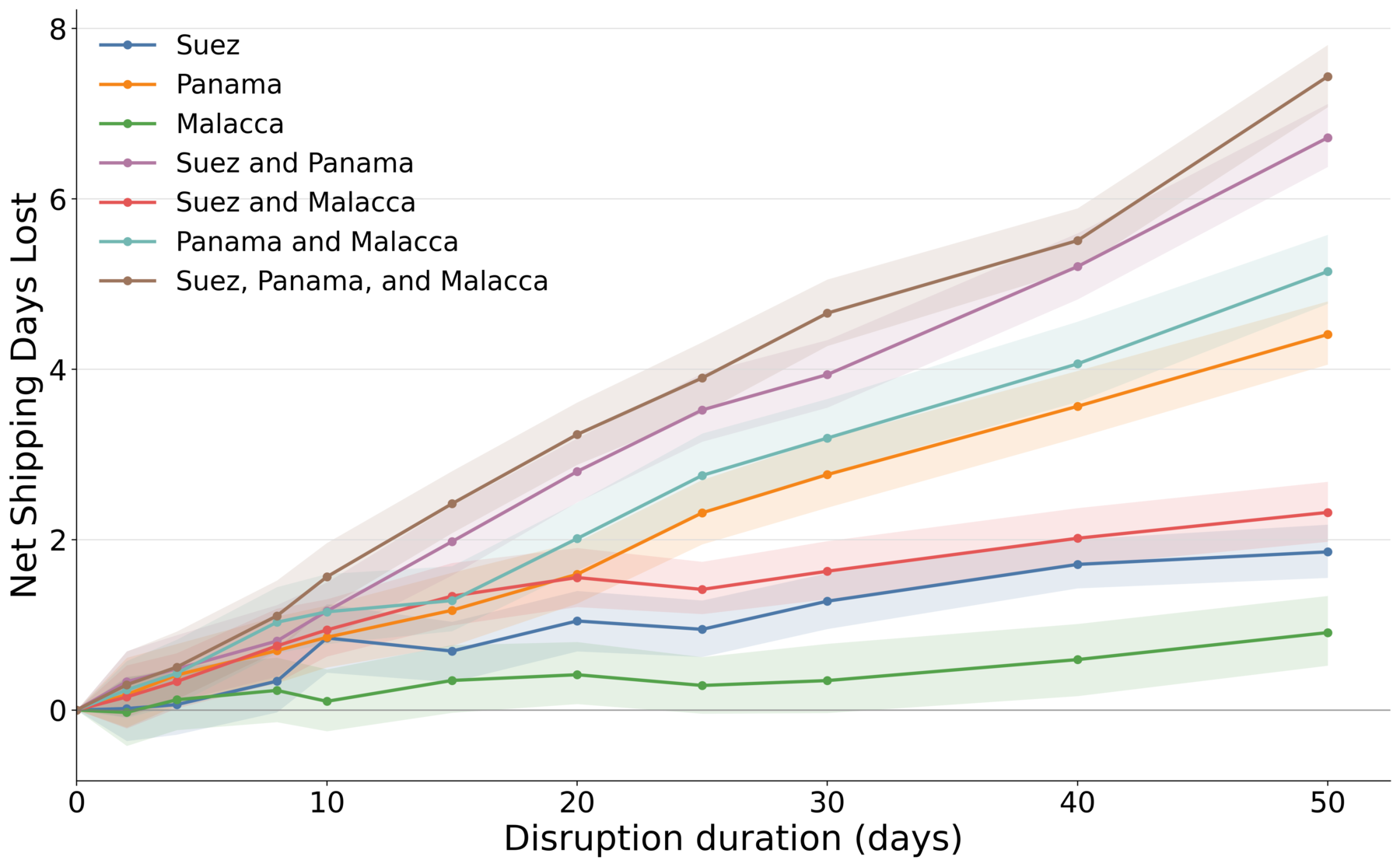}
\scriptsize USA
\end{minipage} &
\begin{minipage}[t]{0.32\linewidth}\centering
\includegraphics[width=\linewidth,trim=8 8 8 8,clip]{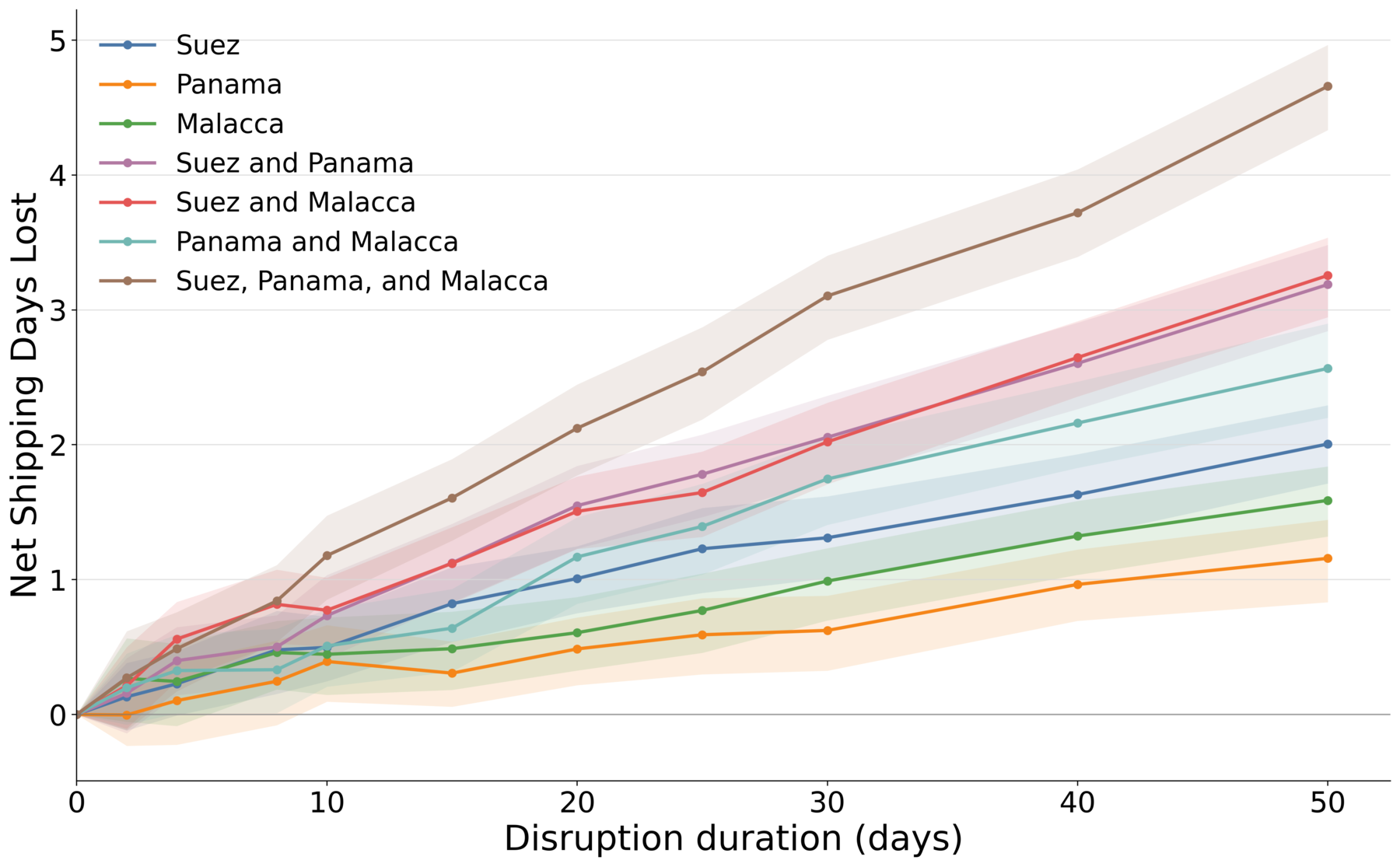}
\scriptsize Mainland China HK Macau
\end{minipage}
\end{tabular}
\caption{Regional net shipping-days lost across closure durations.}
\label{fig:regional_tandem_net_shipping_days_lost}
\end{figure}
\FloatBarrier

\begin{figure}[H]
\centering
\setlength{\tabcolsep}{2pt}
\renewcommand{\arraystretch}{0.78}
\begin{tabular}{ccc}
\begin{minipage}[t]{0.32\linewidth}\centering
\includegraphics[width=\linewidth,trim=8 8 8 8,clip]{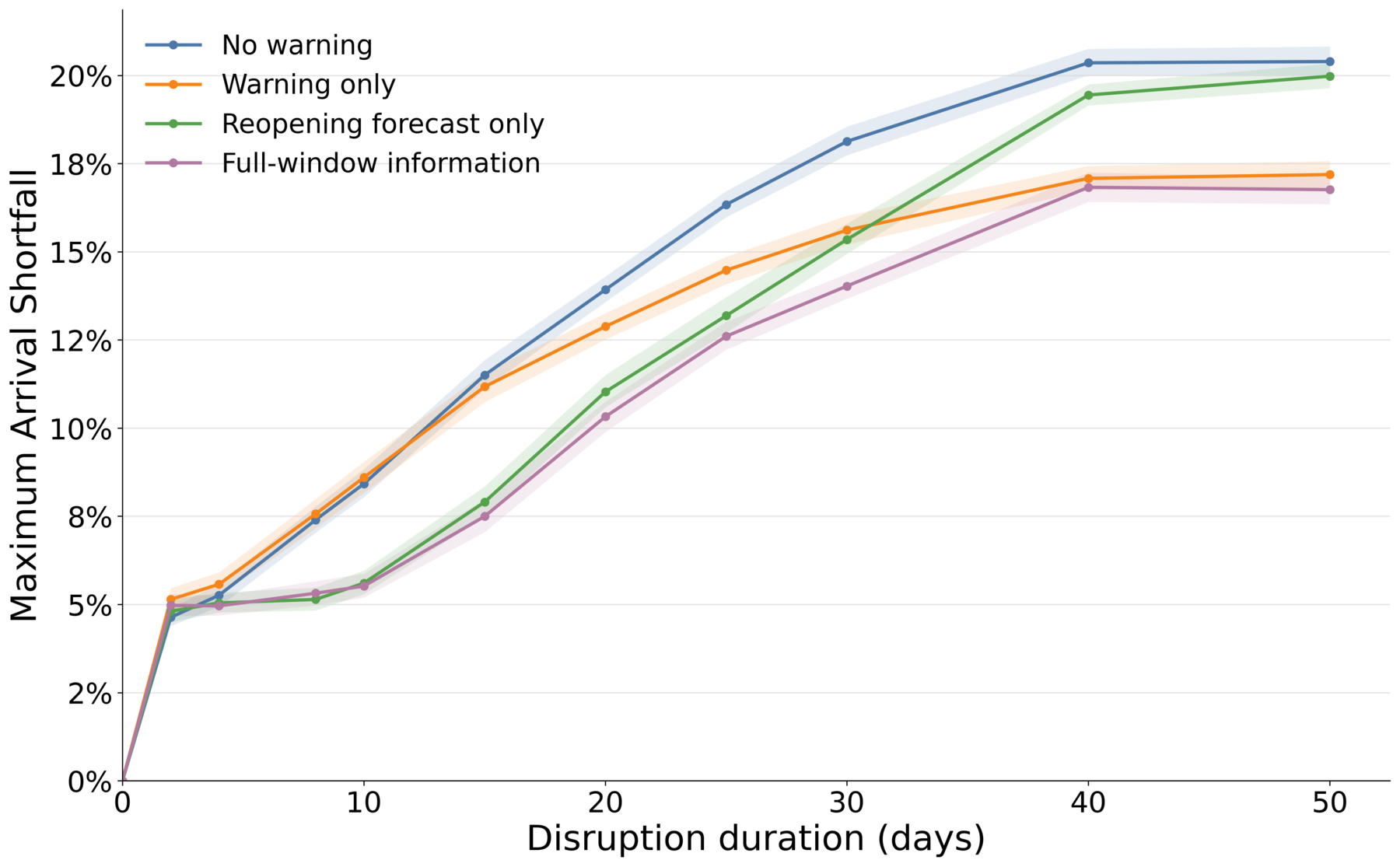}
\scriptsize Americas Atlantic
\end{minipage} &
\begin{minipage}[t]{0.32\linewidth}\centering
\includegraphics[width=\linewidth,trim=8 8 8 8,clip]{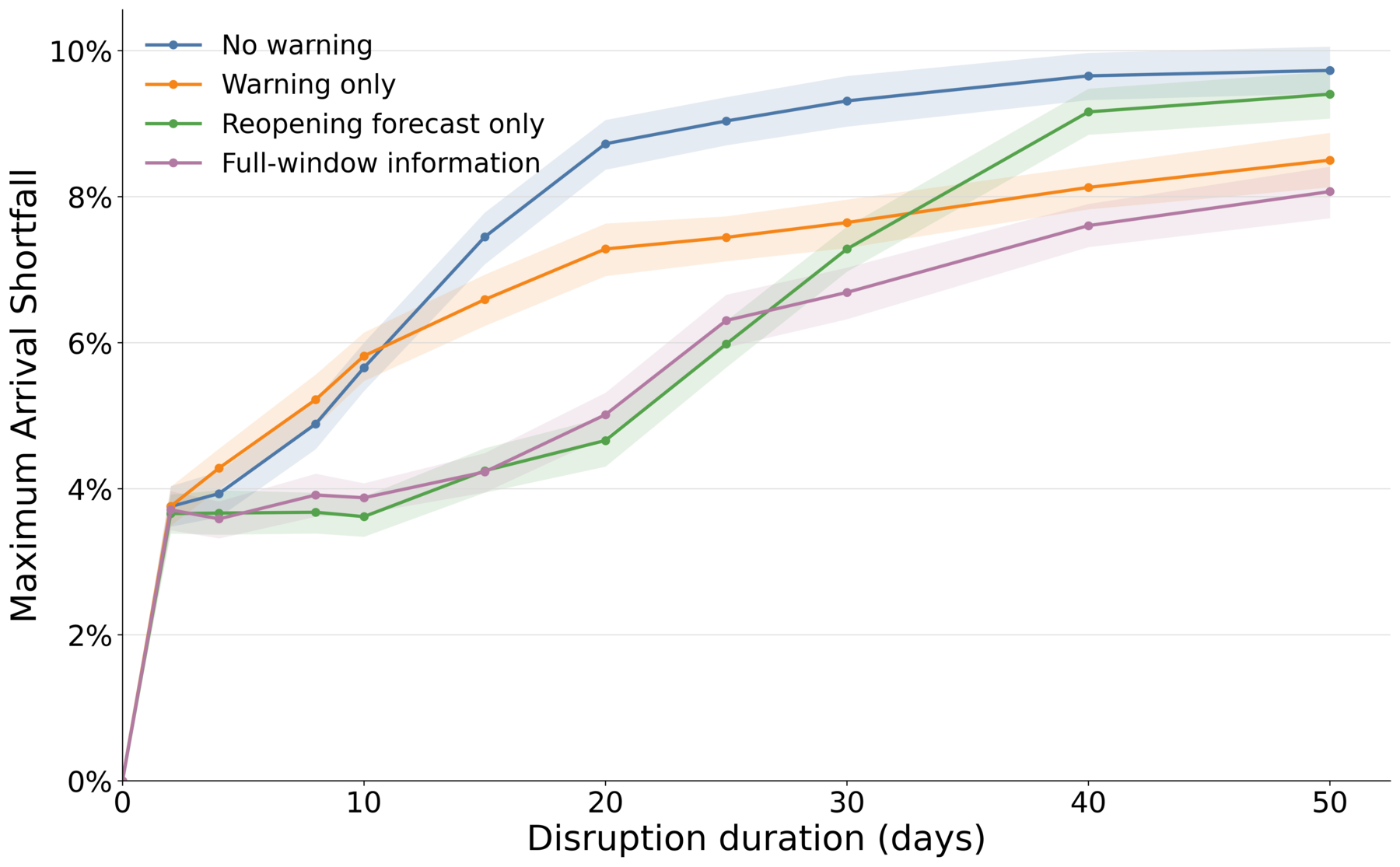}
\scriptsize Europe and Mediterranean
\end{minipage} &
\begin{minipage}[t]{0.32\linewidth}\centering
\includegraphics[width=\linewidth,trim=8 8 8 8,clip]{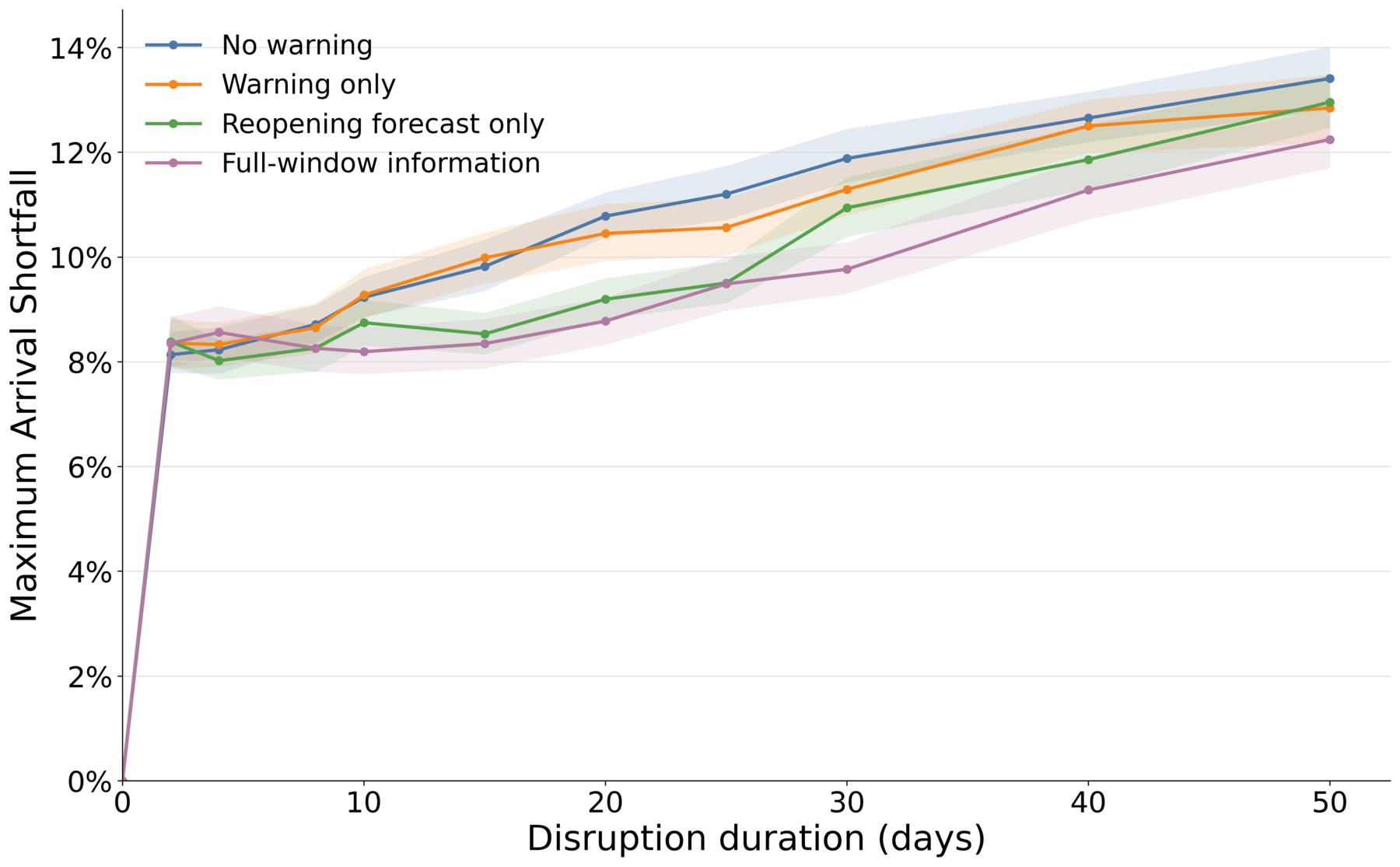}
\scriptsize Africa and Red Sea
\end{minipage} \\
\begin{minipage}[t]{0.32\linewidth}\centering
\includegraphics[width=\linewidth,trim=8 8 8 8,clip]{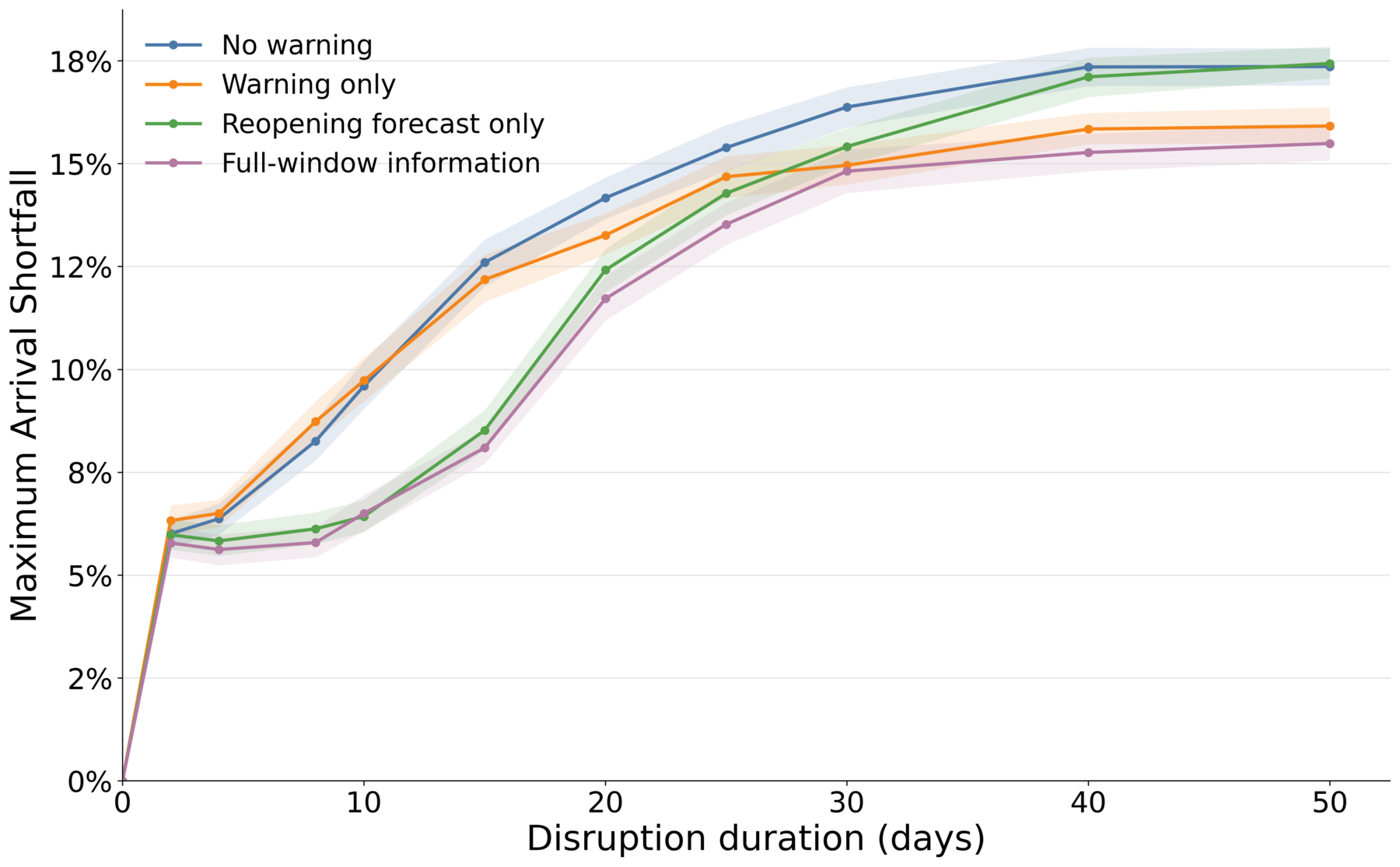}
\scriptsize West Central and South Asia
\end{minipage} &
\begin{minipage}[t]{0.32\linewidth}\centering
\includegraphics[width=\linewidth,trim=8 8 8 8,clip]{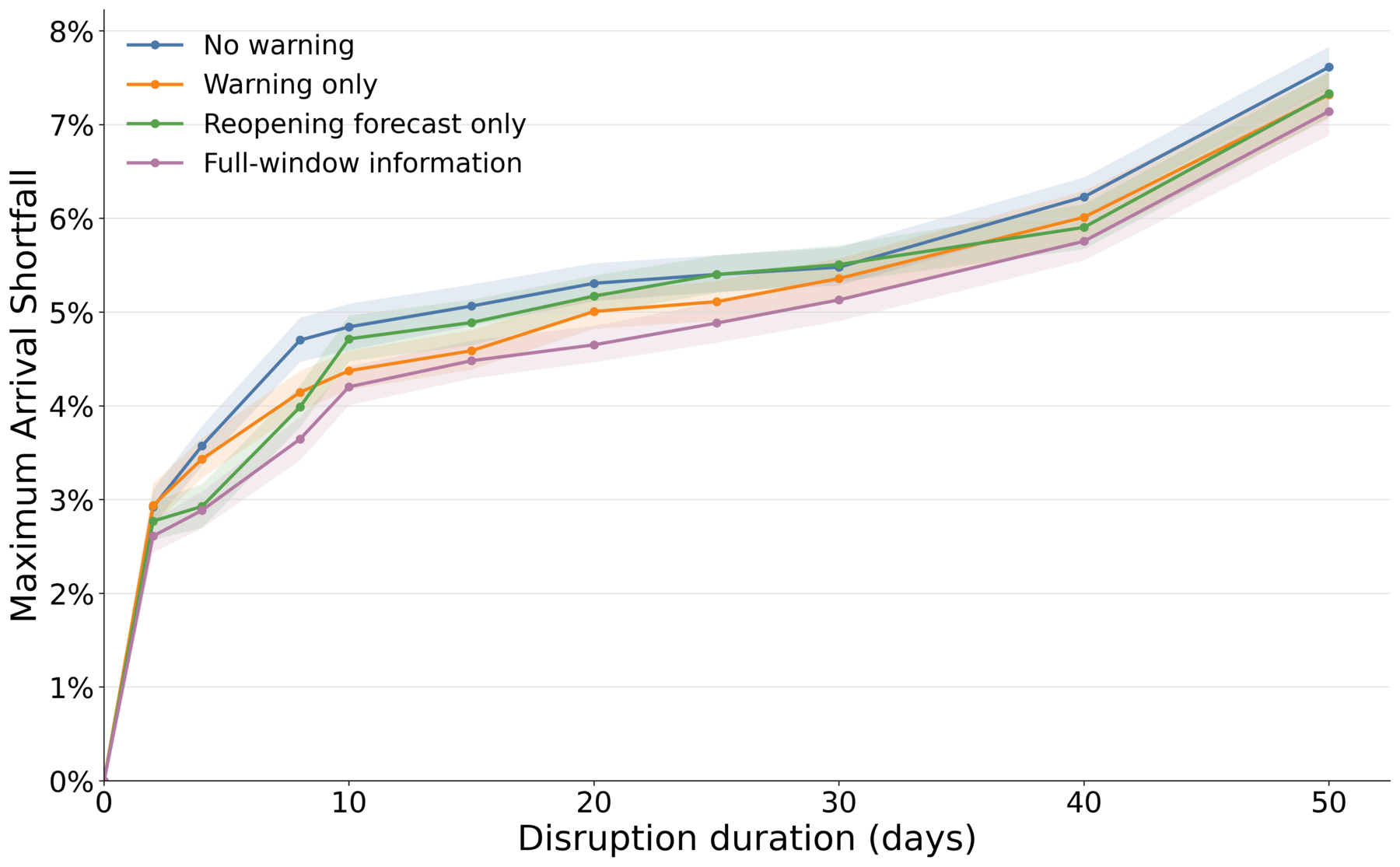}
\scriptsize East and Southeast Asia
\end{minipage} &
\begin{minipage}[t]{0.32\linewidth}\centering
\includegraphics[width=\linewidth,trim=8 8 8 8,clip]{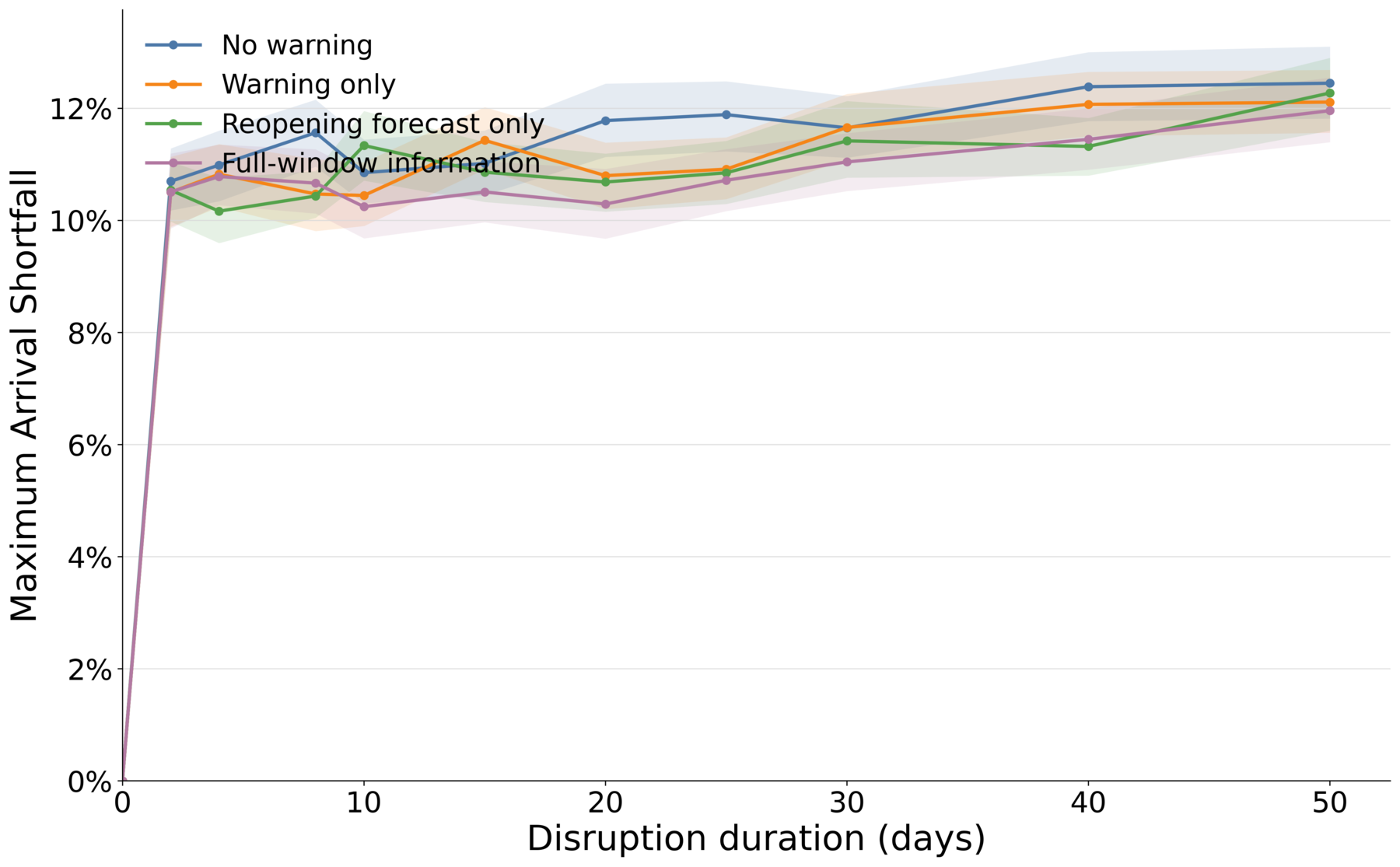}
\scriptsize Oceania and Pacific
\end{minipage} \\
\begin{minipage}[t]{0.32\linewidth}\centering
\includegraphics[width=\linewidth,trim=8 8 8 8,clip]{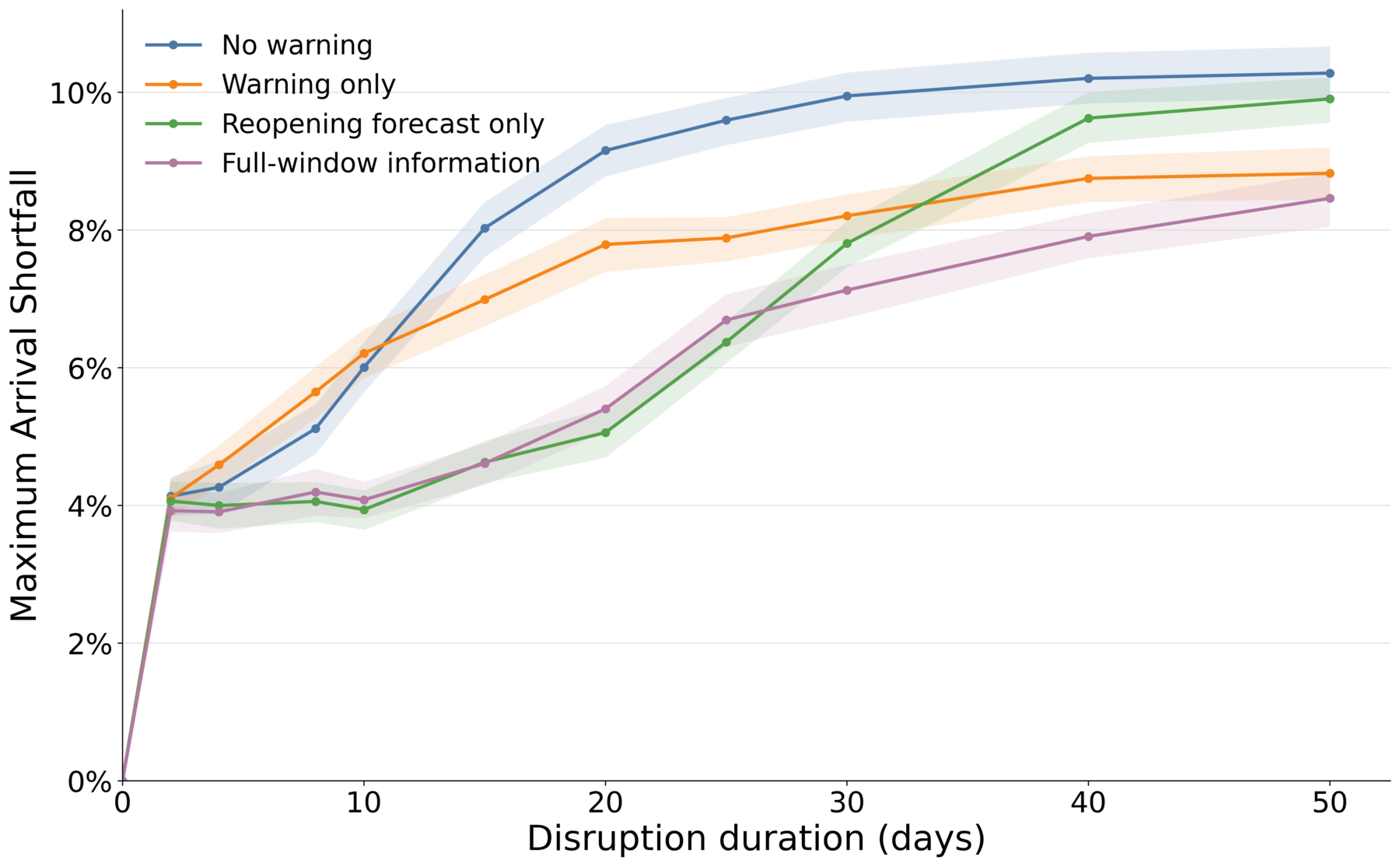}
\scriptsize European Union with UK
\end{minipage} &
\begin{minipage}[t]{0.32\linewidth}\centering
\includegraphics[width=\linewidth,trim=8 8 8 8,clip]{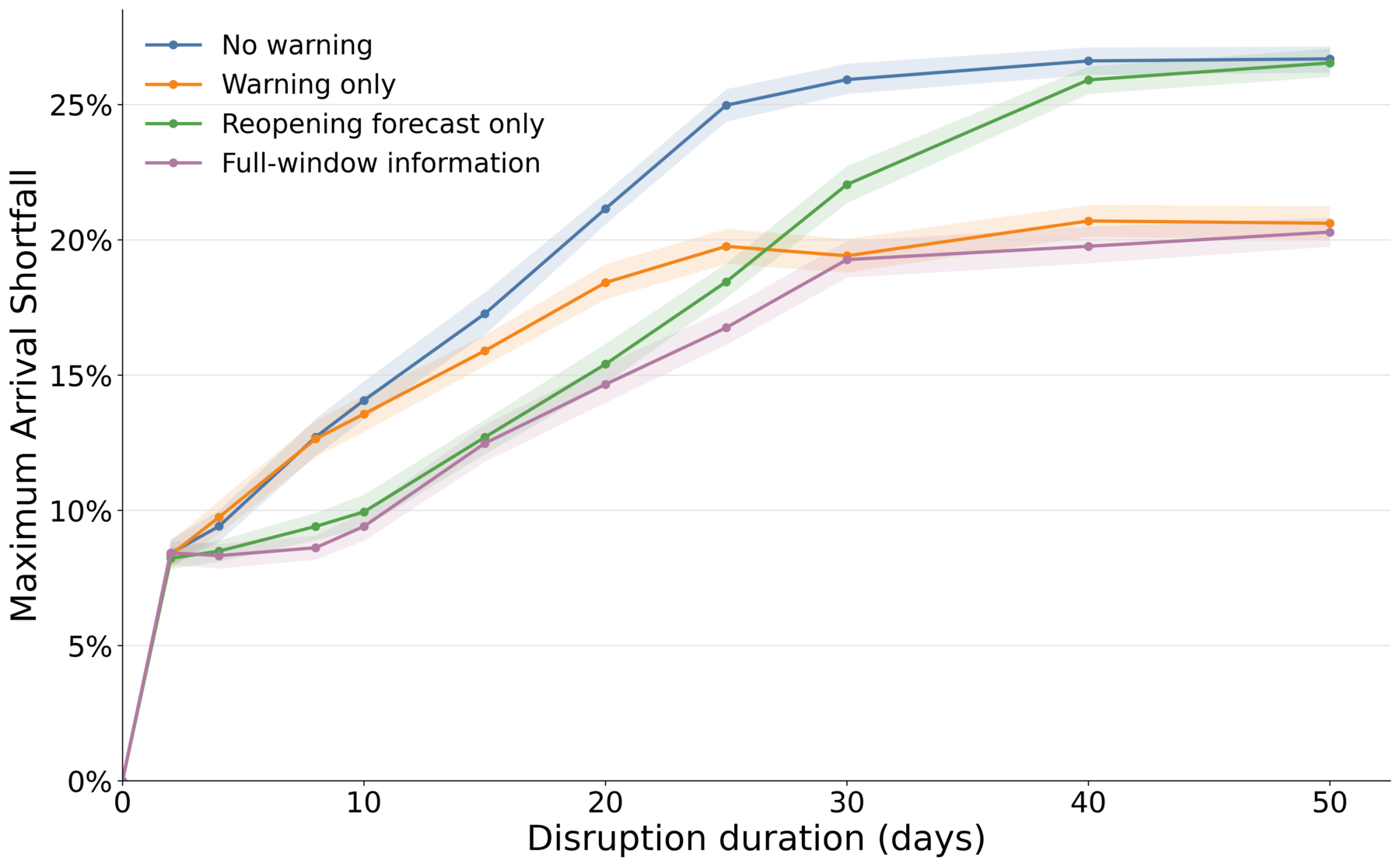}
\scriptsize USA
\end{minipage} &
\begin{minipage}[t]{0.32\linewidth}\centering
\includegraphics[width=\linewidth,trim=8 8 8 8,clip]{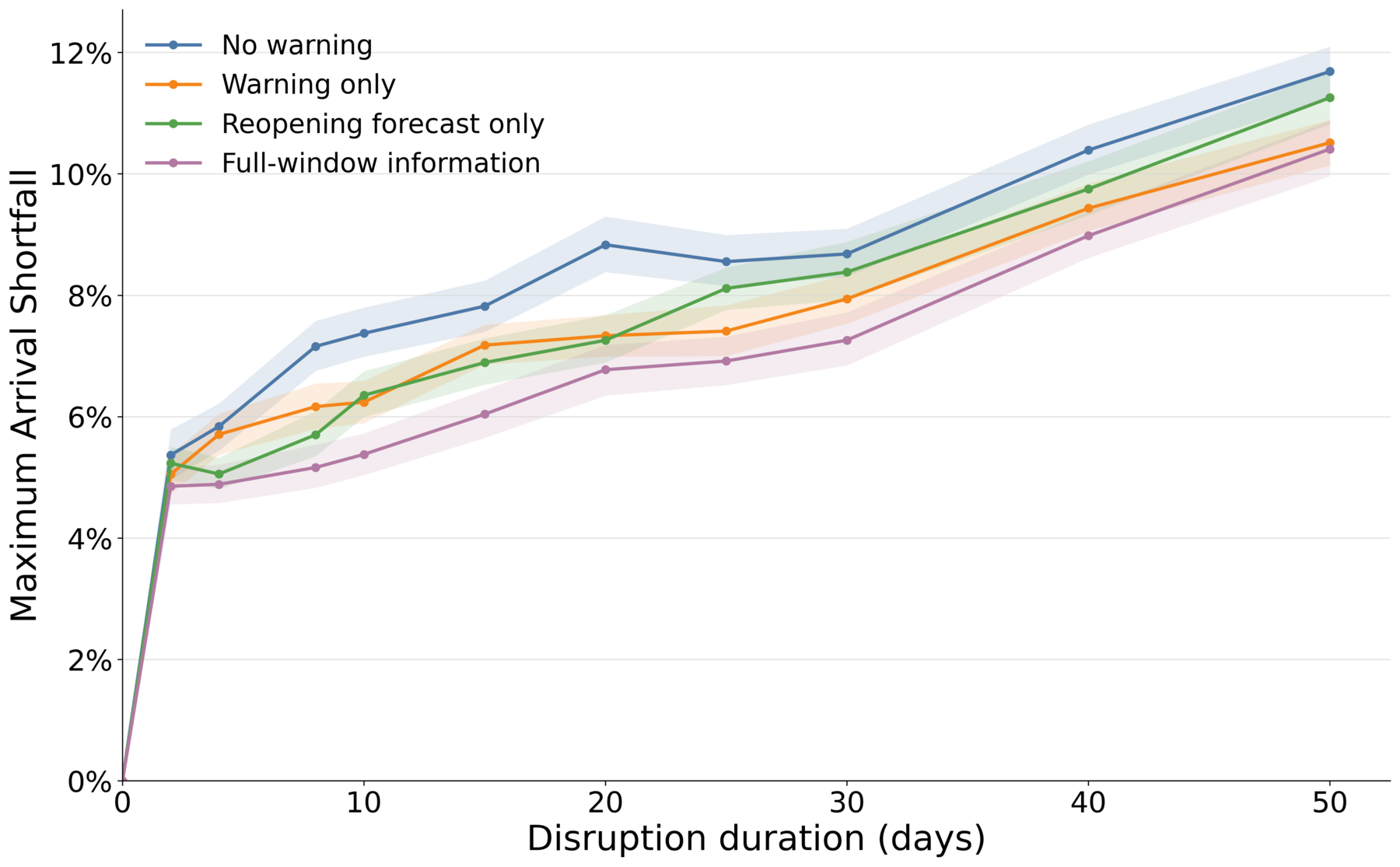}
\scriptsize Mainland China HK Macau
\end{minipage}
\end{tabular}
\caption{Regional maximum arrival shortfall under information regimes.}
\label{fig:regional_predictability_maximum_shortfall}
\end{figure}
\FloatBarrier

\begin{figure}[H]
\centering
\setlength{\tabcolsep}{2pt}
\renewcommand{\arraystretch}{0.78}
\begin{tabular}{ccc}
\begin{minipage}[t]{0.32\linewidth}\centering
\includegraphics[width=\linewidth,trim=8 8 8 8,clip]{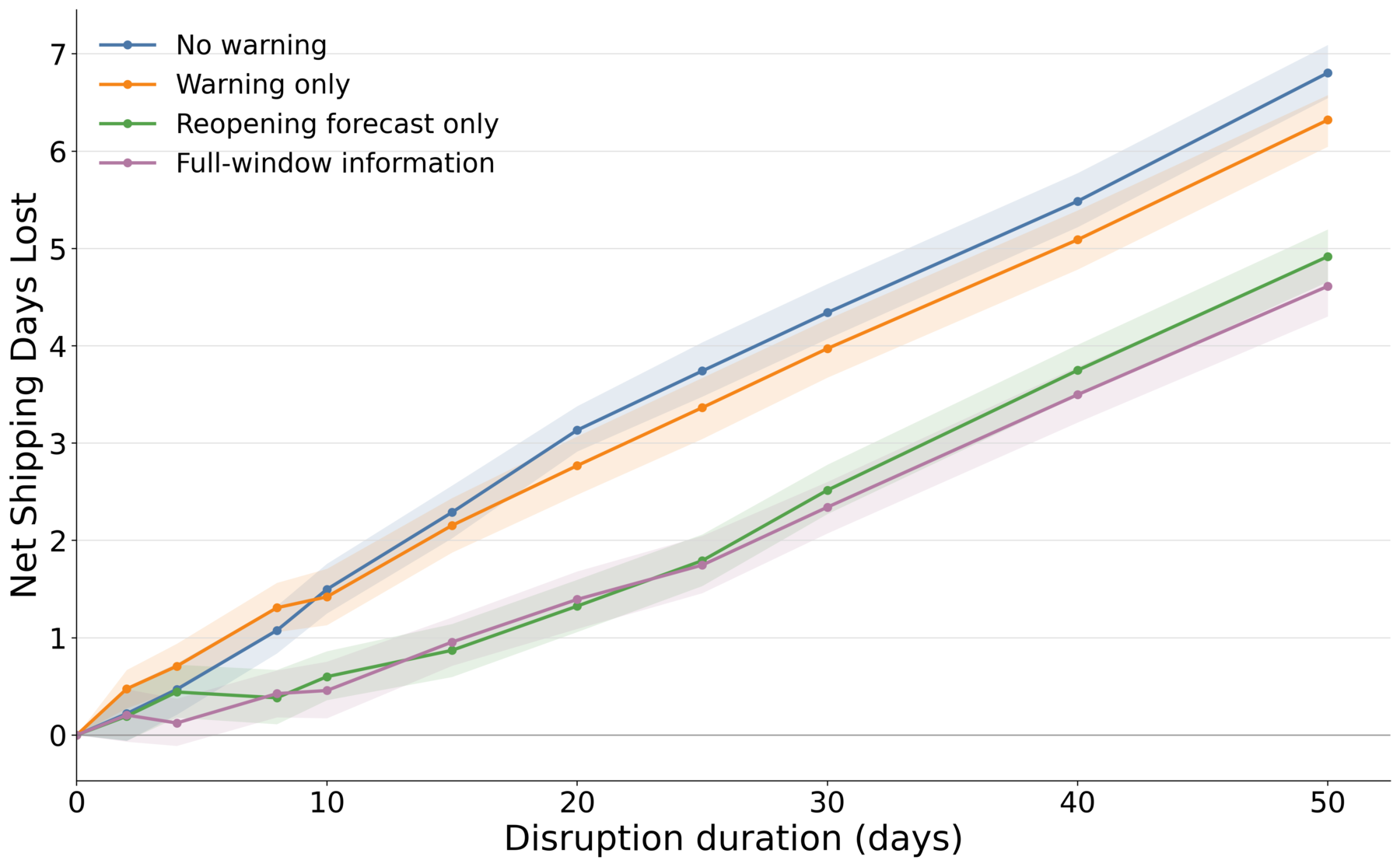}
\scriptsize Americas Atlantic
\end{minipage} &
\begin{minipage}[t]{0.32\linewidth}\centering
\includegraphics[width=\linewidth,trim=8 8 8 8,clip]{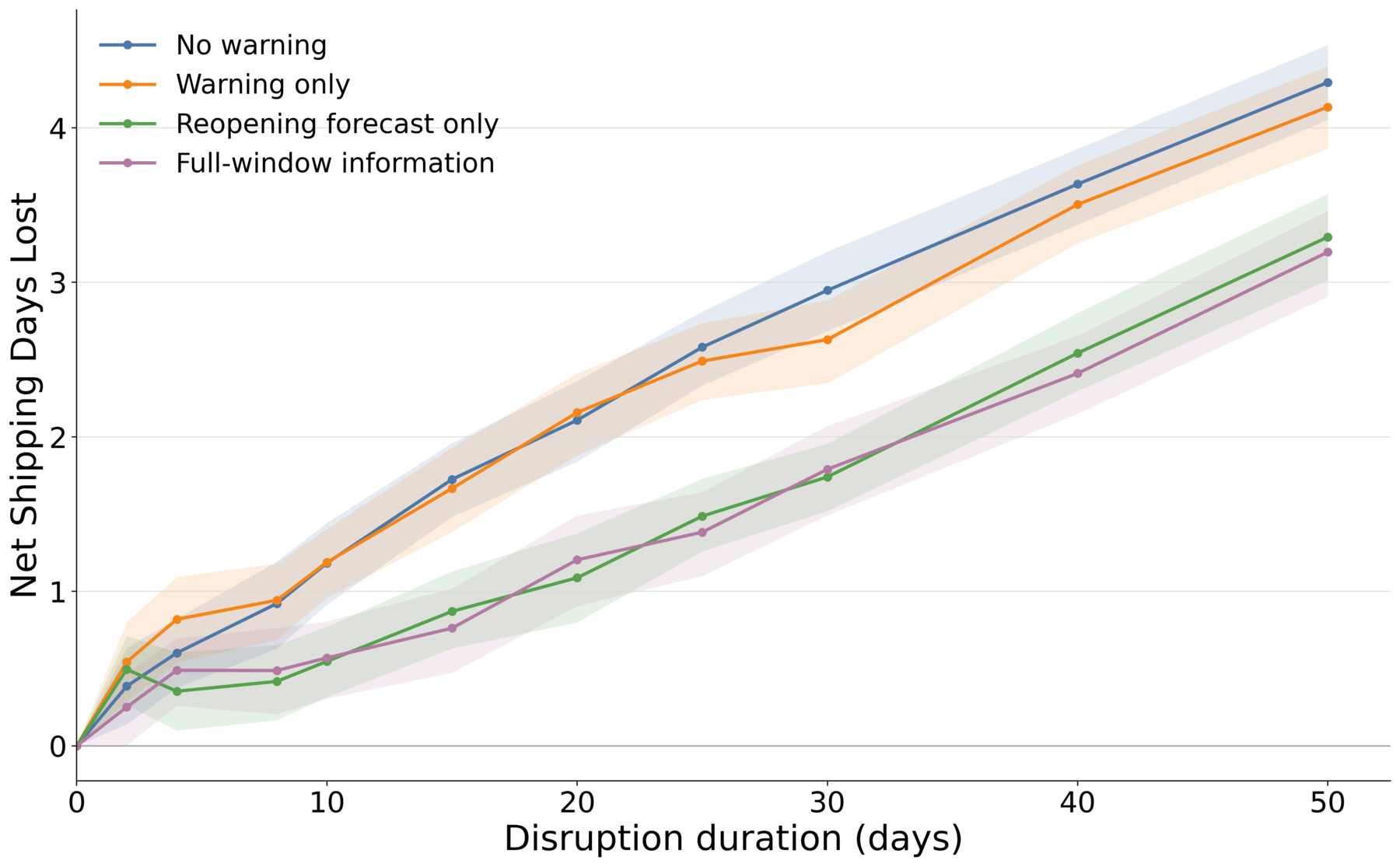}
\scriptsize Europe and Mediterranean
\end{minipage} &
\begin{minipage}[t]{0.32\linewidth}\centering
\includegraphics[width=\linewidth,trim=8 8 8 8,clip]{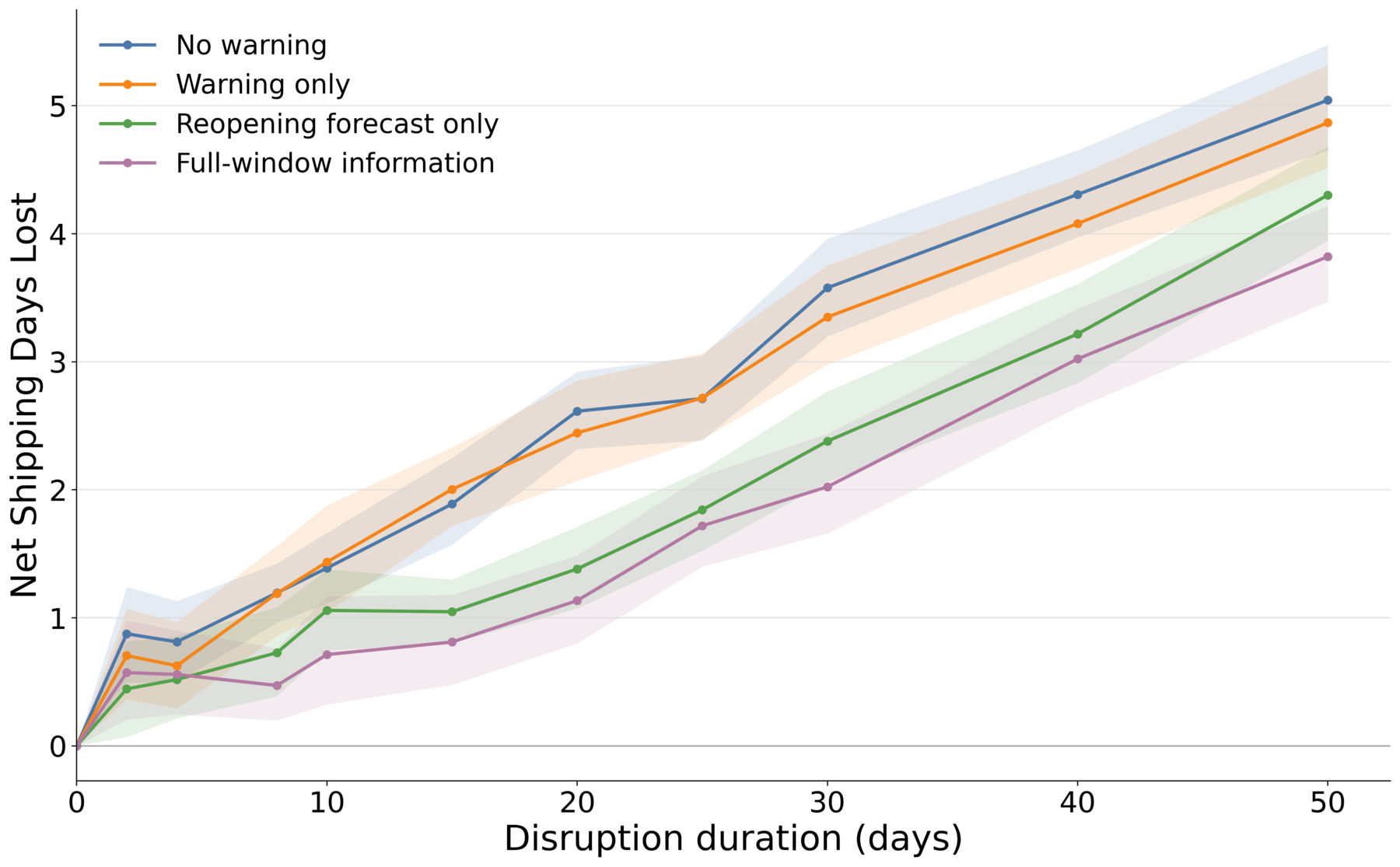}
\scriptsize Africa and Red Sea
\end{minipage} \\
\begin{minipage}[t]{0.32\linewidth}\centering
\includegraphics[width=\linewidth,trim=8 8 8 8,clip]{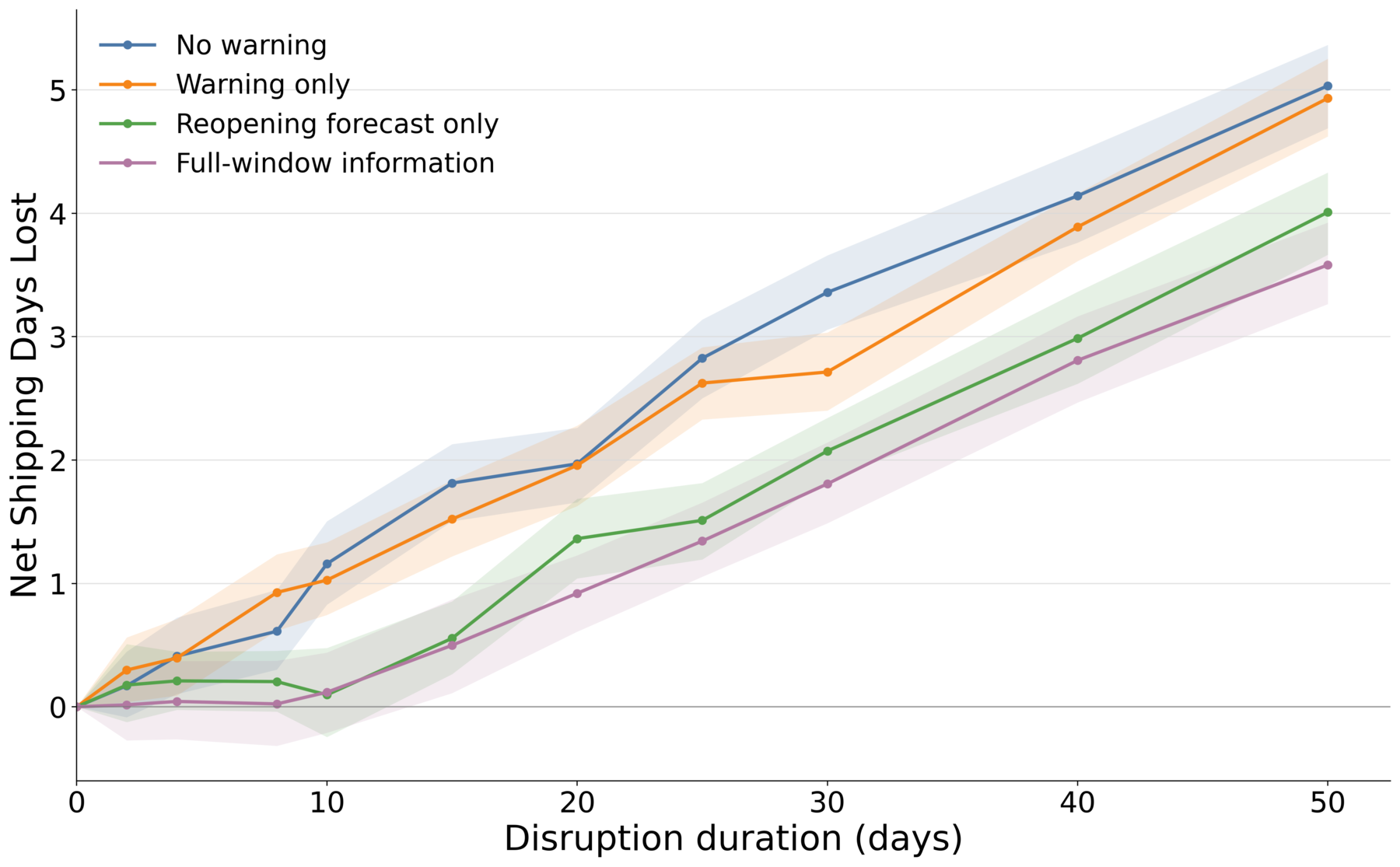}
\scriptsize West Central and South Asia
\end{minipage} &
\begin{minipage}[t]{0.32\linewidth}\centering
\includegraphics[width=\linewidth,trim=8 8 8 8,clip]{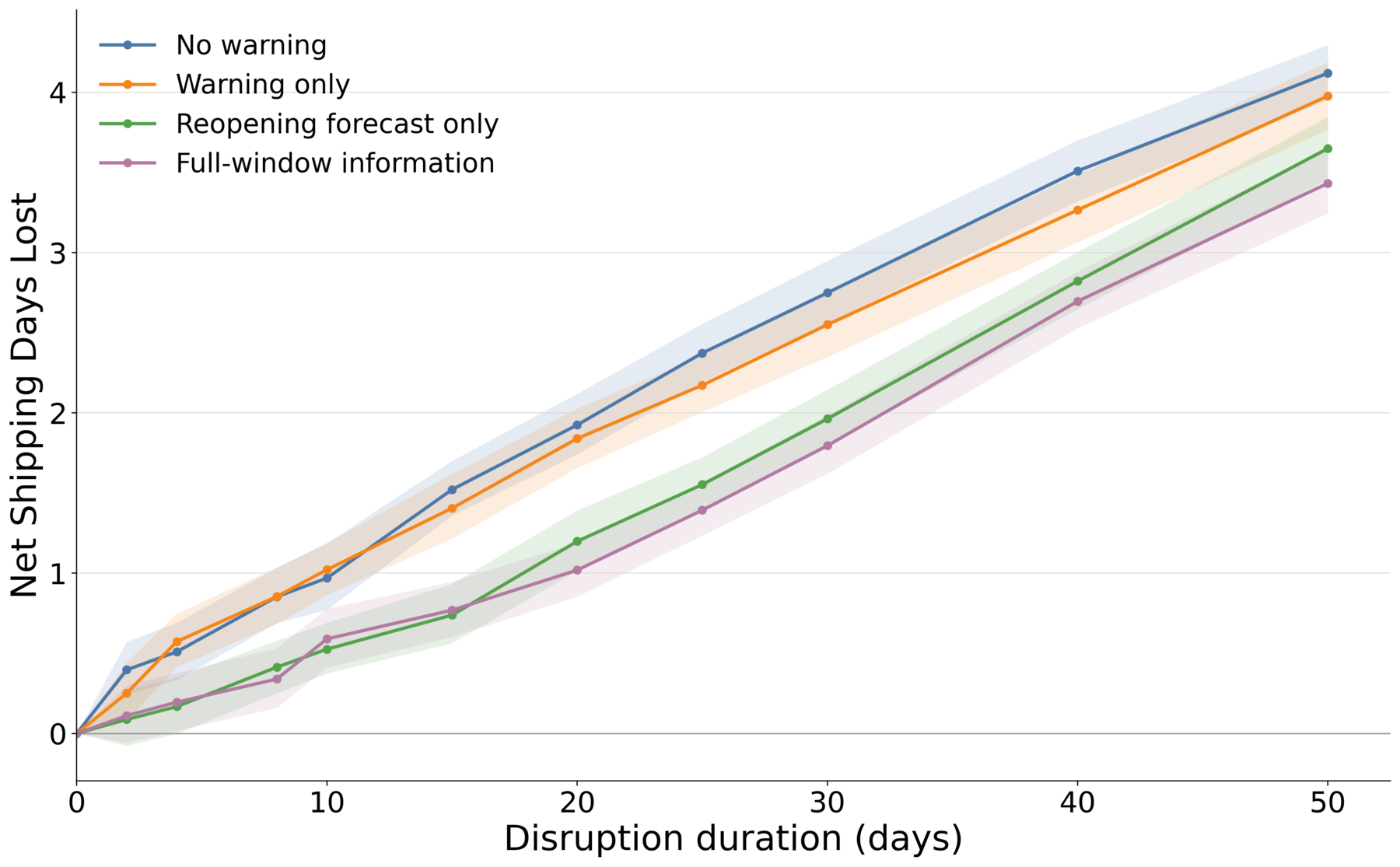}
\scriptsize East and Southeast Asia
\end{minipage} &
\begin{minipage}[t]{0.32\linewidth}\centering
\includegraphics[width=\linewidth,trim=8 8 8 8,clip]{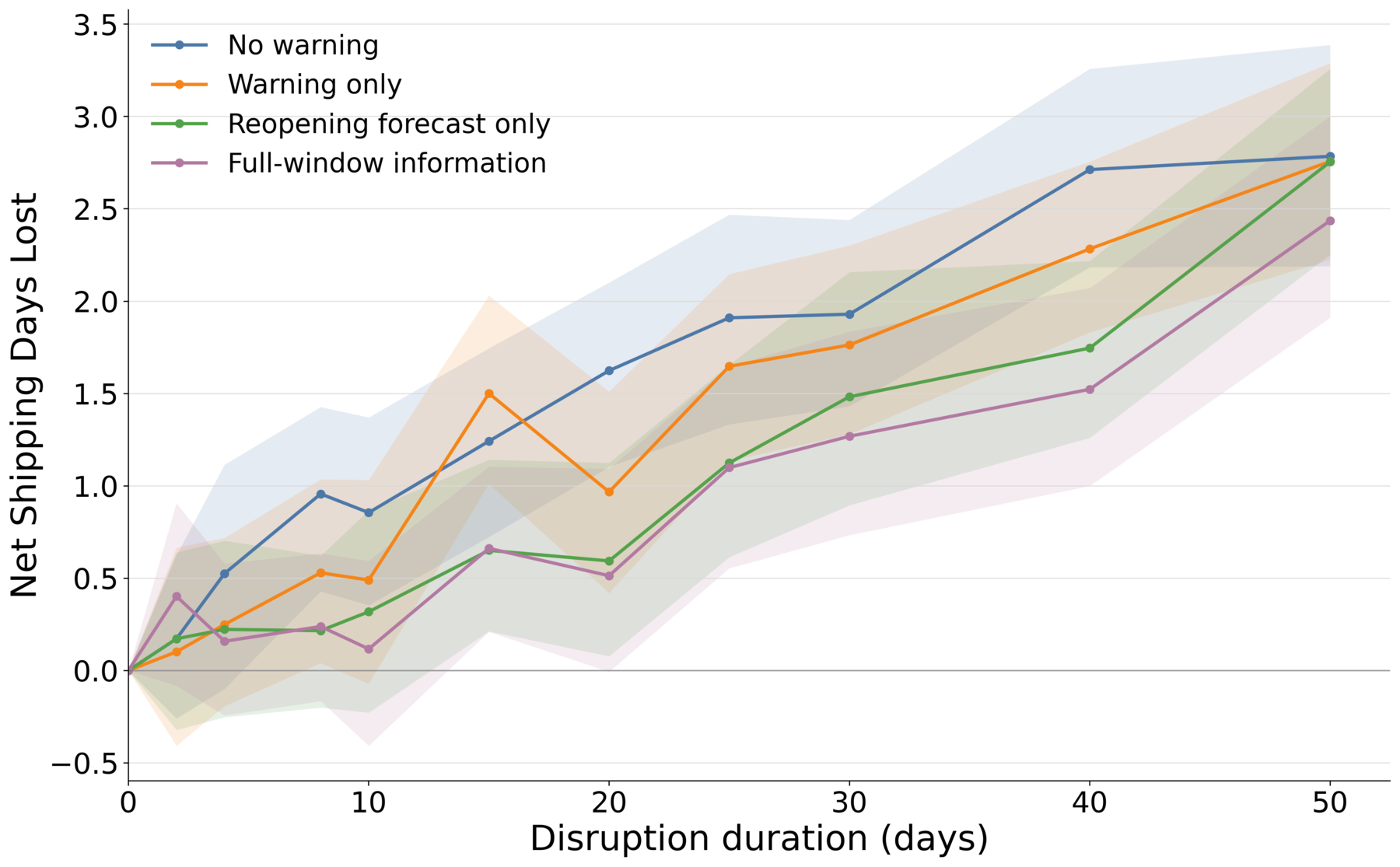}
\scriptsize Oceania and Pacific
\end{minipage} \\
\begin{minipage}[t]{0.32\linewidth}\centering
\includegraphics[width=\linewidth,trim=8 8 8 8,clip]{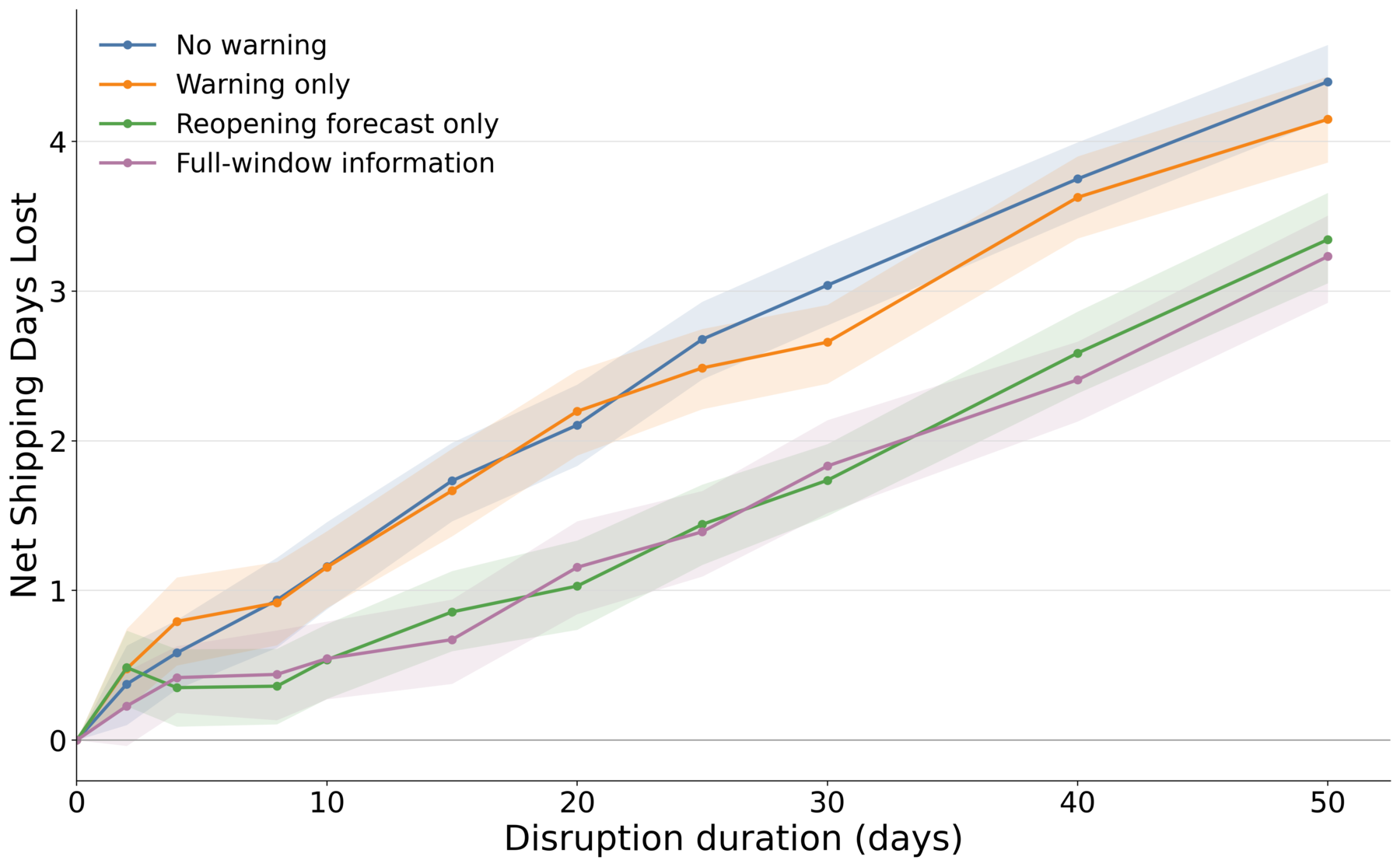}
\scriptsize European Union with UK
\end{minipage} &
\begin{minipage}[t]{0.32\linewidth}\centering
\includegraphics[width=\linewidth,trim=8 8 8 8,clip]{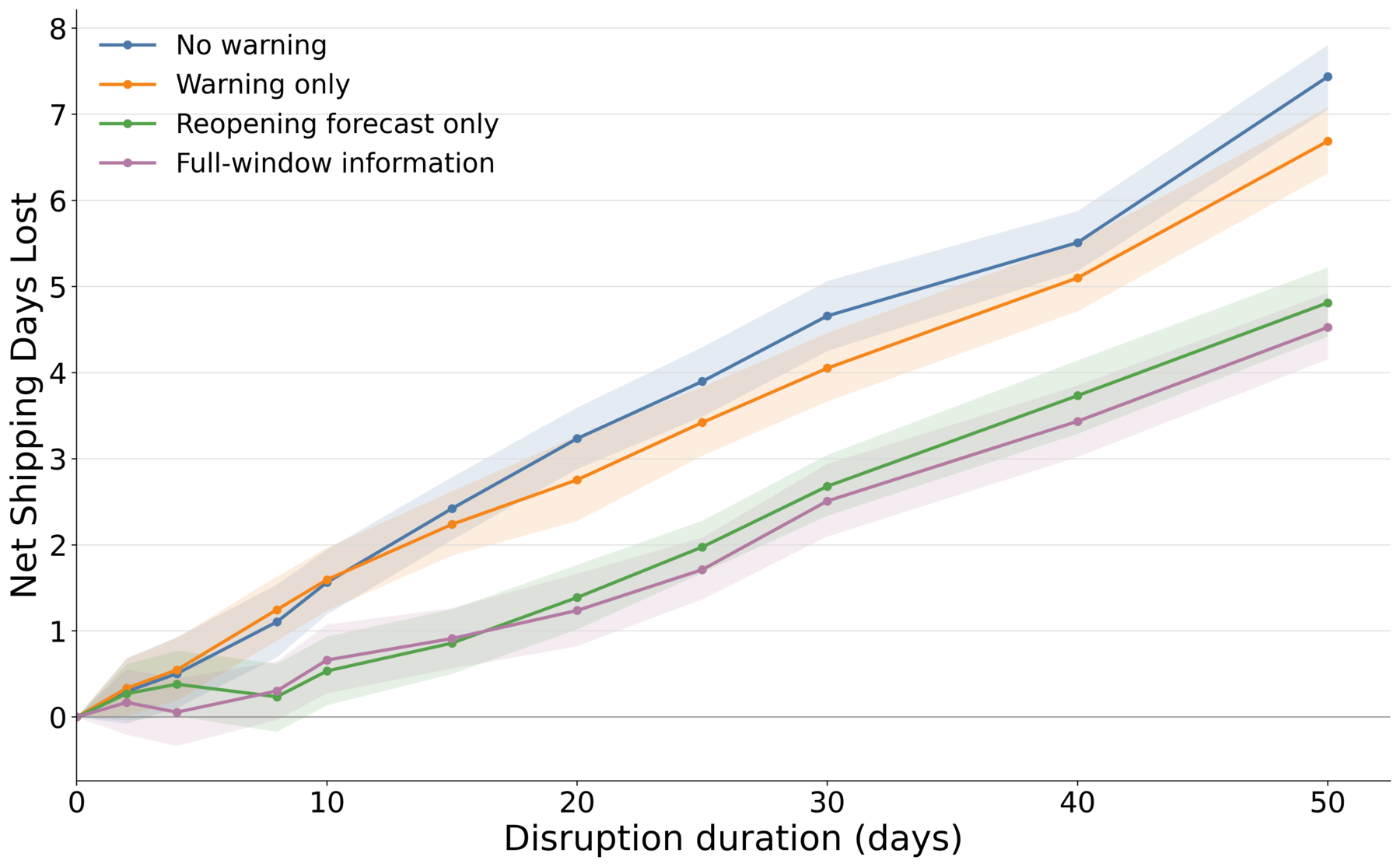}
\scriptsize USA
\end{minipage} &
\begin{minipage}[t]{0.32\linewidth}\centering
\includegraphics[width=\linewidth,trim=8 8 8 8,clip]{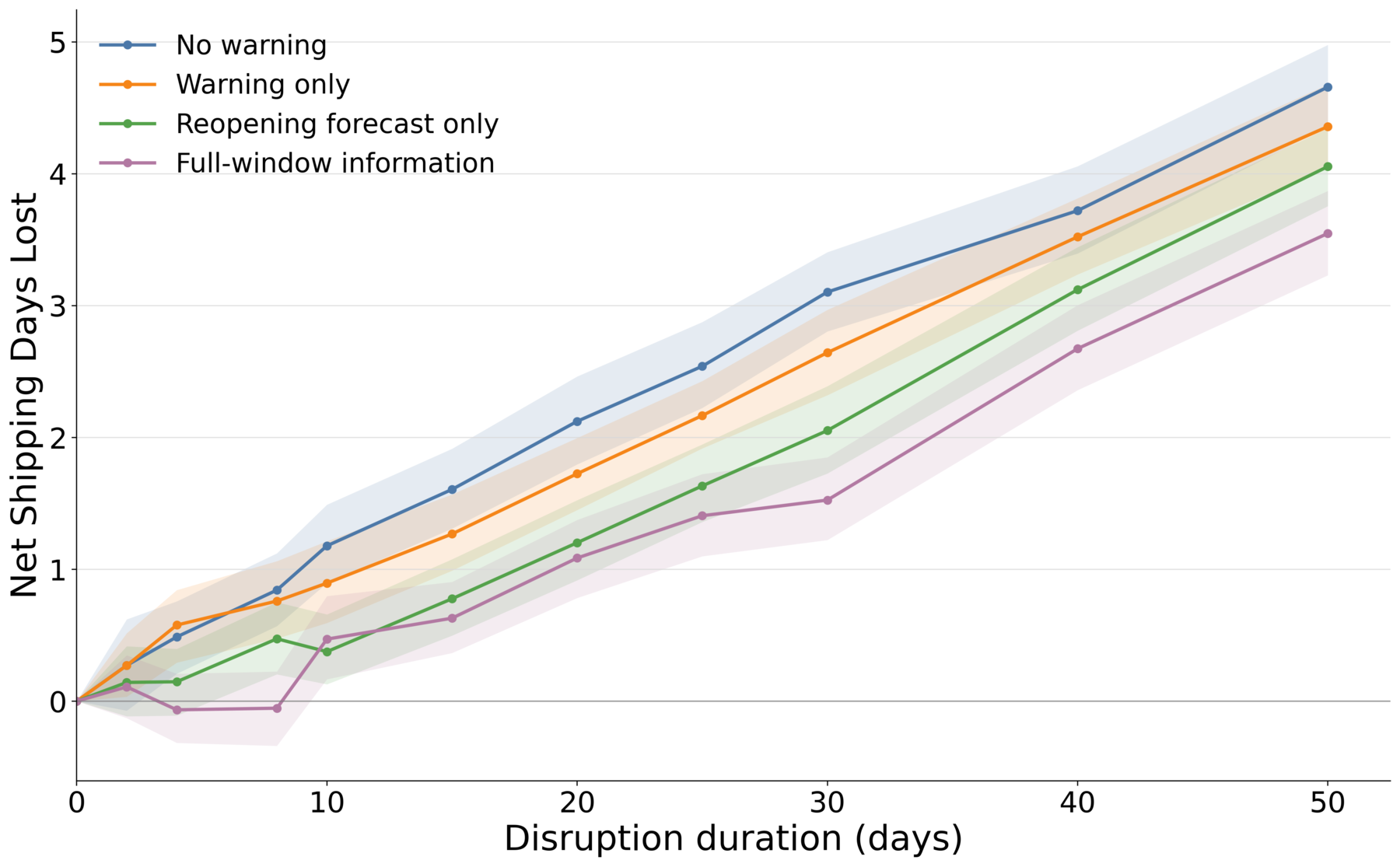}
\scriptsize Mainland China HK Macau
\end{minipage}
\end{tabular}
\caption{Regional net shipping-days lost under information regimes.}
\label{fig:regional_predictability_net_shipping_days_lost}
\end{figure}
\FloatBarrier

\section{Port Level Results}
\begin{figure}[H]
\centering
\setlength{\tabcolsep}{2pt}
\renewcommand{\arraystretch}{0.82}
\begin{tabular}{ccc}
\begin{minipage}[t]{0.32\linewidth}\centering
\scriptsize Singapore\\[-0.2em]
\includegraphics[width=\linewidth,height=0.16\textheight,keepaspectratio,trim=8 8 8 8,clip]{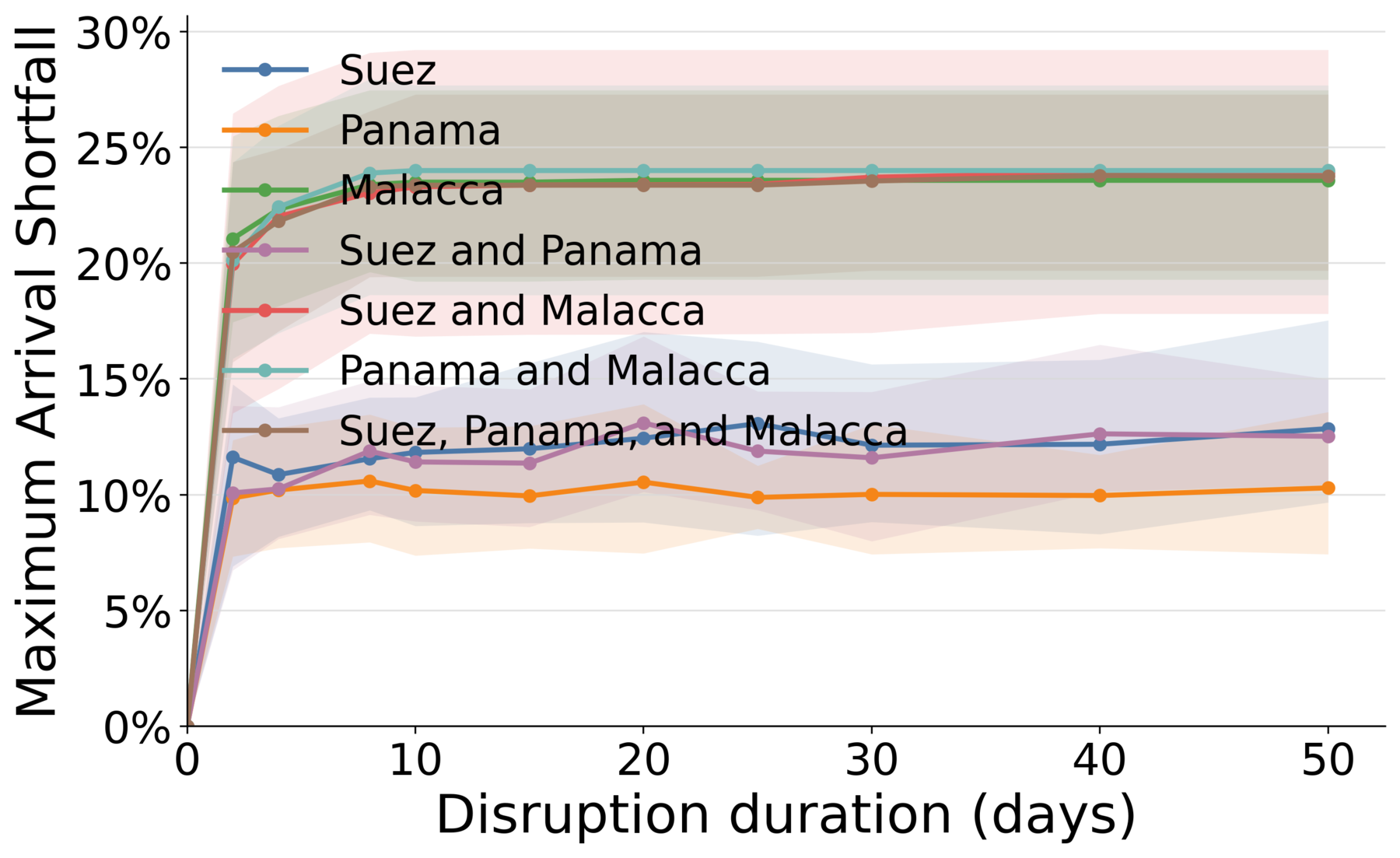}
\end{minipage} &
\begin{minipage}[t]{0.32\linewidth}\centering
\scriptsize Rotterdam\\[-0.2em]
\includegraphics[width=\linewidth,height=0.16\textheight,keepaspectratio,trim=8 8 8 8,clip]{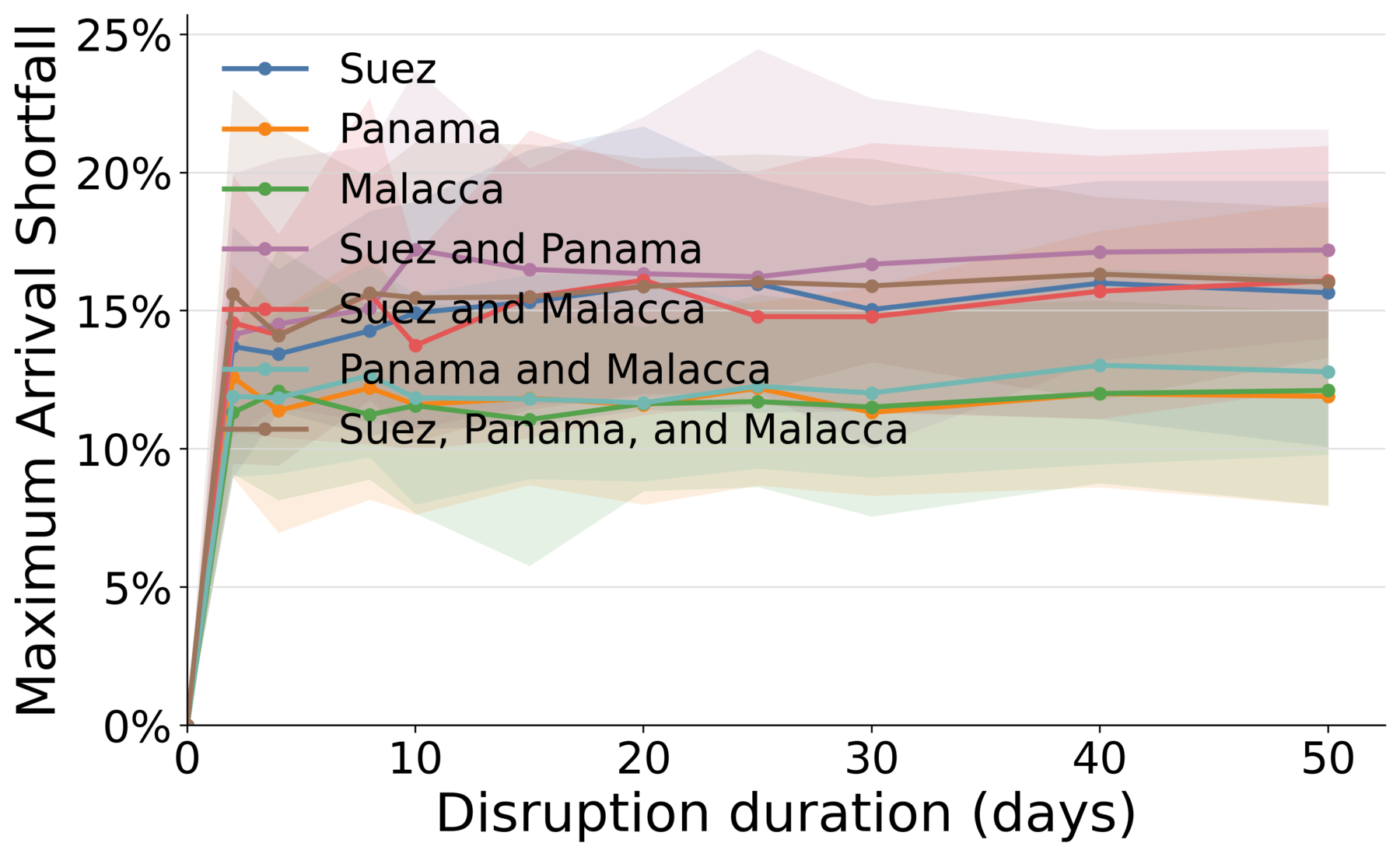}
\end{minipage} &
\begin{minipage}[t]{0.32\linewidth}\centering
\scriptsize Houston\\[-0.2em]
\includegraphics[width=\linewidth,height=0.16\textheight,keepaspectratio,trim=8 8 8 8,clip]{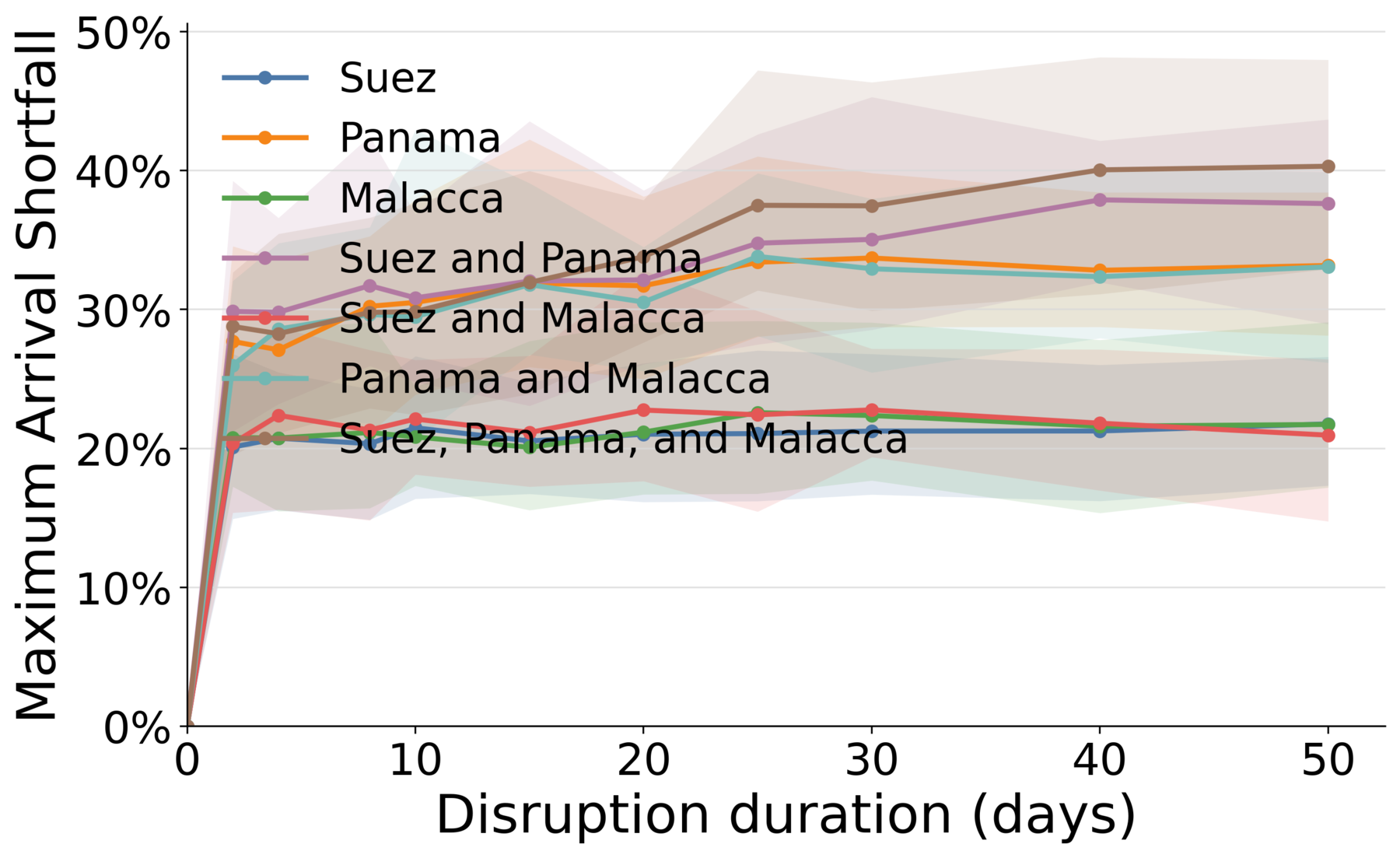}
\end{minipage}
\\[-0.2em]
\begin{minipage}[t]{0.32\linewidth}\centering
\includegraphics[width=\linewidth,height=0.16\textheight,keepaspectratio,trim=8 8 8 8,clip]{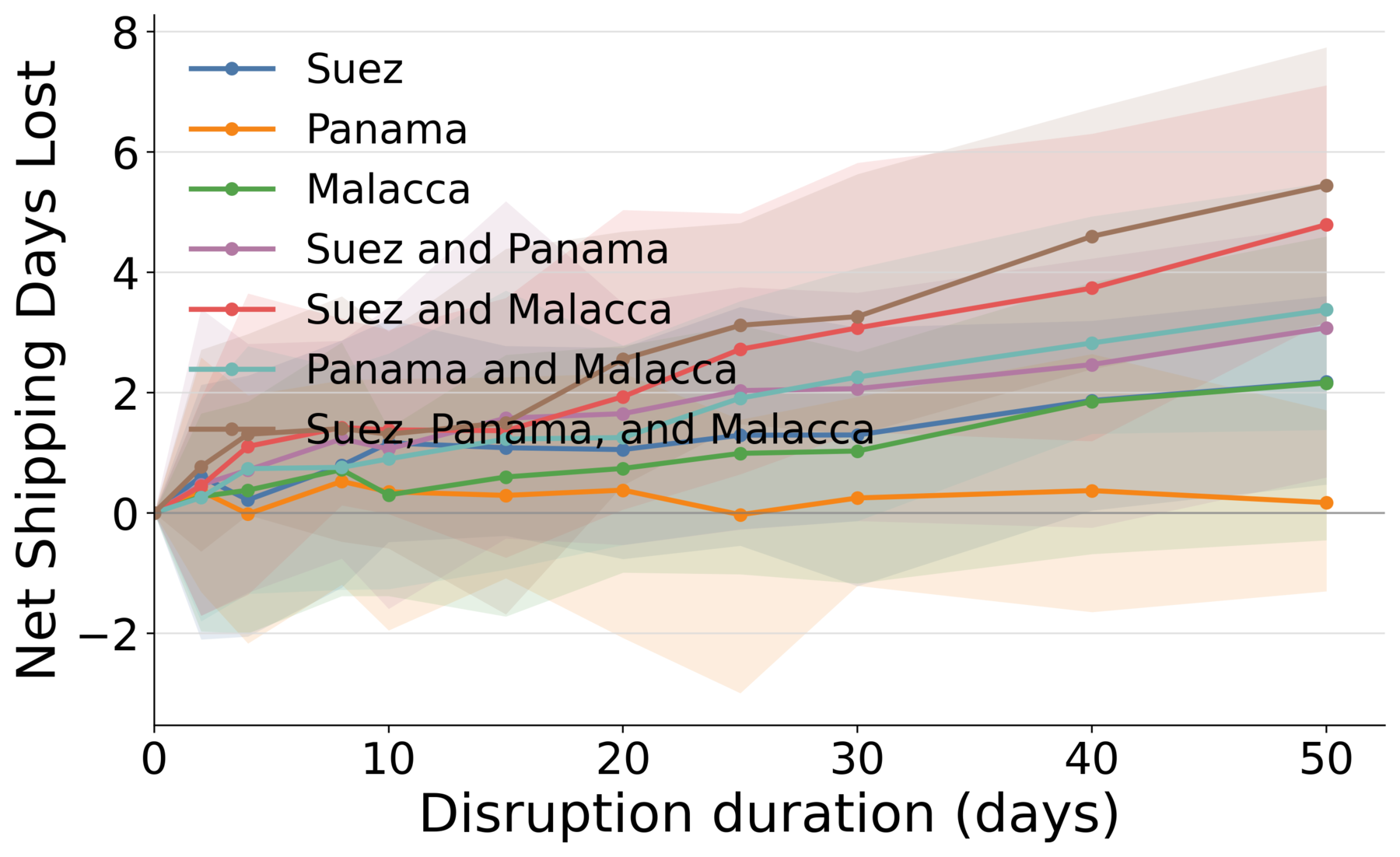}
\end{minipage} &
\begin{minipage}[t]{0.32\linewidth}\centering
\includegraphics[width=\linewidth,height=0.16\textheight,keepaspectratio,trim=8 8 8 8,clip]{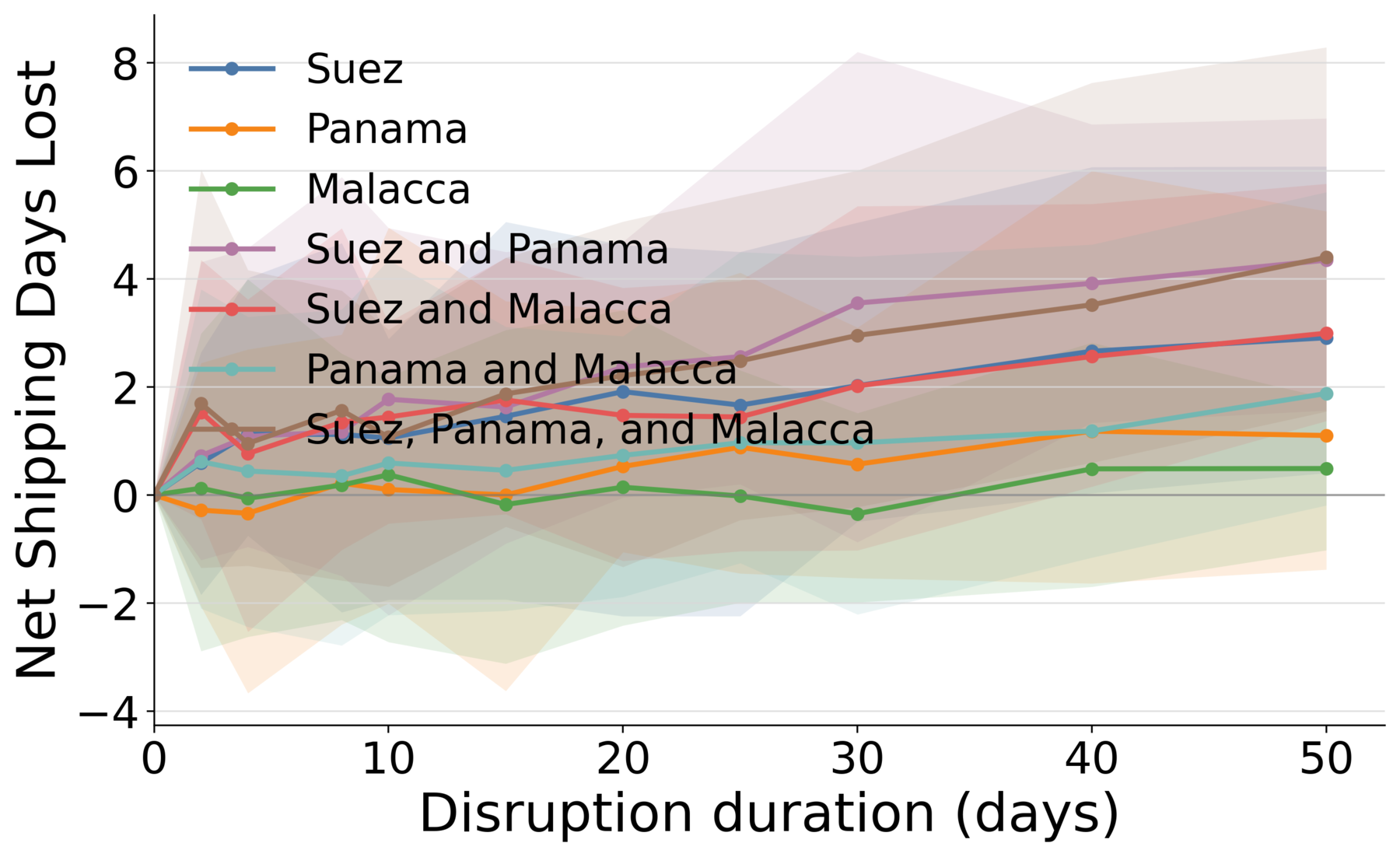}
\end{minipage} &
\begin{minipage}[t]{0.32\linewidth}\centering
\includegraphics[width=\linewidth,height=0.16\textheight,keepaspectratio,trim=8 8 8 8,clip]{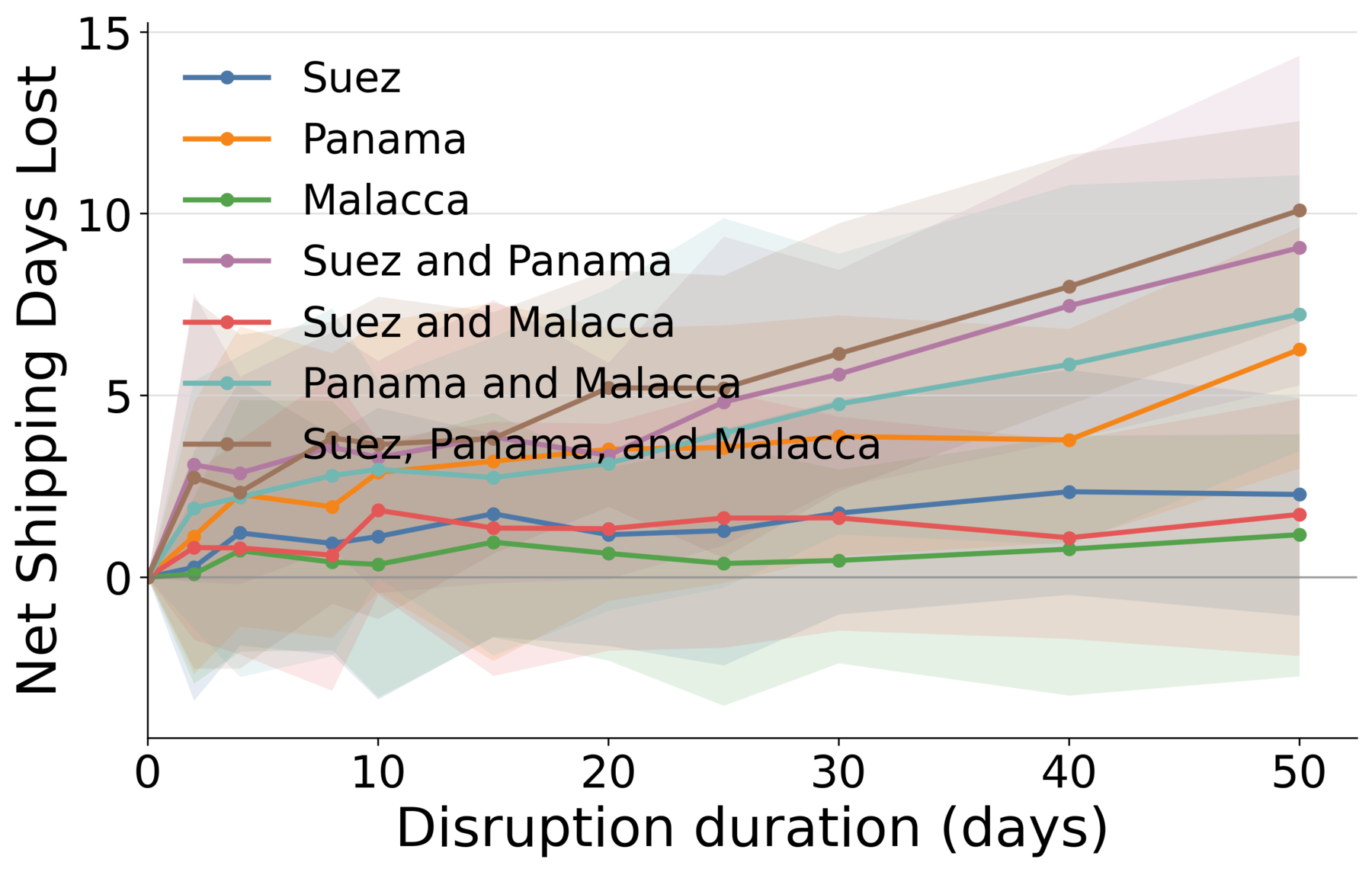}
\end{minipage}
\\[-0.2em]
\begin{minipage}[t]{0.32\linewidth}\centering
\includegraphics[width=\linewidth,height=0.16\textheight,keepaspectratio,trim=8 8 8 8,clip]{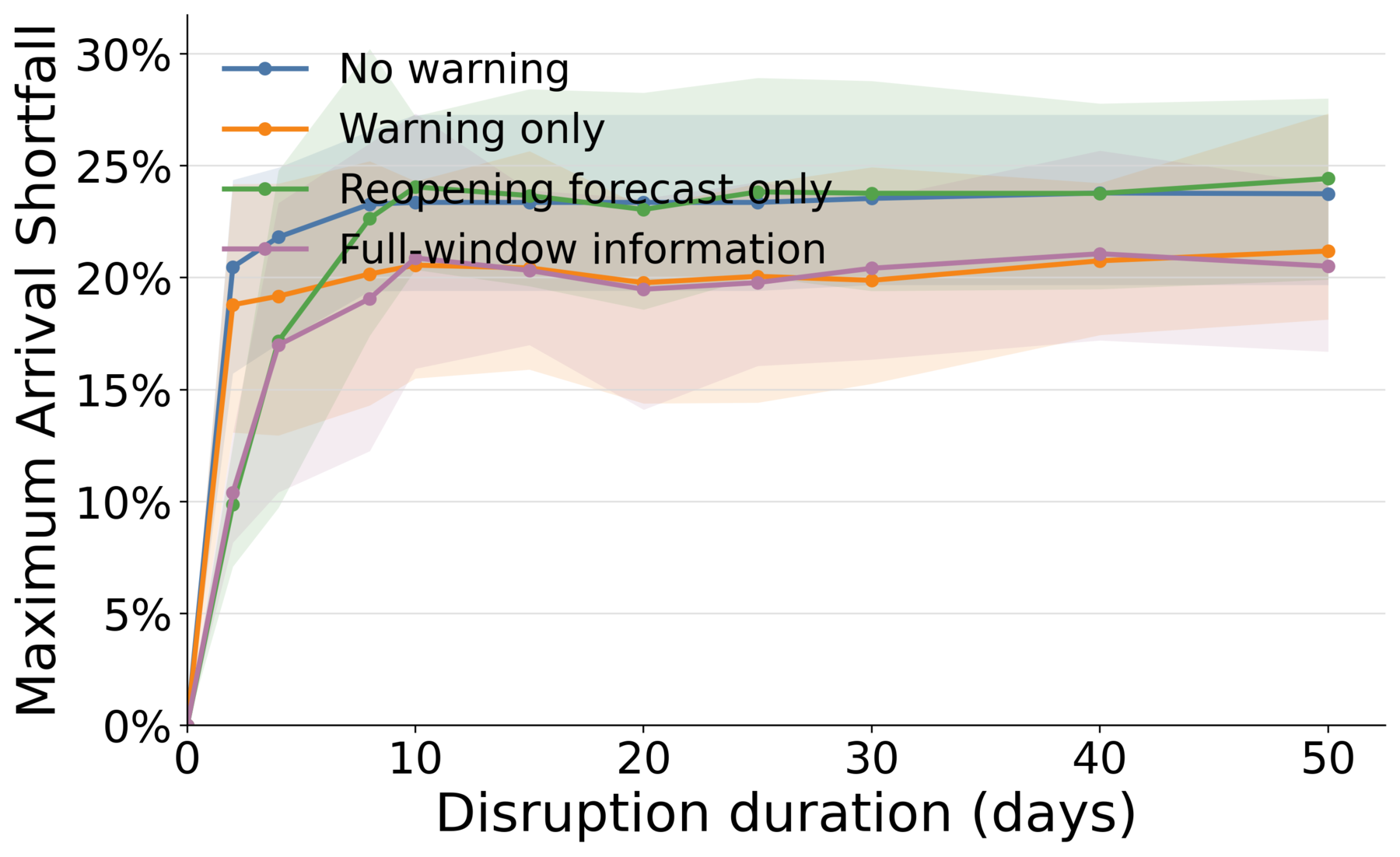}
\end{minipage} &
\begin{minipage}[t]{0.32\linewidth}\centering
\includegraphics[width=\linewidth,height=0.16\textheight,keepaspectratio,trim=8 8 8 8,clip]{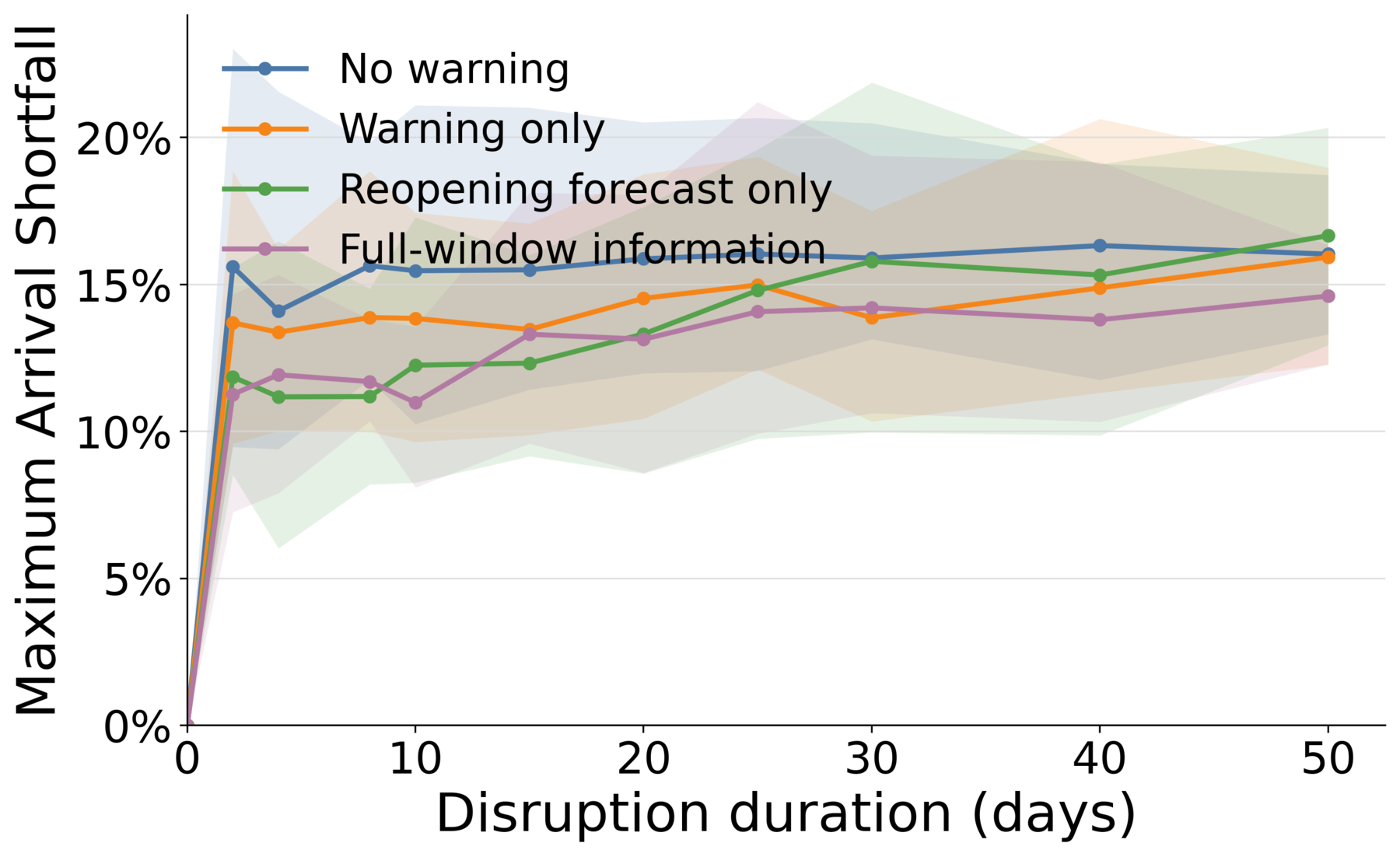}
\end{minipage} &
\begin{minipage}[t]{0.32\linewidth}\centering
\includegraphics[width=\linewidth,height=0.16\textheight,keepaspectratio,trim=8 8 8 8,clip]{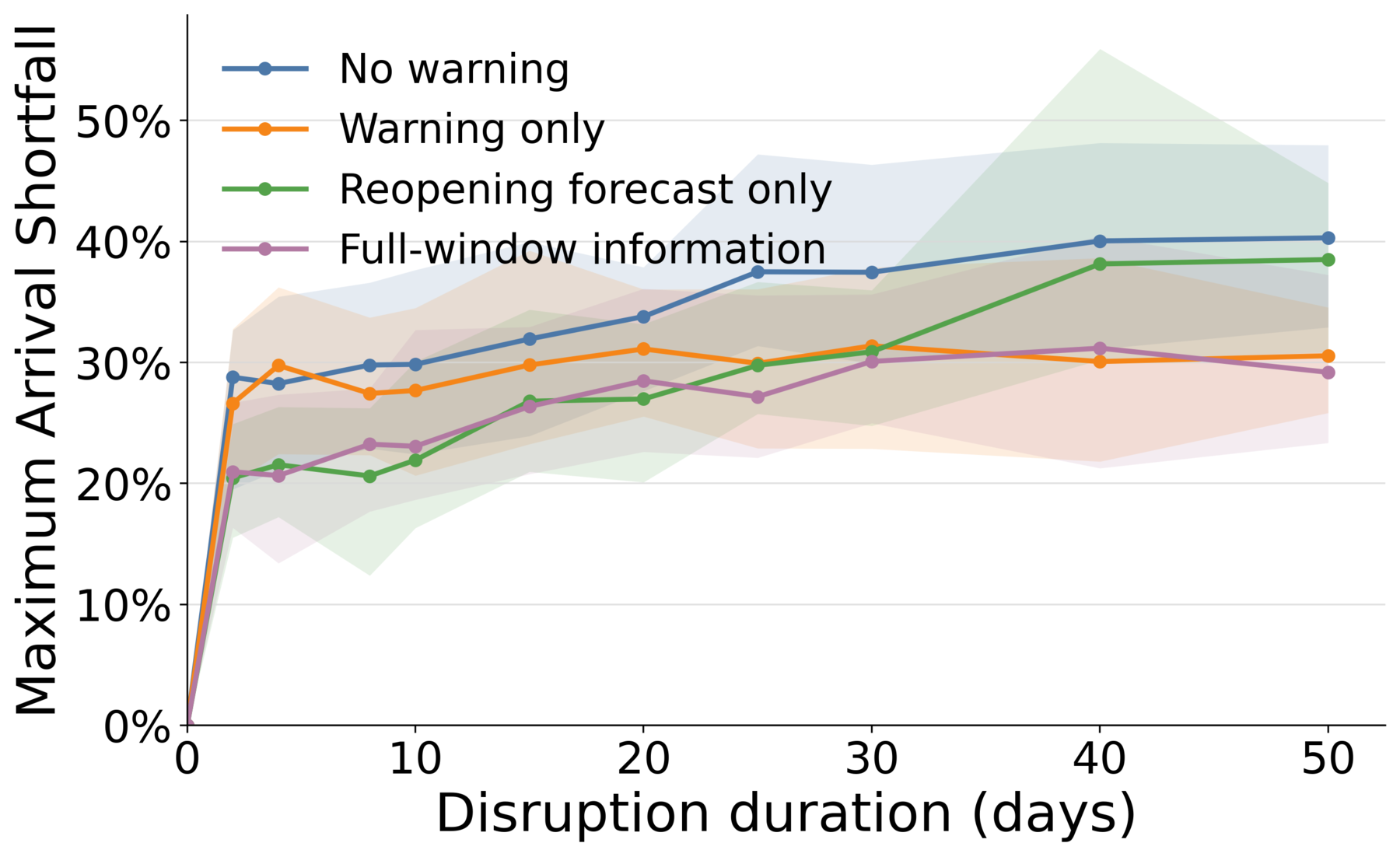}
\end{minipage}
\\[-0.2em]
\begin{minipage}[t]{0.32\linewidth}\centering
\includegraphics[width=\linewidth,height=0.16\textheight,keepaspectratio,trim=8 8 8 8,clip]{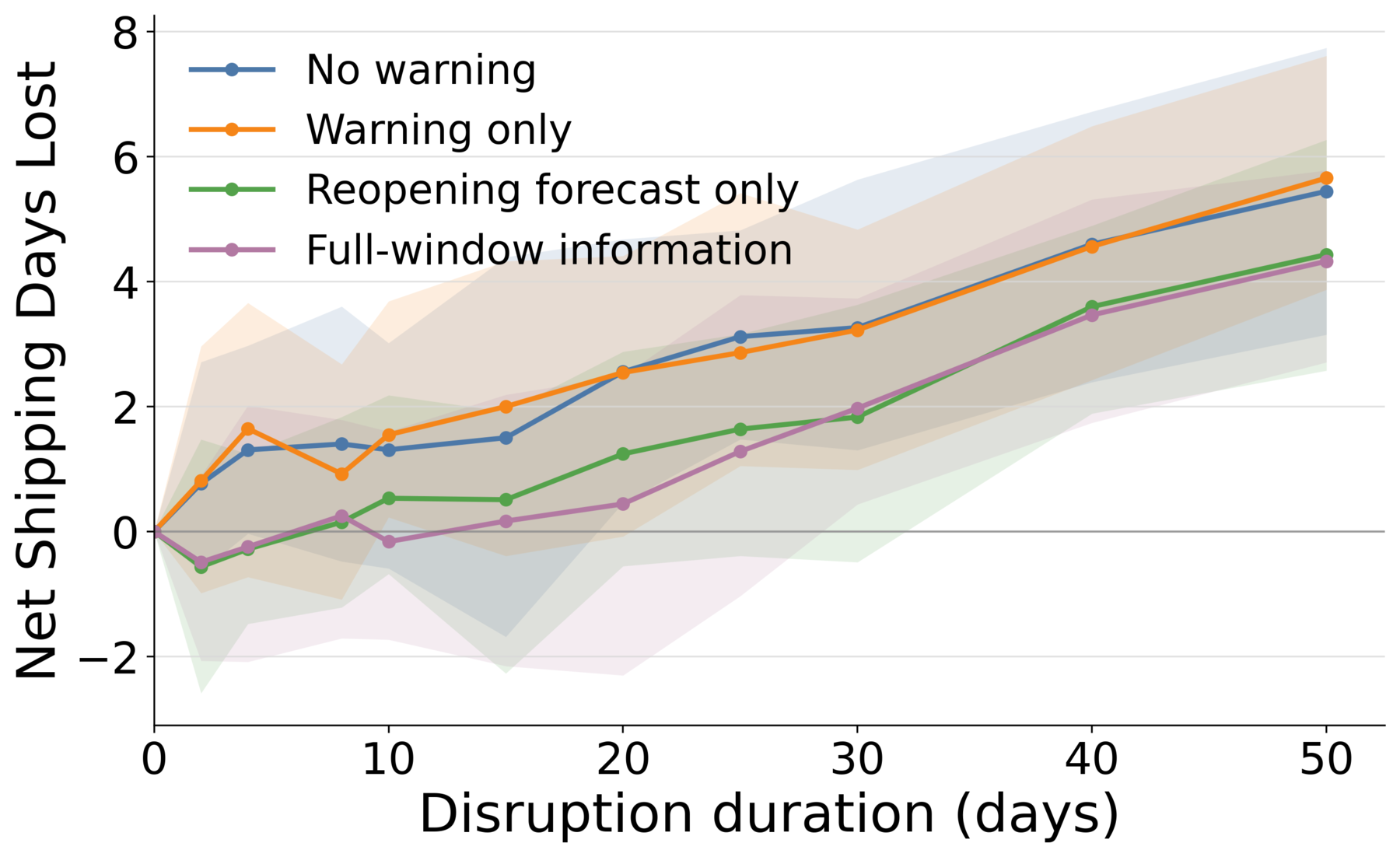}
\end{minipage} &
\begin{minipage}[t]{0.32\linewidth}\centering
\includegraphics[width=\linewidth,height=0.16\textheight,keepaspectratio,trim=8 8 8 8,clip]{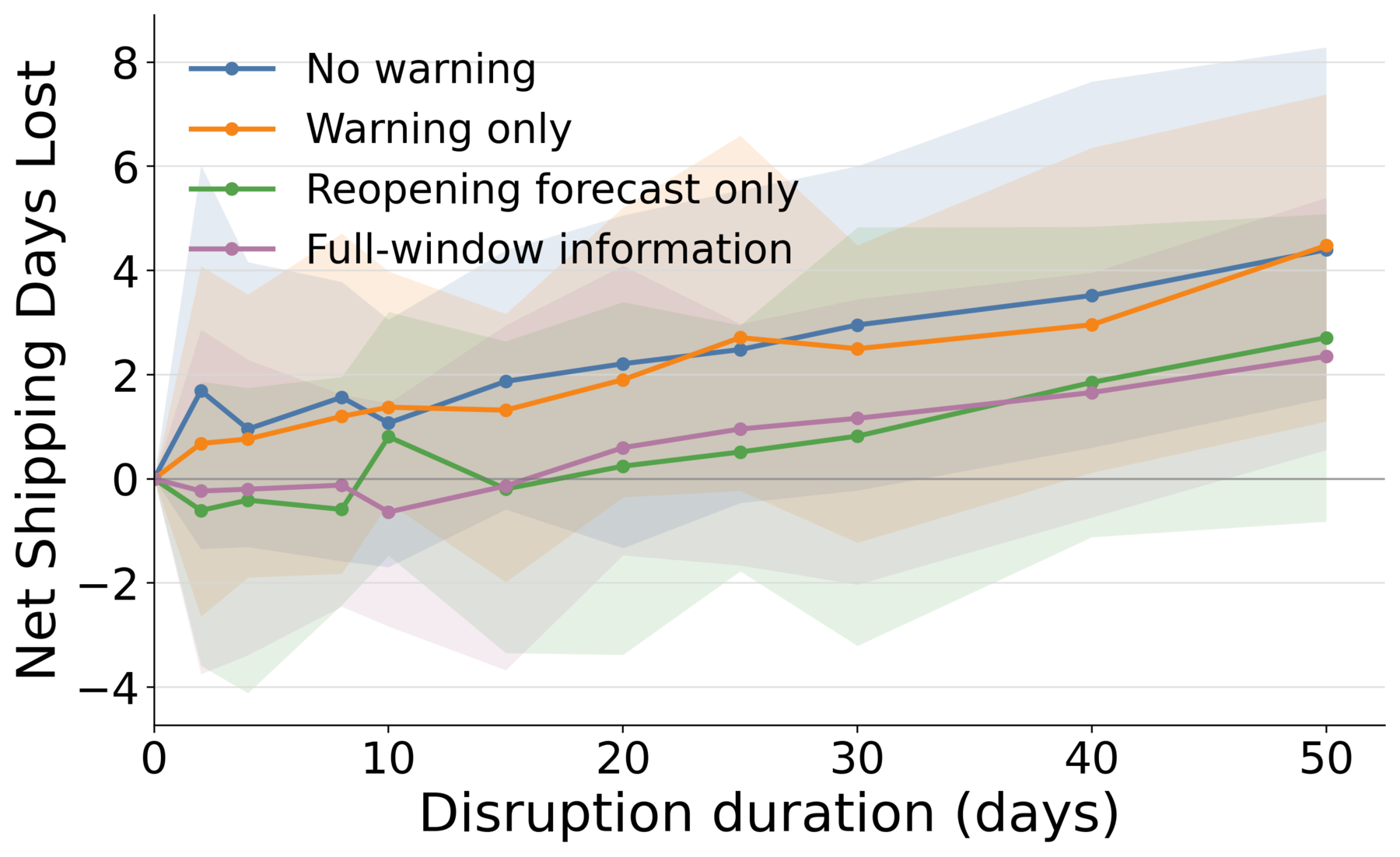}
\end{minipage} &
\begin{minipage}[t]{0.32\linewidth}\centering
\includegraphics[width=\linewidth,height=0.16\textheight,keepaspectratio,trim=8 8 8 8,clip]{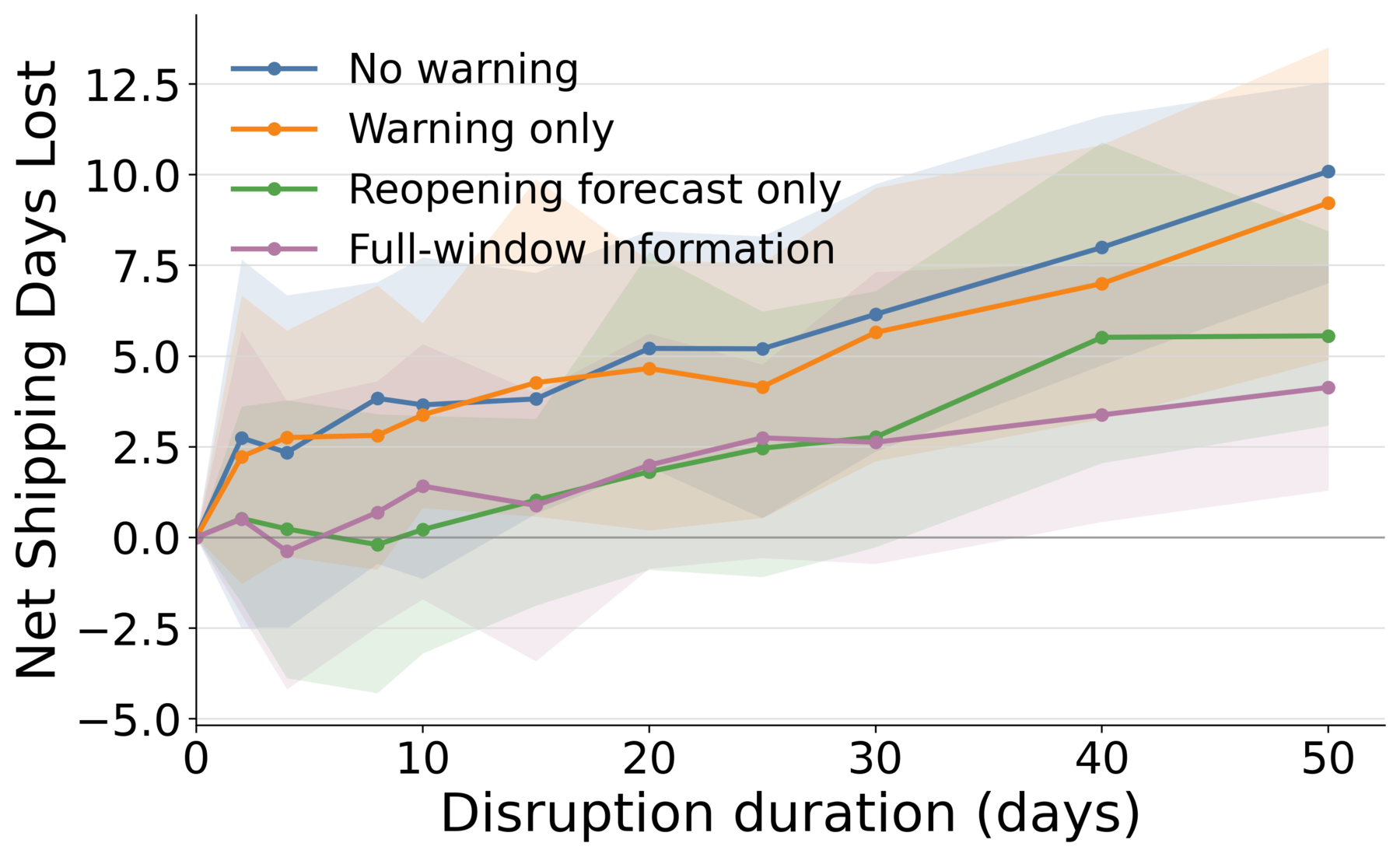}
\end{minipage}
\end{tabular}
\caption{Selected port-level disruption metrics for Singapore, Rotterdam, and
Houston. Rows show, from top to bottom, maximum arrival shortfall across closure
durations, net shipping-days lost across closure durations, maximum arrival
shortfall under information regimes, and net shipping-days lost under
information regimes.}
\label{fig:selected_ports_results}
\end{figure}
\FloatBarrier

\section{Static Slope Comparison}

The static benchmark measures structural route exposure from the two-step
Markov model: expected next-leg flows from previous port $i$ and current port
$j$ to destination port $k$ are assigned to undisrupted shortest paths. The
adaptive simulations measure realized arrival losses after rerouting, queues,
longer trips, and delayed vessel cycles. We use static exposure as a structural
benchmark for the adaptive arrival-loss estimates.

For the duration-slope comparison, we fit adaptive normalized net shipping-day
loss over the long-duration window of 30, 40, and 50 days. We then compare this
fit with the corresponding second-order total static daily exposure. The
adaptive net-loss slope in percentage-point units is shown in the main text;
here we report the corresponding dimensionless adaptive/static ratio.

\begin{figure}[p]
\centering
\includegraphics[width=0.96\linewidth]{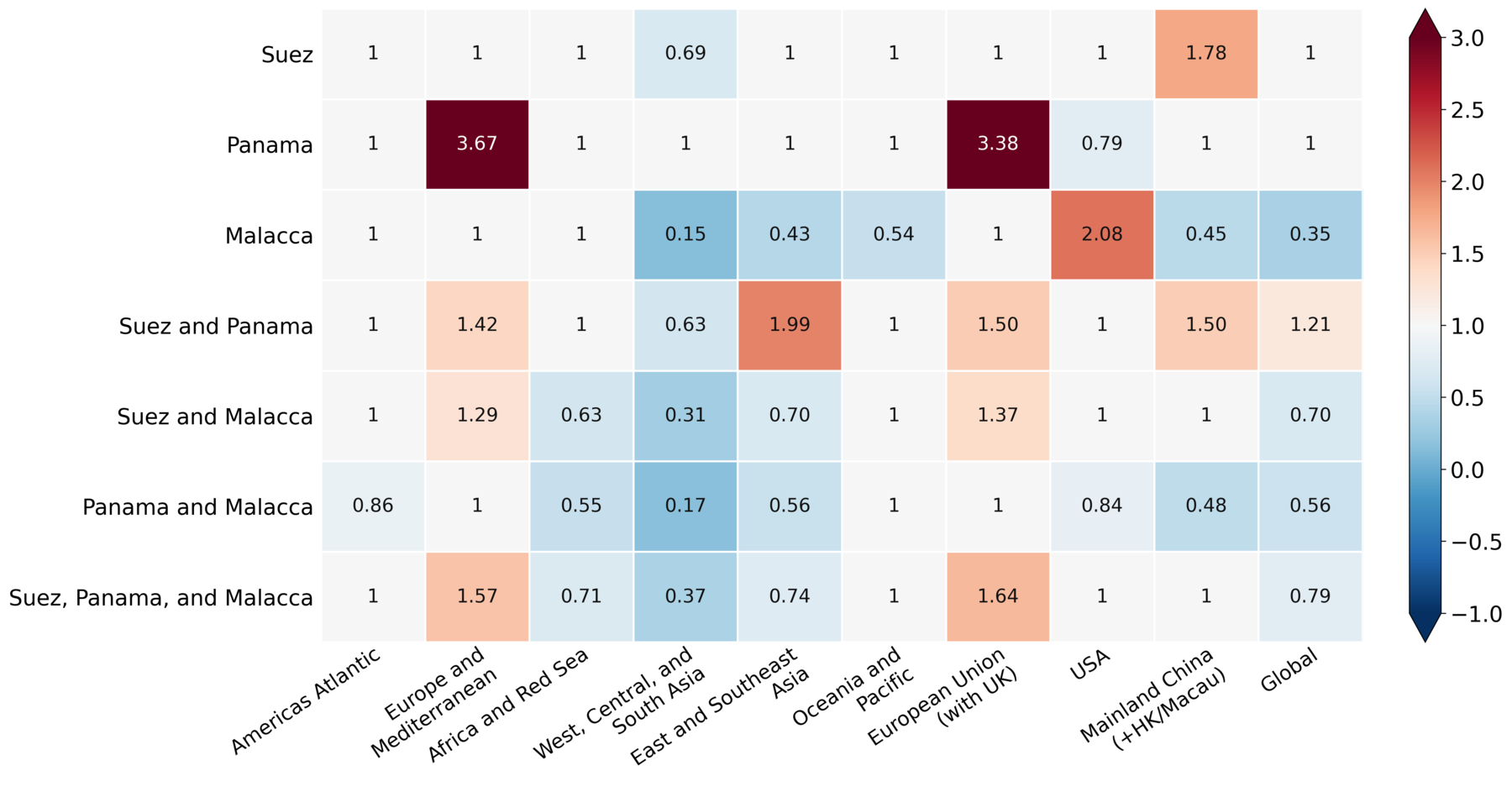}
\caption{Regional adaptive/static net-loss slope ratio over the 30-, 40-, and
50-day duration window using the second-order total static benchmark. Values
below one indicate that static exposure is larger than the adaptive net-loss
slope; values above one indicate larger realized adaptive losses than static
exposure. Negative values indicate a net catch-up or surplus slope over the
30-, 40-, and 50-day window.}
\label{fig:static-slope-ratio-si}
\end{figure}

\FloatBarrier

\begin{table}[p]
\centering
\small
\caption{Global endpoint values for the main scenario curves. Values report the
50-day endpoint mean with the 5th--95th percentile seed interval in brackets.
Maximum arrival shortfall is reported as a percentage; net loss is reported in
net shipping days.}
\label{tab:global-endpoint-values}
\resizebox{\textwidth}{!}{%
\begin{tabular}{llrr}
\toprule
Panel & Scenario or information regime & Maximum arrival shortfall & Net shipping-days lost \\
\midrule
Baseline & Suez & 5.26 [5.15, 5.36] & 2.18 [2.11, 2.25] \\
Baseline & Panama & 3.36 [3.26, 3.44] & 1.57 [1.49, 1.65] \\
Baseline & Malacca & 3.30 [3.19, 3.40] & 1.29 [1.21, 1.36] \\
Baseline & Suez--Panama & 7.87 [7.75, 7.98] & 3.71 [3.63, 3.78] \\
Baseline & Suez--Malacca & 7.27 [7.15, 7.38] & 3.27 [3.19, 3.34] \\
Baseline & Panama--Malacca & 5.77 [5.63, 5.91] & 2.70 [2.61, 2.79] \\
Baseline & Suez--Panama--Malacca & 9.73 [9.60, 9.85] & 4.62 [4.55, 4.69] \\
Information & No warning & 9.73 [9.60, 9.85] & 4.62 [4.55, 4.69] \\
Information & Warning only & 8.55 [8.44, 8.66] & 4.43 [4.34, 4.52] \\
Information & Reopening forecast only & 9.67 [9.55, 9.80] & 3.78 [3.69, 3.87] \\
Information & Full-window information & 8.40 [8.24, 8.55] & 3.54 [3.46, 3.64] \\
\bottomrule
\end{tabular}%
}
\end{table}
\FloatBarrier

\section{Regions}

\begingroup
\scriptsize
\setlength{\tabcolsep}{3pt}
\renewcommand{\arraystretch}{0.92}
\begin{longtable}{p{0.21\linewidth} p{0.24\linewidth} p{0.47\linewidth}}
\caption{Regional groupings used in the regional panels. The six broad regions aggregate United Nations M49 subregions. The European Union with the United Kingdom, USA, and Mainland China with Hong Kong and Macau are non-exclusive custom aggregates reported separately.}
\label{tab:regional_groupings}\\
  \hline
  \textbf{Used region} & \textbf{Constituent region} & \textbf{ISO3 codes} \\
  \hline
\endfirsthead
  \hline
  \textbf{Used region} & \textbf{Constituent region} & \textbf{ISO3 codes} \\
  \hline
\endhead
  Americas Atlantic
  & Northern America
  & BMU, CAN, GRL, SPM, USA \\

  & Latin America and the Caribbean
  & ABW, AIA, ARG, ATG, BES, BHS, BLM, BLZ, BOL, BRA, BRB, BVT, CHL, COL, CRI, CUB, CUW, CYM, DMA, DOM, ECU, FLK, GLP, GRD, GTM, GUF, GUY, HND, HTI, JAM, KNA, LCA, MAF, MEX, MSR, MTQ, NIC,
  PAN, PER, PRI, PRY, SGS, SLV, SUR, SXM, TCA, TTO, URY, VCT, VEN, VGB, VIR \\
  \hline

  Europe and Mediterranean
  & Northern Europe
  & ALA, DNK, EST, FIN, FRO, GBR, GGY, IMN, IRL, ISL, JEY, LTU, LVA, NOR, SJM, SWE \\

  & Western Europe
  & AUT, BEL, CHE, DEU, FRA, LIE, LUX, MCO, NLD \\

  & Eastern Europe
  & BGR, BLR, CZE, HUN, MDA, POL, ROU, RUS, SVK, UKR \\

  & Southern Europe
  & ALB, AND, BIH, ESP, GIB, GRC, HRV, ITA, MKD, MLT, MNE, PRT, SMR, SRB, SVN, VAT \\
  \hline

  Africa and Red Sea
  & Northern Africa
  & DZA, EGY, ESH, LBY, MAR, SDN, TUN \\

  & Sub-Saharan Africa
  & AGO, ATF, BDI, BEN, BFA, BWA, CAF, CIV, CMR, COD, COG, COM, CPV, DJI, ERI, ETH, GAB, GHA, GIN, GMB, GNB, GNQ, IOT, KEN, LBR, LSO, MDG, MLI, MOZ, MRT, MUS, MWI, MYT, NAM, NER, NGA, REU,
  RWA, SEN, SHN, SLE, SOM, SSD, STP, SWZ, SYC, TCD, TGO, TZA, UGA, ZAF, ZMB, ZWE \\
  \hline

  West, Central, and South Asia
  & Central Asia
  & KAZ, KGZ, TJK, TKM, UZB \\

  & Southern Asia
  & AFG, BGD, BTN, IND, IRN, LKA, MDV, NPL, PAK \\

  & Western Asia
  & ARE, ARM, AZE, BHR, CYP, GEO, IRQ, ISR, JOR, KWT, LBN, OMN, PSE, QAT, SAU, SYR, TUR, YEM \\
  \hline

  East and Southeast Asia
  & Eastern Asia
  & CHN, HKG, JPN, KOR, MAC, MNG, PRK, TWN \\

  & South-eastern Asia
  & BRN, IDN, KHM, LAO, MMR, MYS, PHL, SGP, THA, TLS, VNM \\
  \hline

  Oceania and Pacific
  & Australia and New Zealand
  & AUS, CCK, CXR, HMD, NFK, NZL \\

  & Melanesia
  & FJI, NCL, PNG, SLB, VUT \\

  & Micronesia
  & FSM, GUM, KIR, MHL, MNP, NRU, PLW, UMI \\

  & Polynesia
  & ASM, COK, NIU, PCN, PYF, TKL, TON, TUV, WLF, WSM \\
  \hline

  European Union (with UK)
  & Custom
  & AUT, BEL, BGR, CYP, CZE, DEU, DNK, ESP, EST, FIN, FRA, GBR, GRC, HRV, HUN, IRL, ITA, LTU, LUX, LVA, MLT, NLD, POL, PRT, ROU, SVK, SVN, SWE \\
  \hline

  USA
  & Custom
  & USA \\
  \hline

  Mainland China (+HK/Macau)
  & Custom
  & CHN, HKG, MAC \\

  \hline
\end{longtable}
\endgroup

\section{Model Limitations Discussion}

The model is deliberately parsimonious. It assumes a constant vessel speed, so
it cannot capture speed adjustments due to weather, fuel management, fuel
prices, operator strategy, or schedule recovery. It assumes stationary
transition probabilities, so costs, delays, congestion signals, and policy
signals do not feed back into the Markov kernel that generates destination
choices. Port operations remain stylized through a single service-capacity
parameter and a single shared queue at each port, with service and idle times
represented by calibrated stochastic distributions. Short-run operational
responses such as overtime, berth reassignment, priority rules, tug and pilot
constraints, or carrier-specific dispatch decisions are therefore not
represented.

The results should be read as short-run arrival-flow disruptions, not as
welfare losses, trade-value losses, supply-chain value-added losses, or long-run
market adjustments. The model treats maritime mobility as weakly coupled to
trade demand. It does not represent freight prices \cite{pirrong_contracting_1993,brancaccio_geography_2020},
cargo values, inventories, production responses, endogenous demand, or
seasonality \cite{teo_unveiling_2025}. Over longer horizons, market forces can
move ships, cargo, and supply chains onto new routes \cite{tavasszy_strategic_2011,brancaccio_geography_2020,kosowska-stamirowska_network_2020}.
The fixed Markov destination model is not designed to capture that structural
adaptation.

\section{Data Analysis and Filtering}

\subsection{Historical Port-Call Data}

Historical port-call data were derived from vessel track data, Automatic
Identification System (AIS), spanning January 1, 2019, to April 27, 2025. The
AIS data processing, port-call inference, port identifier assignment, vessel
type classification, and handling of missing or erroneous ship dimensions
followed the PortWatch methodology \cite{verschuur2021global,verschuur_ports_2022}.
The dataset comprises commercial vessels engaged in international trade,
including five categories: tankers, dry bulk vessels, general cargo vessels,
container vessels, and ro-ro vessels. For the purpose of the analysis we
aggregate general cargo, container, and ro-ro vessels under a generic cargo
ship type.

We clean the data by first removing all ships whose IMO and MMSI number is
missing. This leaves 70,457 unique vessels. We then filter out all vessels with
less than 10 port calls in order to remove low-activity vessels. The size
analysis also excludes 27 vessels with missing width or length information. The
remaining 59,725 ships are split into three size categories based on area,
computed from the width and length provided in the AIS data. We categorize the
smallest third of ships as small and the largest third of ships as large. The
rest is categorized as medium.

\begin{table}[htbp]
\caption{Aggregate unique ships and port calls per ship type and size across
the full January 2019--April 2025 historical dataset. Ships are deduplicated
across years.}
\label{tab:si_type_size_all_years_dedup}
\centering
\begin{tabular}{llrr}
\toprule
\textbf{Type} & \textbf{Size} & \textbf{Number of Ships} & \textbf{Number of port calls} \\
\midrule
Cargo & Small & 8,793 & 1,619,157 \\
Cargo & Medium & 7,873 & 2,070,504 \\
Cargo & Large & 6,757 & 1,700,230 \\
Cargo & \textbf{Total} & \textbf{23,423} & \textbf{5,389,891} \\
\midrule
Dry Bulk & Small & 723 & 191,162 \\
Dry Bulk & Medium & 3,459 & 380,964 \\
Dry Bulk & Large & 13,193 & 960,753 \\
Dry Bulk & \textbf{Total} & \textbf{17,375} & \textbf{1,532,879} \\
\midrule
Tanker & Small & 5,429 & 1,455,793 \\
Tanker & Medium & 5,128 & 821,063 \\
Tanker & Large & 8,370 & 692,160 \\
Tanker & \textbf{Total} & \textbf{18,927} & \textbf{2,969,016} \\
\midrule
Combined & \textbf{Total} & \textbf{59,725} & \textbf{9,891,786} \\
\bottomrule
\end{tabular}
\end{table}

Table~\ref{tab:si_type_size_all_years_dedup} describes the all-years historical
sample used for estimating routing and descriptive size categories. The
scenario simulations use a smaller active fleet, because only a subset of ships
is active in any single year. For the number of ships present in the simulation
we use the 2024 active fleet: ships with at least 10 port calls in the reference
year 2024. This fleet size is shown in Table~\ref{tab:si_unique_ships}.

\begin{table}[htbp]
\caption{Active 2024 fleet size used in scenario simulations.}
\label{tab:si_unique_ships}
\centering
\begin{tabular}{lr}
\toprule
\textbf{Type} & \textbf{Count} \\
\midrule
Cargo & 14,621 \\
Dry Bulk & 10,303 \\
Tanker & 11,030 \\
\midrule
Total & 35,954 \\
\bottomrule
\end{tabular}
\end{table}

\section{Markov Validation}

\subsection{Markov Modelling of Port Sequences}

Starting from the cleaned port-call sequences described above, we model
next-port choices with finite-order discrete-time Markov chains
\cite{guo_trajectory_2018,marten2020scalable,spadon_learning_2025,fu_multi-scale_2024}.
For each ship type we fit orders $k\in\{0,1,2,3\}$. Each configuration first
defines an evaluation port set $S$: either the top 50 ports or the full
observed set. The top ports are
computed based on average ship arrivals per day. All predictive losses are
computed only on transitions whose realized next port lies in $S$ so that rare
or low-support ports contribute consistently across model orders.

Let $s_1,\dots,s_T$ be a sequence after filtering to $S$. An order-$k$ model
specifies
\[
P(s_t \mid s_{t-1},\dots,s_{t-k}) =
\frac{N(s_{t-k},\dots,s_{t-1},s_t) + \alpha\,\pi(s_t)}
{\sum_{x\in S} N(s_{t-k},\dots,s_{t-1},x) + \alpha},
\]
where $N(\cdot)$ are transition counts extracted for that slice. We set
$\alpha=1.0$, and $\pi(\cdot)$ is the uniform distribution on $S$. This keeps
support identical across models within a scenario. By convention, a zero-order
model predicts the order-0 marginal distribution on $S$.

The ABM uses the fitted vessel-type order-2 transition tables directly. The
tables are loaded, normalized by row, and sampled directly. Histories outside
the fitted order-2 table use the current-port marginal options for that vessel
type.

\paragraph*{Sequence-to-evaluation alignment.}
Raw trajectories may contain ports not in $S$ when we only consider the most
frequented ports. In that case we drop ports outside $S$ but keep the remaining
chain if it still yields transitions in $S$, meaning that we skip intermediate
ports.

\paragraph*{Scoring and back-off.}
Higher orders may face unseen histories. At scoring time we therefore back off
from order $k$ to $k-1$, then to shorter histories, and finally to the
order-0 marginal distribution on $S$. Back-off is applied after alignment and
only to assign nonzero probabilities to evaluated transitions.

\paragraph*{Estimation window.}
Models are trained on trajectories up to and including 2023, using the
alignment rule, evaluation set $S$, and order $k$. We estimate in parallel
models fitted on a single ship type, a pooled model fitted on all types but
evaluated on the same $S$.

\paragraph*{Validation logic.}
The validation asks whether recent port history improves next-port prediction
relative to simpler alternatives. For each held-out ship, we compare the
probability assigned to the actually observed next port under models with
different memory lengths. If an order-1 model improves over an order-0 model,
then knowing the current port helps. If an order-2 model improves over an
order-1 model, then the previous two ports contain additional predictive
information. We repeat this comparison for the top-50-port evaluation set and
for the full observed port set, and we separately compare vessel-type-specific
models against a pooled model fitted across vessel types. This separates three
questions: how much memory is useful, whether results depend on restricting the
score to high-support ports, and whether vessel-type conditioning improves
prediction.

\paragraph*{Cross-validation.}
Predictive performance is measured with ship-level $K$-fold cross-validation
($K=20$). Ships are partitioned, models are refitted on $K-1$ folds, and the
held-out fold is scored transition-wise only when the realized next port is in
$S$. The metric is the per-transition predictive log-likelihood
\[
\text{PLL} = \frac{1}{N} \sum_{i=1}^{N} \log p(x_i \mid h_i), \quad x_i \in S,
\]
with $h_i$ the aligned history and $N$ the number of evaluated transitions. We
retain PLL per ship for bootstrap resampling.

\paragraph*{Temporal validation.}
To test forward performance, models trained on 2023 data are evaluated on 2024
and on January--April 2025 sequences, using the same $S$, alignment rule, and
back-off. A season-matched variant trains on January--April 2024 and evaluates
on January--April 2025. Perplexity $\exp(-\text{PLL})$ is reported from the
same scores.

\paragraph*{Pooled vs. type-specific effects.}
For each held-out ship $s$ with $n_s$ evaluated transitions we compute
\[
LL_{m,k}(s) = \sum_{(h,x)\in s,\, x\in S} \log
\frac{c_m(h,x)+\alpha\,\pi(x)}{c_m(h,\cdot)+\alpha},
\]
for $m$ equal to the model fitted on that ship type or the pooled model. The
per-transition gain is
\[
d_s = \frac{LL_{\text{type},k}(s) - LL_{\text{pooled},k}(s)}{n_s}.
\]
We average $d_s$ across ships using $n_s$ as weights and obtain percentile
confidence intervals and two-sided $p$-values from $B=100$ bootstrap resamples
of ships. Holm correction is applied within each scenario, order, strategy, and
$\alpha$.

\paragraph*{Markov order comparison.}
Adjacent orders $(0,1)$, $(1,2)$, and $(2,3)$ are compared on the same
predictive log-likelihood values, always restricted to $x\in S$. For ships with
scores under both orders we form the predictive log-likelihood difference per
transition and bootstrap across ships ($B=100$). Holm correction is again
applied to control for multiple comparisons. This identifies the highest order
that yields detectable predictive gains once low-support ports are excluded
from the loss. The scenario simulations use vessel-type-specific order-2
Markov models.

\begin{figure}[htbp]
\centering
\begin{minipage}{0.48\textwidth}
\centering
\includegraphics[width=\linewidth]{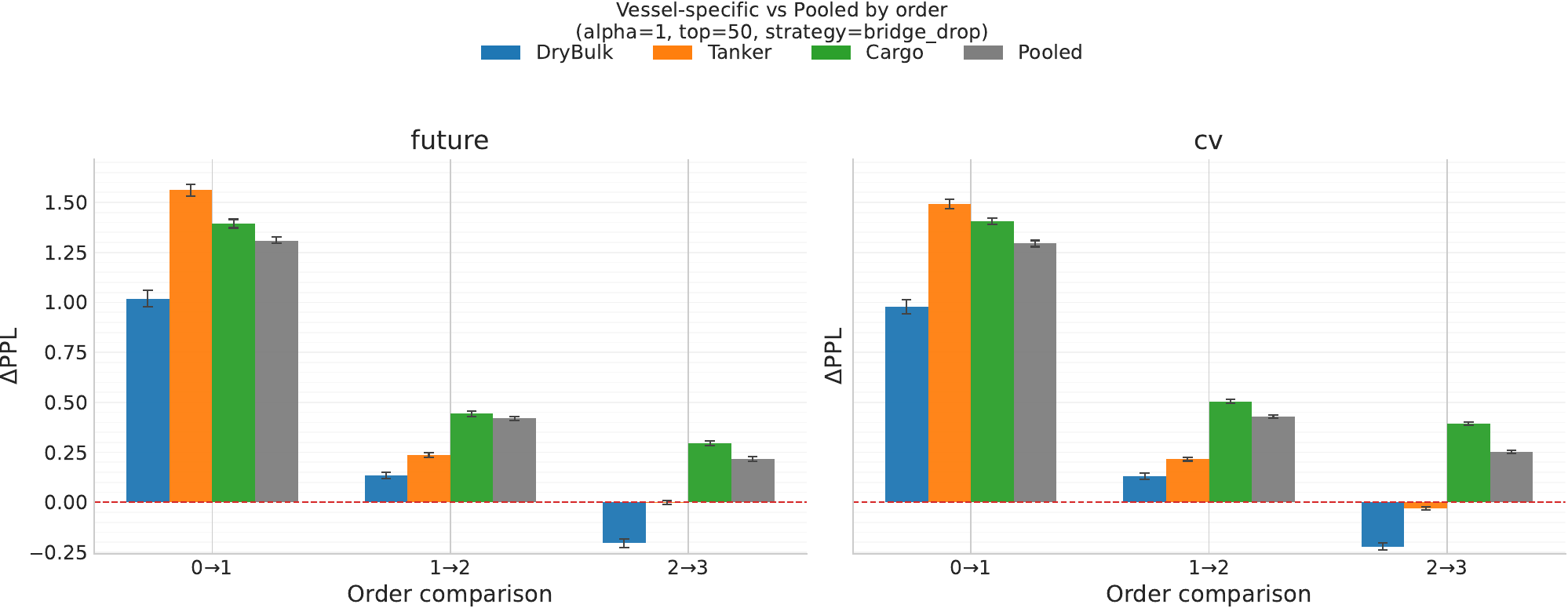}
\small A. Order gains for the top 50 ports
\end{minipage}\hfill
\begin{minipage}{0.48\textwidth}
\centering
\includegraphics[width=\linewidth]{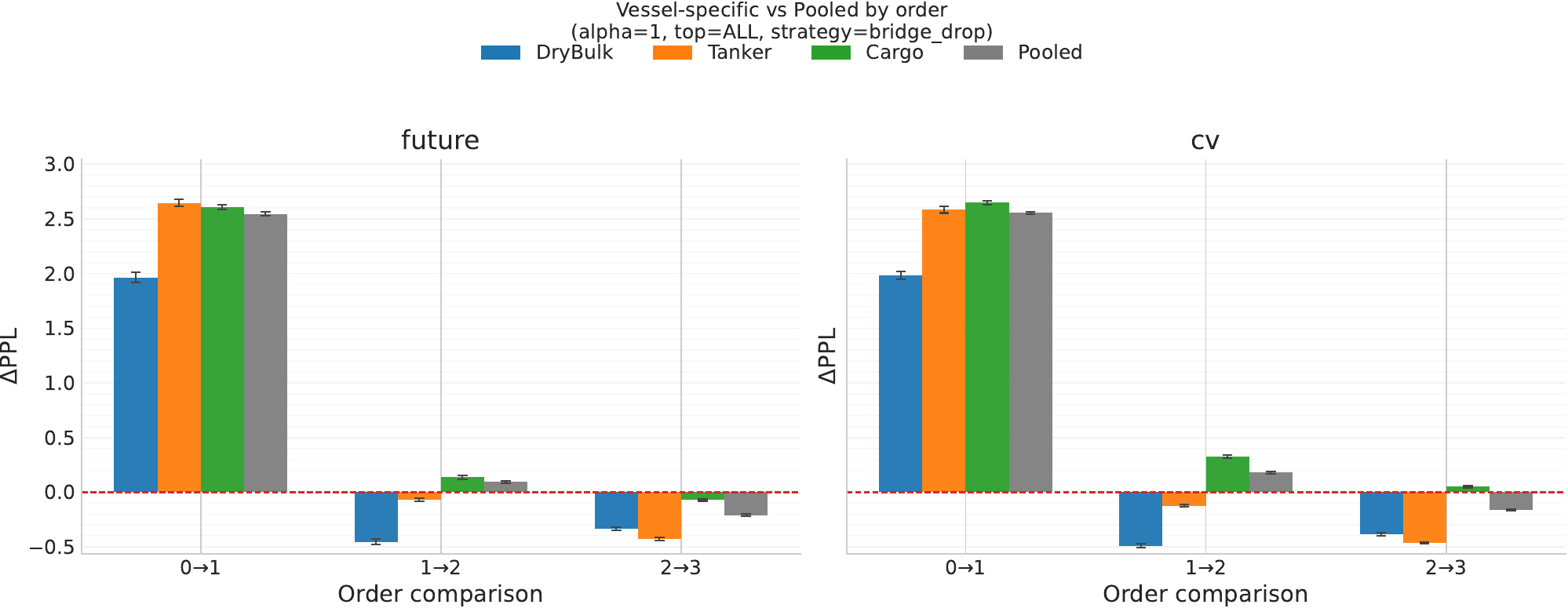}
\small B. Order gains for all ports
\end{minipage}

\vspace{0.6em}

\begin{minipage}{0.48\textwidth}
\centering
\includegraphics[width=\linewidth]{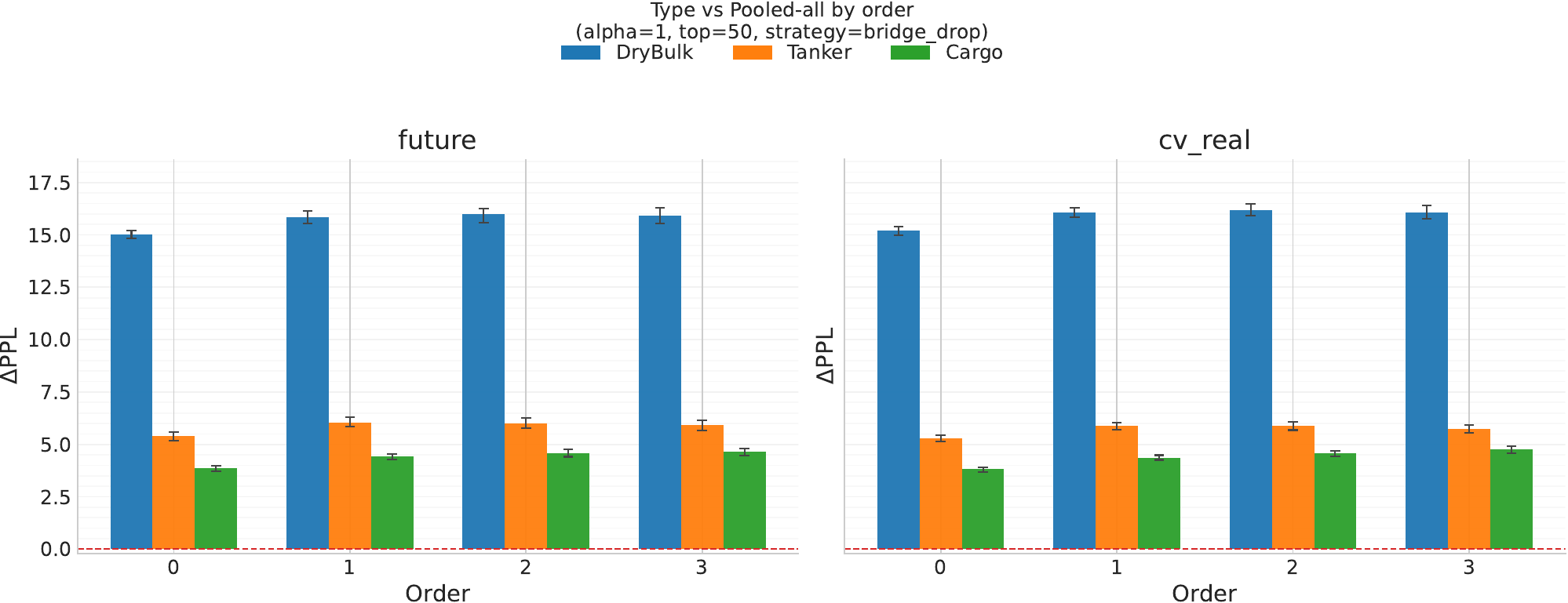}
\small C. Perplexity by Markov order
\end{minipage}\hfill
\begin{minipage}{0.48\textwidth}
\centering
\includegraphics[width=\linewidth]{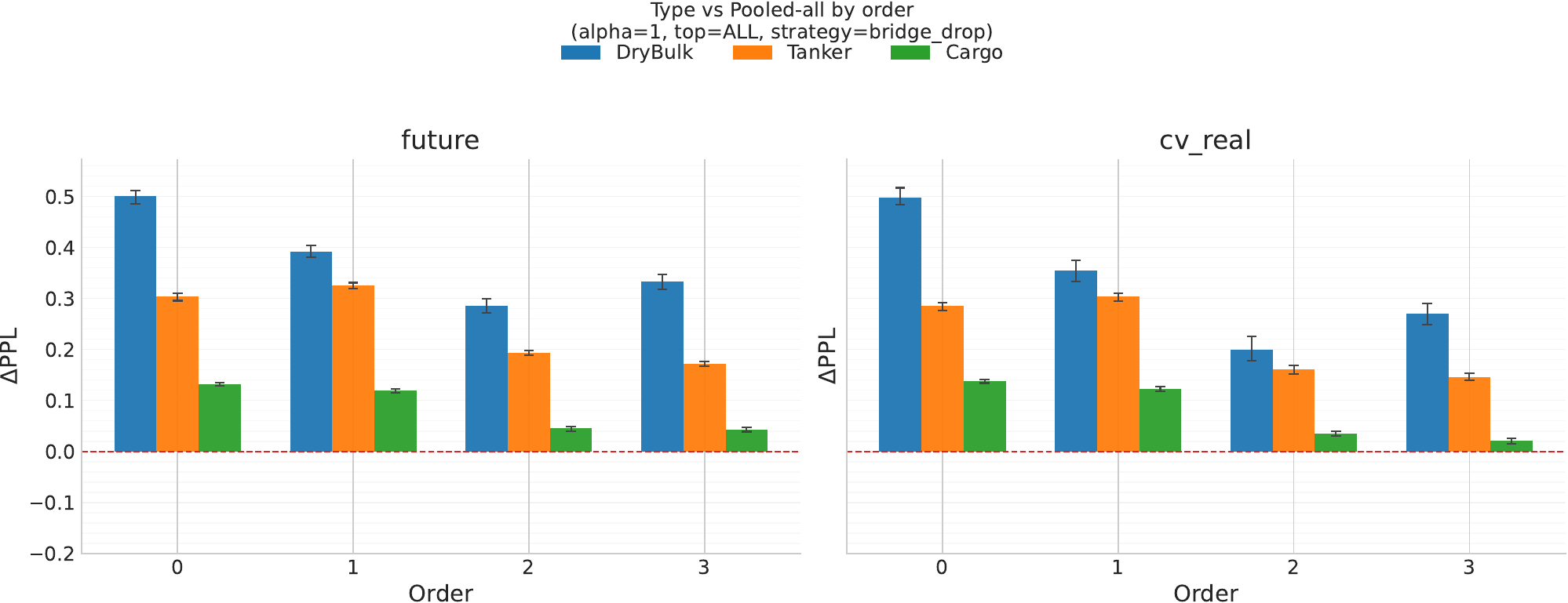}
\small D. Type-specific gain relative to pooled
\end{minipage}

\caption{Markov order and vessel-type effects. Panels A and B show the change
in predictive performance when increasing Markov order, evaluated for forward
evaluation and cross-validation, and for vessel-type-specific models and a
pooled model. Panel A restricts evaluation to the 50 largest ports. Panel B
evaluates on the full port set. Panel C shows perplexity by Markov order for
vessel-type-specific models. Panel D shows the improvement of
vessel-type-specific models relative to a pooled model across orders. Error bars
denote uncertainty from ship-level resampling as described above.}
\label{fig:si_markov_validation}
\end{figure}

\FloatBarrier
\section{Model Details}

\subsection{Agent-Based Model}

\paragraph*{Network representation.}
The environment of the model is an undirected weighted network. The network is
based on SeaRoute, with the addition of global ports
\cite{eurostat_searoute,verschuur2021global}. Since all ships are assumed to
have the same speed, the weight of each edge is the traversal time in days.

\paragraph*{Port dynamics.}
Each port has a single service-capacity parameter and a single shared queue.
Capacity is calibrated from the historical 90th percentile of the total number
of ships present in a day, where a ship is considered present if it is between
port calls, including the days of arrival and departure. Additional ships wait
for a service slot to become available before being processed. The time a ship
takes to be processed at a port is sampled from an exponential distribution
calibrated so that its mean matches the mean observed port-stay duration for
that ship type and port combination. We compute per-port, per-type service
times from AIS data by measuring the time between port arrival and departure
timestamps.

\paragraph*{Ship behaviour.}
The agents in our model are individual ships. We assume all ships have the same
speed, but allow for three different ship types: cargo, tanker, and dry bulk.
All ships behave in the same manner; the difference is the Markov model they
use and the type-specific service and idle-time parameters they encounter.
Ships are initialized from Markov-consistent order-2 histories sampled from the
stationary distribution over observed two-port histories. The simulator uses
the most recent sampled port as the current port, samples the initial
destination from the vessel-type Markov model, computes the corresponding A*
route, and places the ship along that route. Initial movement times are drawn
uniformly from the first 30 model days.

Once at a port, a ship is processed by the port and then generates its next
destination from the vessel-type-specific order-2 Markov model conditional on
its immediate history. If the sampled destination is the current port, the ship
idles at that port. Idling time is sampled from a calibrated stochastic
distribution, and the ship does not occupy a queue space while idling. If the
next destination is different from the current port, the ship finds the
shortest path using the A* algorithm on the marine network, with geographical
distance as the heuristic. Ships recompute routes immediately after network
capacity changes and also periodically during movement, using a 20-model-day
recalculation cadence while the ship is en route.

\paragraph*{Edge shock.}
Edge shocks temporarily remove edges from the network, making them impassable.
For chokepoint closures, we remove the edges corresponding to the Suez Canal,
Panama Canal, and Strait of Malacca. The shock has a duration, after which the
edge is returned to the network. After each closure is applied or lifted,
available ship trajectories are recomputed so that ships take the changed
marine network into account.

\subsection{Experiments}

\paragraph*{Parameters.}
Vessel speed is fixed at 10 nautical miles per hour in all simulations. The
fleet size used in scenario simulations is shown in
Table~\ref{tab:si_unique_ships}.

\paragraph*{Closures.}
For the blockade experiments, we impose temporary closures of the three major
maritime chokepoints, Suez, Panama, and Malacca, by removing the corresponding
edges from the network at model day $t=1200$. For chokepoints represented by
multiple edges, all relevant segments are removed. We sweep closure durations
of 1, 2, 4, 5, 6, 8, 10, 12, 15, 20, 30, 40, and 50 days. Each run starts at
model day 0 and is executed until model day 1500. Output analysis uses the
pre-shock baseline window from model day 800 to 1000. By default, each scenario
is replicated over 50 random seeds. The same list of random seeds is reused
across scenarios when the information-regime specification is fixed.

For the simultaneous Suez, Panama, and Malacca closure, we also consider four
information regimes: no information on the closure window, known beginning,
known end, and known beginning plus known end. Single-chokepoint and
pairwise-chokepoint scenarios use the no-information regime by default.

\subsection{Output Analysis and Metrics}

\paragraph*{Smoothing and normalization.}
The output of the model is daily arrivals of ships to individual ports. In the
reported analyses, an arrival is counted on the model day when the ship
completes service at the destination port. Due to the discrete nature of port
arrivals, these time series have high levels of day-to-day variance. We run
multiple simulations with different random seeds.
For each seed, we smooth the time series separately for pre-shock and
post-shock periods using a 7-day moving-average window. For windows that would
overlap the shock start or end, we use a smaller window that does not cross the
shock boundary. We then normalize each time series of port arrivals by dividing
by the steady-state mean daily arrivals $\bar A_i$ computed over the pre-shock
period, days 800 to 1000,
\[
x_i(d)=\frac{A_i(d)}{\bar A_i}.
\]
Regional smoothed time series are computed by aggregating all ports in a region
and then applying the same smoothing and normalization procedure.

\paragraph*{Maximum arrival shortfall.}
We compute maximum arrival shortfall as the largest positive shortfall below
the normalized baseline,
\[
\max\left(0, 1-\min_{d\in\mathcal{T}}x_i(d)\right),
\]
which captures the largest proportional shortfall over the evaluation horizon
$\mathcal{T}$ and is reported as a positive percentage loss.

\paragraph*{Net cumulative loss.}
For each normalized time series $x_i(d)$, we sum deficits below
$1-\sigma_i$ and subtract surpluses above $1+\sigma_i$, where $\sigma_i$ is
the standard deviation of the normalized pre-shock series. The reported unit is
net shipping days,
\[
L_i=
\sum_{d\in\mathcal{T}:x_i(d)<1-\sigma_i} \left(1-x_i(d)\right)
-
\sum_{d\in\mathcal{T}:x_i(d)>1+\sigma_i} \left(x_i(d)-1\right).
\]
This metric is the time-integrated normalized arrival deficit after later
arrival surpluses within the evaluation horizon are accounted for.

\paragraph*{Uncertainty and null comparisons.}
For ratio, slope, and arrival-shortfall analyses, we use simple null
hypotheses tied to the relevant baseline value. For adaptive--static ratios,
the null hypothesis is equality between the adaptive and static estimates,
$H_0:R=1$. For duration-response slopes, the null hypothesis is zero slope,
$H_0:\beta=0$; when dynamic slopes are compared with static per-day losses, the
corresponding null is
$H_0:\beta=s_i$, where $s_i$ is the static per-day loss estimate. For maximum
arrival shortfall, the null hypothesis is baseline-level disruption,
$H_0:D=0$. Uncertainty is summarized using the seed-level distribution for
ratios and normal 95\% intervals for slope and shortfall estimates. Null
hypotheses are treated as unsupported when the corresponding interval excludes
the null value.

\paragraph*{Static counterfactual.}
As a structural exposure benchmark, we compute static chokepoint exposure from
the two-step Markov model. Using transition probabilities $P^{(c)}_{ijk}$ from
previous port $i$ and current port $j$ to destination port $k$ for vessel class
$c$, we compute the stationary distribution $\pi^{(c)}_{ij}$ over two-port
histories. Each expected next-leg flow $\pi^{(c)}_{ij}P^{(c)}_{ijk}$ is
assigned to the shortest travel-time path from current port $j$ to destination
port $k$ on the undisrupted maritime network. For chokepoint $q$, vessel class
$c$, and destination port $k$, this gives
\[
E^{(c)}_{kq}
=
100
\frac{
\sum_{i,j:j\ne k}
\pi^{(c)}_{ij} P^{(c)}_{ijk}
\mathbf{1}\{q\in r(j,k)\}
}{
\sum_{i,j:j\ne k}
\pi^{(c)}_{ij} P^{(c)}_{ijk}
},
\]
where $r(j,k)$ is the shortest route from $j$ to $k$. This benchmark measures
the baseline share of expected arrivals whose fixed shortest route depends on
each chokepoint.

\section{Vessel Speed Table}

We use SeaRoute \cite{eurostat_searoute} to compute the sea distance between
consecutive port calls. Based on the elapsed time between port calls, we
estimate end-to-end speeds for ships by type. This end-to-end speed captures
sailing time plus any waiting outside port boundaries, such as anchorage areas,
that is not explicitly recorded in AIS data as port service time. Given the
limited variance between speeds of different vessel types, we use a constant
speed of 10 nautical miles per hour for all ships in the simulation as a coarse
approximation. Port delays, including queues and service times at ports, are
modeled separately through port capacity constraints and service-time
distributions. We leave explicit variable speeds to future research.

\begin{table}[htbp]
\caption{Average vessel speeds by type and distance range.}
\label{tab:si_speed_summary}
\centering
\begin{tabular}{lcc lcc}
\toprule
\textbf{Vessel Type} & \textbf{nm/h} & \textbf{km/h} &
\textbf{Distance Range} & \textbf{nm/h} & \textbf{km/h} \\
\midrule
Cargo & 11.14 & 20.63 & $<1000$ km & 9.8 & 18.15 \\
Tanker & 8.9 & 16.48 & 1000--4999 km & 11.0 & 20.37 \\
Dry Bulk & 9.0 & 16.67 & 5000--9999 km & 11.0 & 20.37 \\
 & & & $\geq10000$ km & 11.0 & 20.37 \\
\midrule
\textbf{Overall Average} & \textbf{10.0} & \textbf{18.52} & & & \\
\bottomrule
\end{tabular}
\end{table}

\FloatBarrier

\end{document}